\documentclass[epj]{svjour}
\usepackage{graphicx}
\usepackage{hyperref}
\begin{document}
\title{Heavy-ion collisions at the Large Hadron Collider: a review of the results from Run 1}
\titlerunning {Heavy-ion collisions at the Large Hadron Collider}
\author{N\'estor Armesto\inst{1} \and Enrico Scomparin\inst{2}% etc
}                     % Do not remove
\institute{Departamento de F\'{i}õsica de Part\'{i}culas and IGFAE, Universidade de Santiago de Compostela, 15706 Santiago de Compostela, Galicia-Spain \and  INFN, Sezione di Torino, 10125 Torino, Italy}
\date{Received: date / Revised version: date}
%\linenumbers
\abstract{
We present an overview of the results obtained in pPb and PbPb collisions at the Large Hadron Collider during Run 1. We first discuss the results for global characteristics: cross sections, hadron multiplicities, azimuthal asymmetries, correlations at low transverse momentum, hadrochemistry, and femtoscopy. We then review hard and electromagnetic probes: particles with high transverse momentum, jets, heavy quarks, quarkonium, electroweak bosons and high transverse momentum photons, low transverse momentum photons and dileptons, and ultraperipheral collisions. We mainly focus on the experimental results, and  present very briefly the main current theoretical explanations.
\PACS{
      {PACS-key}{discribing text of that key}   \and
      {PACS-key}{discribing text of that key}
     } % end of PACS codes
} %end of abstract
\maketitle
\section{Introduction}
\label{intro}

Quantum Chromodynamics (QCD) \cite{Fritzsch:1973pi} is the quantum field theory of the strong interaction, see the recent books \cite{book1,book2}. While its perturbative domain \cite{Gross:1973id,Politzer:1973fx} has been widely explored and calculations are constantly being improved towards higher orders in the expansion in the strong coupling constant, our knowledge on the non-perturbative regime is far from being satisfactory, see e.g. the review \cite{Greensite:2011zz}. Numerical lattice computations are currently able to describe the hadronic spectrum \cite{Fodor:2012gf} and even light nuclei, but not to deal with dynamic quantities like transport coefficients or cross sections, or with nuclei composed of more than a very few nucleons. Therefore, an analytic understanding of some basic features of QCD matter in its usual state, such as confinement and chiral symmetry breaking, is badly needed. One way of getting information about these aspects of QCD is by studying matter in phases where quarks and gluons are no more confined into hadrons and where chiral symmetry is restored \cite{Lee:1974ma,Collins:1974ky,Hofmann:1975by,Freedman:1976ub,Cabibbo:1975ig,Shuryak:1978ij}.

Such a phase of matter, named the Quark-Gluon Plasma (QGP, see the reviews \cite{qgp1,qgp2,qgp3,qgp4,Chesler:2015lsa,Ghiglieri:2015zma,Jeon:2015dfa,Ding:2015ona,Blaizot:2015lma,Gelis:2015gza,jqqgp5}), where the degrees of freedom are quarks and gluons, can be created by colliding heavy ions at the Large Hadron Collider (LHC) at CERN. Its detailed characterization should provide insight into the mentioned, yet unexplained features of QCD that are crucial for understanding hadron and nuclear properties, both  static (mass generation and spectra) and dynamic (cross sections). In addition, new high-density regimes of QCD are theoretically expected and signatures of their existence can be explored in both proton-nucleus and nucleus-nucleus collisions. Furthermore, such studies offer the possibility of investigating basic physics questions, like the number of interactions between particles required to consider a collection of them as a true medium and the corresponding transition from a microscopic to a macroscopic description, in a relativistic quantum system.

Experiments performed since 2000 at the Relativistic Heavy Ion Collider (RHIC) at BNL, where heavy-ions are collided up to a center-of-mass energy per nucleon-nucleon  interaction $\sqrt{s_{\rm NN}}=200$ GeV, have shown that a system is created~\cite{Gyulassy:2004zy,Muller:2006ee} in which particles are produced  with a large degree of coherence, 
far from a mere superposition of nucleon-nucleon collisions. The obtained energy densities well exceed those above which
lattice QCD calculations predict the formation of a QGP, $\epsilon \sim 1$ GeV/fm$^3$~\cite{Karsch:2001cy}. The measured particle spectra are reproduced by viscous relativistic hydrodynamics, if the hydrodynamic regime is assumed to hold at short times after the collision and the shear viscosity over entropy density ratio is small. The yields and expected back-to-back correlations of high-energy partons are strongly suppressed, compared to expectations from extrapolations from pp collisions and from perturbative QCD calculations that include nuclear effects not due to the presence of a hot deconfined medium. Such cold nuclear matter effects (CNM) include in primis the nuclear modification of parton densites (nPDF) \cite{Eskola:2009uj,deFlorian:2011fp}.
From these findings, it was argued \cite{Gyulassy:2004zy,Muller:2006ee} that the formed deconfined partonic medium behaves like an 
almost ideal liquid very shortly after the collision, and is very opaque to the high-energy partons traversing it.

\begin{figure}[htbp]
\begin{center}
\includegraphics[width=0.5\textwidth]{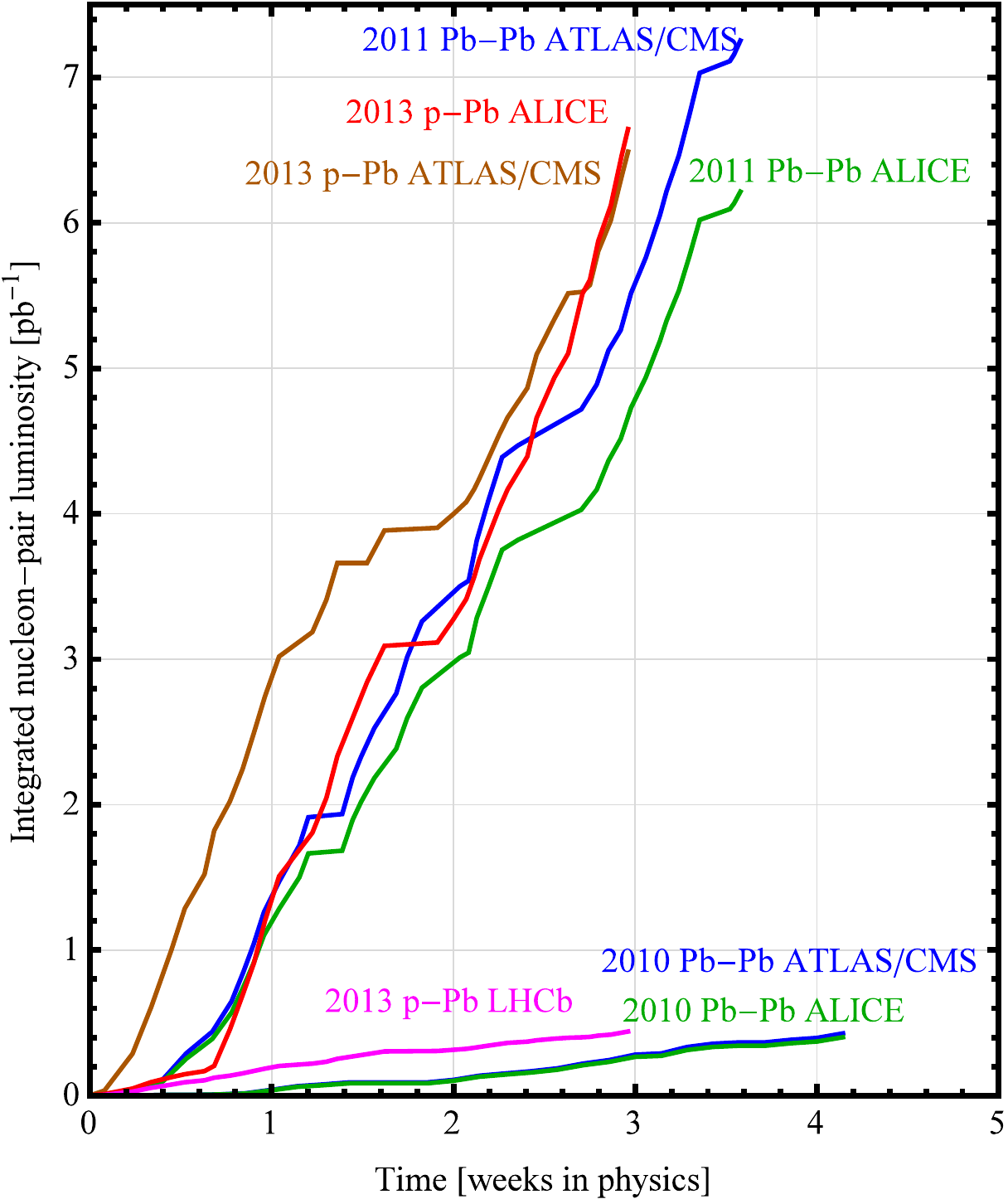}
\end{center}
\caption{Delivered integrated luminosity per nucleon-nucleon pair at the LHC for collisions involving ions, during Run 1. Taken from  \cite{Fischer:2014wfa}.} 
\label{fig:introduction1}
\end{figure}

At the LHC, the much larger $\sqrt{s_{\rm NN}}$ (up to 2.76 TeV for PbPb collisions) is expected to lead to the creation of a denser, hotter, longer-lived medium with a much more abundant rate of various rare particles that can be used to probe it. Previous to the first PbPb and pPb runs, predictions and proposals were compiled in
\cite{Abreu:2007kv,Armesto:2009ug,Salgado:2011wc,Albacete:2013ei}. In 2010 and 2011 two PbPb runs took place, and (after a pilot run in September 2012) a pPb run, at $\sqrt{s_{\rm NN}}=5.02$ TeV in early 2013. The corresponding delivered integrated luminosities per nucleon-nucleon pair are shown in Fig. \ref{fig:introduction1}, taken from \cite{Fischer:2014wfa}.

In this paper we review the main experimental results from the full Run 1  of the LHC with ions (other reviews can be found in
\cite{Muller:2012zq,Roland:2014jsa,Norbeck:2014loa}), as well as some of the theoretical discussions triggered by the measurements. The manuscript is organised as follows. In the next Section we discuss soft probes of the QGP: cross sections, hadron multiplicities and global event characteristics, azimuthal asymmetries, correlations at low transverse momentum, hadrochemistry, and intensity interferometry between identical particles (femtoscopy). In Section \ref{hard}, we review hard and electromagnetic probes at the LHC, namely: particle production at large transverse momentum, jets, heavy flavours, quarkonia, electroweak bosons and large transverse momentum photons, low transverse momentum photons and dileptons, and ultraperipheral collisions. Note that some of these processes, particularly when studied in pPb collisions, provide the required benchmarking for such hard probes. Needless to say, the organisation that we have chosen is not unique and the discussion of some observables will extend across different Subsections. Due to the experimental  findings that indicate that pPb collisions cannot be simply considered a reference system but show features in common with PbPb, we  address pPb results inside each item related to an observable as an integral part of the ion programme, without separating a specific Section for proton-nucleus. We conclude with a summary.

\section{Soft probes}
\label{soft}

One possibility to characterise  the medium created in high-energy heavy-ion collisions is through the use of external probes, which we will address in Section \ref{hard}. Nevertheless, properties of the medium can also be extracted from the behaviour of the bulk of particle production i.e. global characteristics like multiplicities or cross sections, correlations, chemical composition$\dots$ 
Such observables, that will be addressed in this Section, are called soft probes, as they are dominated by momentum scales of the order of the typical ones of the particles that compose it, as average transverse momentum\footnote{It is a standard practice to use temperature but it should be kept in mind that the question of the thermodynamical equilibration of the medium is still an open issue.}. We refer the reader to the reviews \cite{qgp1,qgp2,qgp3,qgp4,Abreu:2007kv,Armesto:2009ug} for extensive information.

%Information about the nature of the medium created in high-energy heavy-ion collisions can be obtained from the behaviour and properties of the medium itself e.g. global characteristics like multiplicities or cross sections, its dynamical behaviour, correlations, composition,$\dots$ These are the so-called {\it soft probes} of the medium, that we address in this Section, as they are observables dominated by momentum scales of the order of the ones of the medium, typically temperature - if the medium can be characterized by such a quantity We refer the reader to the reviews \cite{qgp1,qgp2,qgp3,qgp4,Abreu:2007kv,Armesto:2009ug} for extensive information.

\subsection{Cross sections}
\label{xsec}

A key quantity to be measured is the total inelastic (or at least non-single diffractive, NSD) cross section that affects the normalisation of many different observables, the translation from yields to cross sections and the determination of the centrality of the collisions. Such a measurement is challenging already in pp collisions. ALICE has measured~\cite{Abelev:2014epa} the visible cross section (i.e. the cross section for events with activity in the kinematic region covered by the experiment) in \mbox{pPb} collisions via van de Meer scans for various trigger conditions. In particular, asking for a coincidence between  particle production on both sides of the interaction point, in the pseudorapidity regions $2.8<\eta<5.1$ and $-3.7<\eta<-1.7$, the value $\sigma_{\rm pPb}= 2.09\pm0.07$ b has been obtained. This quantity was shown to correspond, within less than 1\%, to the NSD cross section~\cite{ALICE:2012xs}. CMS~\cite{Khachatryan:2015zaa} has performed an extrapolation to the total inelastic cross section, obtaining 2061 $\pm$ 3 (stat.) $\pm$ 34 (syst.) $\pm$ 72 (lumi.) mb. This estimate was shown to be in agreement with calculations of the same quantity performed in the frame of the Glauber model, starting from the pp inelastic cross section $\sigma_{\rm pp}= 70.0\pm1.5 $ mb~\cite{Antchev:2013iaa}.

Moving to \mbox{PbPb} collisions at $\sqrt{s_{\rm NN}}=2.76$ TeV, ALICE~\cite{ALICE:2012aa} has measured the cross section for electromagnetic dissociation (at large impact parameters beyond the range of the strong interaction) with neutron emission, a very important measurement for luminosity calibration and for understanding the lifetime of the ion beams. The obtained value, $\sigma_{\rm single\,EMD}=187.4 \pm 0.2 ({\rm stat}) ^{+13.2}_{-11.2}({\rm syst})$ b is much larger than the corresponding hadronic cross section, $\sigma_{\rm PbPb}=$ 7.7 $\pm$ 0.1 (stat.) $^{+0.6} _{-0.5}$ (syst.) b. The latter measurement has been used for centrality determination and validation of the Glauber model in \cite{Abelev:2013qoq}.

\subsection{Hadron multiplicities and global characteristics}
\label{multi}

\begin{figure}[htbp]
\centering
\resizebox{0.47\textwidth}{!}
{\includegraphics*{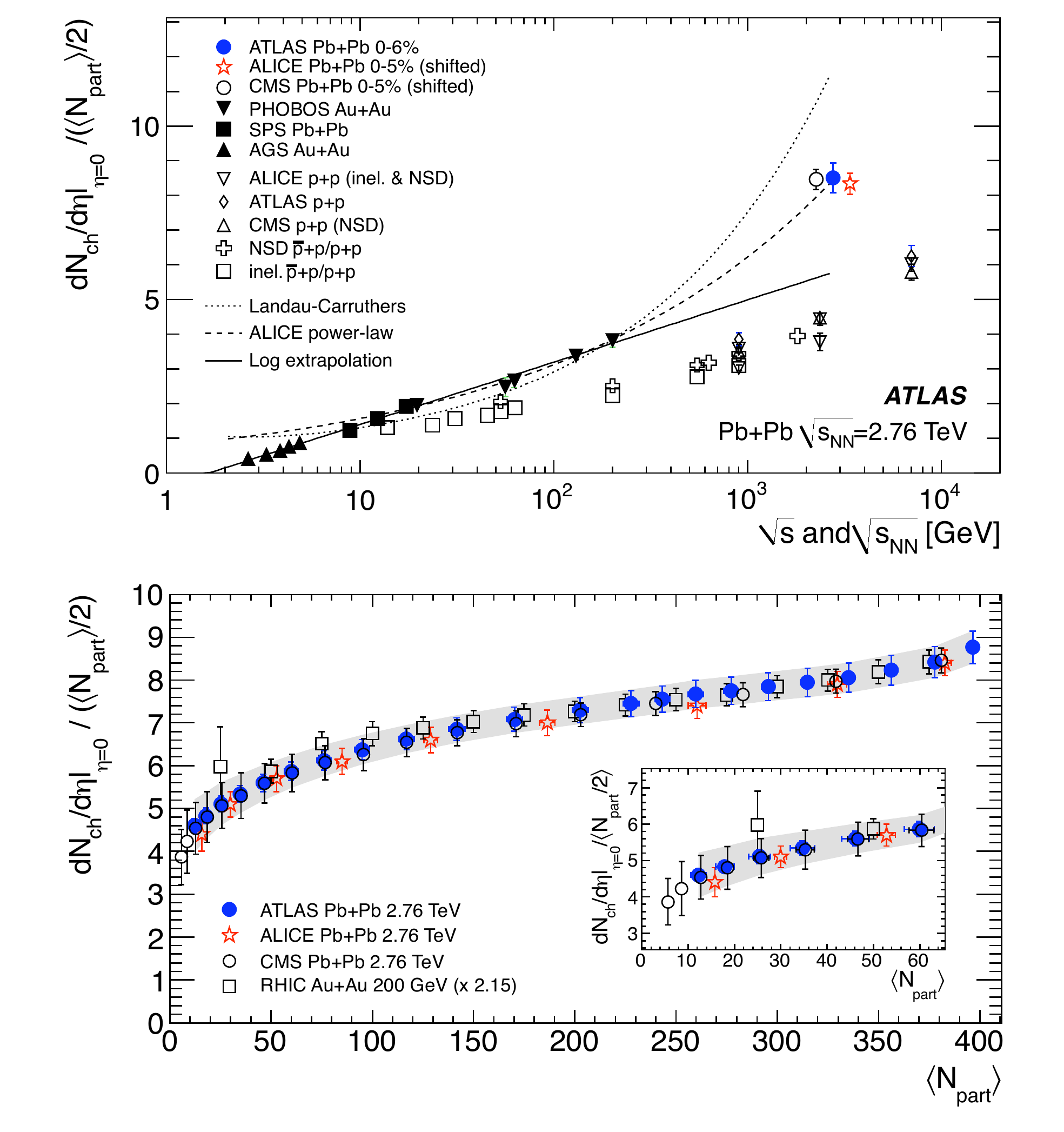}}
\resizebox{0.52\textwidth}{!}
{\includegraphics*{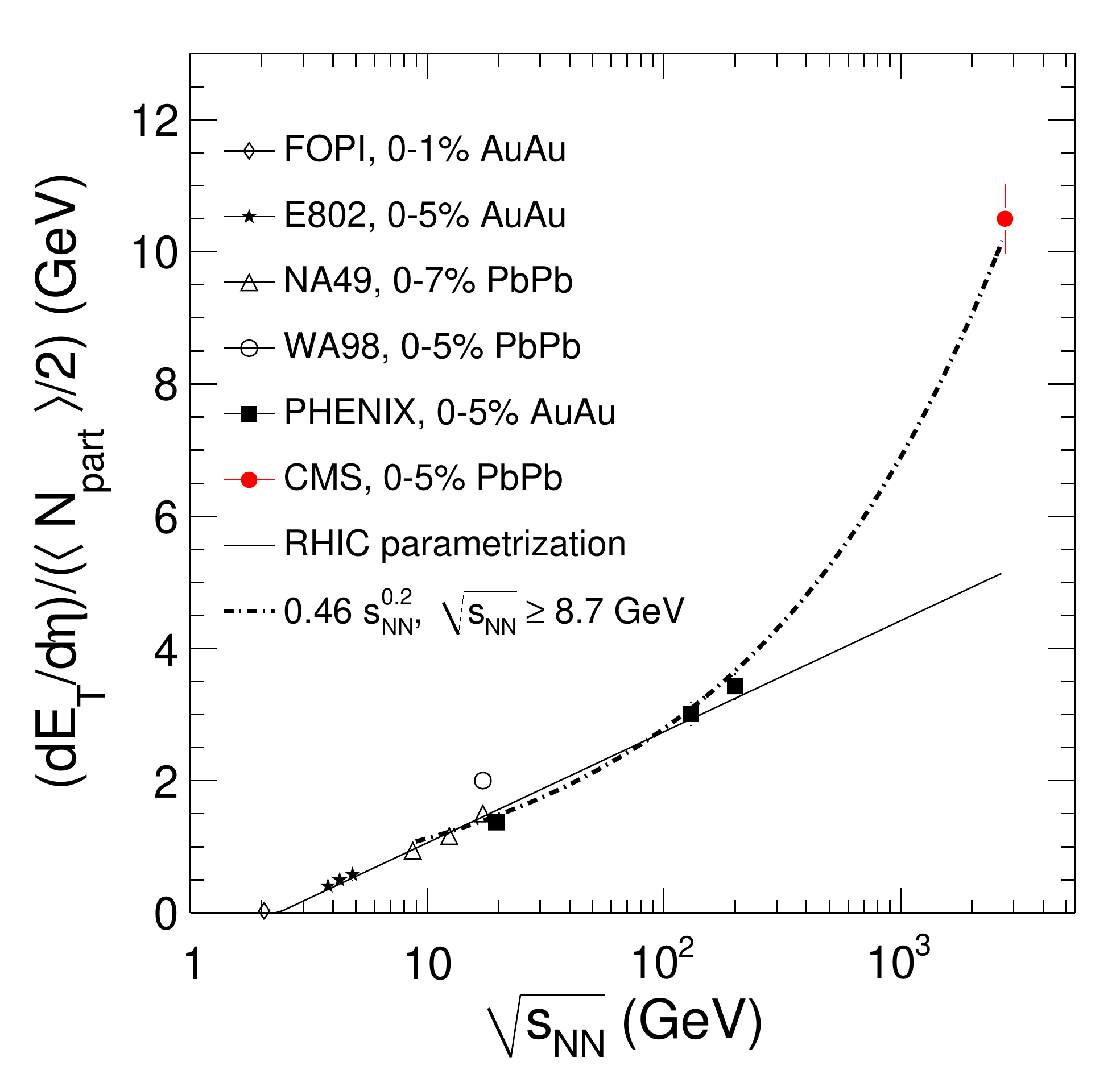}}
\caption{Left: (top) $\sqrt{s}$-dependence of the charged particle ${\rm d}N_{\rm ch}/{\rm d}\eta$ per 
colliding nucleon pair  from a variety of measurements in
pp and ${\rm p}\overline{\rm p}$ and central nucleus-nucleus collisions,
including the ATLAS 0-6\% centrality measurement for $|\eta|<0.5$ and the 0-5\% centrality ALICE and CMS measurements (points shifted 
horizontally for clarity). The curves show different expectations for
the $\sqrt{s}$-dependence in nucleus-nucleus collisions:
results of a Landau hydrodynamics calculation~\cite{Carruthers:1973ws} (dotted line), an $s^{0.15}$ extrapolation of RHIC and SPS data (dashed line), a logarithmic extrapolation of RHIC and SPS data from \cite{Busza:2007ke,Back:2005hs} (solid line); (bottom) $({\rm d}N_{\rm ch}/{\rm d}\eta)|_{\eta =0}/(\langle N_{\rm part}\rangle/2)$ vs. $N_{\rm part}$ for 2\% centrality intervals over 0-20\% and 5\% centrality intervals over 20-80\%. Error bars represent combined statistical and systematic uncertainties on the ${\rm d}N_{\rm ch}/{\rm d}\eta$ measurements, whereas the shaded band indicates the total systematic uncertainty including $N_{\rm part}$ uncertainties. The RHIC measurements have been multiplied by 2.15 to allow comparison with the $\sqrt{s_{\rm NN}} = 2.76$ TeV results. The inset shows the $N_{\mathrm{part}}<60$ region in more detail. 
Right: Transverse energy density normalized by ($\langle N_{\rm{part}} \rangle/2$) for central collisions at $\eta=0$ versus $\sqrt{s_{_\mathrm{NN}}}$. The vertical bar on the CMS point is the full systematic uncertainty. The statistical uncertainty is negligible. The solid line is a parametrization of  lower-energy data compiled by the PHENIX Collaboration~\cite{Adler:2004zn}. The dashed line corresponds to a power law fit $s_{_\mathrm{NN}}^{n}$ for $\sqrt{s_{_\mathrm{NN}}} \geq8.7$ GeV.} 
\label{fig:chmultET}
\end{figure}

The study of the observables connected with the global characteristics of the event is usually among the first measurements that are performed in the study of
ultrarelativistic hevy-ion collisions when a new energy domain opens up. In particular, the measurements of the charged hadron multiplicity as a function of the centrality of the collision and of the pseudorapidity are sensitive to the relative contribution from hard scatterings, which is expected to be proportional to the number of nucleon-nucleon collisions, and soft processes, usually assumed to be proportional to the number of participant nucleons. They also represent an interesting test for theoretical calculations, as they are sensitive to the modelling of the initial state of the collision in terms of gluon saturation.
Results from ALICE~\cite{Aamodt:2010pb}, ATLAS~\cite{ATLAS:2011ag} and CMS~\cite{Chatrchyan:2011pb} shown in Fig.~\ref{fig:chmultET}(left,top) indicate a steep rise in the charged hadron multiplicity density per participant pair $({\rm d}N_{\rm ch}/{\rm d}\eta)|_{\eta =0}/(\langle N_{\rm part}\rangle/2)$ when moving from top RHIC energy to the LHC, a feature possibly related to the significant increase of the contribution of hard processes. In particular, the approximately logarithmic increase observed from AGS up to RHIC does not hold up to LHC energy, while a power-law dependence ($s^{0.15}$) approximately reproduces the data. As it was the case at lower energy, ${\rm d}N_{\rm ch}/{\rm d}\eta|_{\eta =0}/(\langle N_{\rm part}\rangle/2)$ is systematically higher for heavy-ion collisions with respect to pp, showing that nuclear collisions cannot be described as a mere superposition of nucleon-nucleon scatterings. Another observable connected with the global properties of the event is the pseudorapidity distribution of the transverse energy $E_{\rm T}$~\cite{Chatrchyan:2012mb}, shown in Fig.~\ref{fig:chmultET}(right). It reaches values up to $\sim$2.1 TeV for central \mbox{PbPb} collisions and it has a stronger increase (a factor $3.07\pm0.24$) than the charged particle multiplicity (2.17$\pm$0.15) when moving from RHIC to LHC energy, implying that there is a significant increase of the mean transverse energy per particle. Estimates of the energy density according to the Bjorken formula~\cite{Bjorken:1982qr} $\epsilon= ({\rm d}E_{\rm T}/{\rm d}\eta)|_{\eta =0}/(A\cdot\tau_{\rm 0})$, where $A$ is the overlap area of the incident nuclei and $\tau_{\rm 0}$ is the parton formation time, yield $\epsilon = 14$ GeV/fm$^3$ for a time $\tau_{\rm 0}=1$ fm/c. Such a value is larger by more than one order of magnitude than the estimates for the critical energy density for the phase transition to a deconfined state~\cite{Karsch:2001cy}.

\begin{figure}[htbp]
\centering
\resizebox{0.45\textwidth}{!}
{\includegraphics*{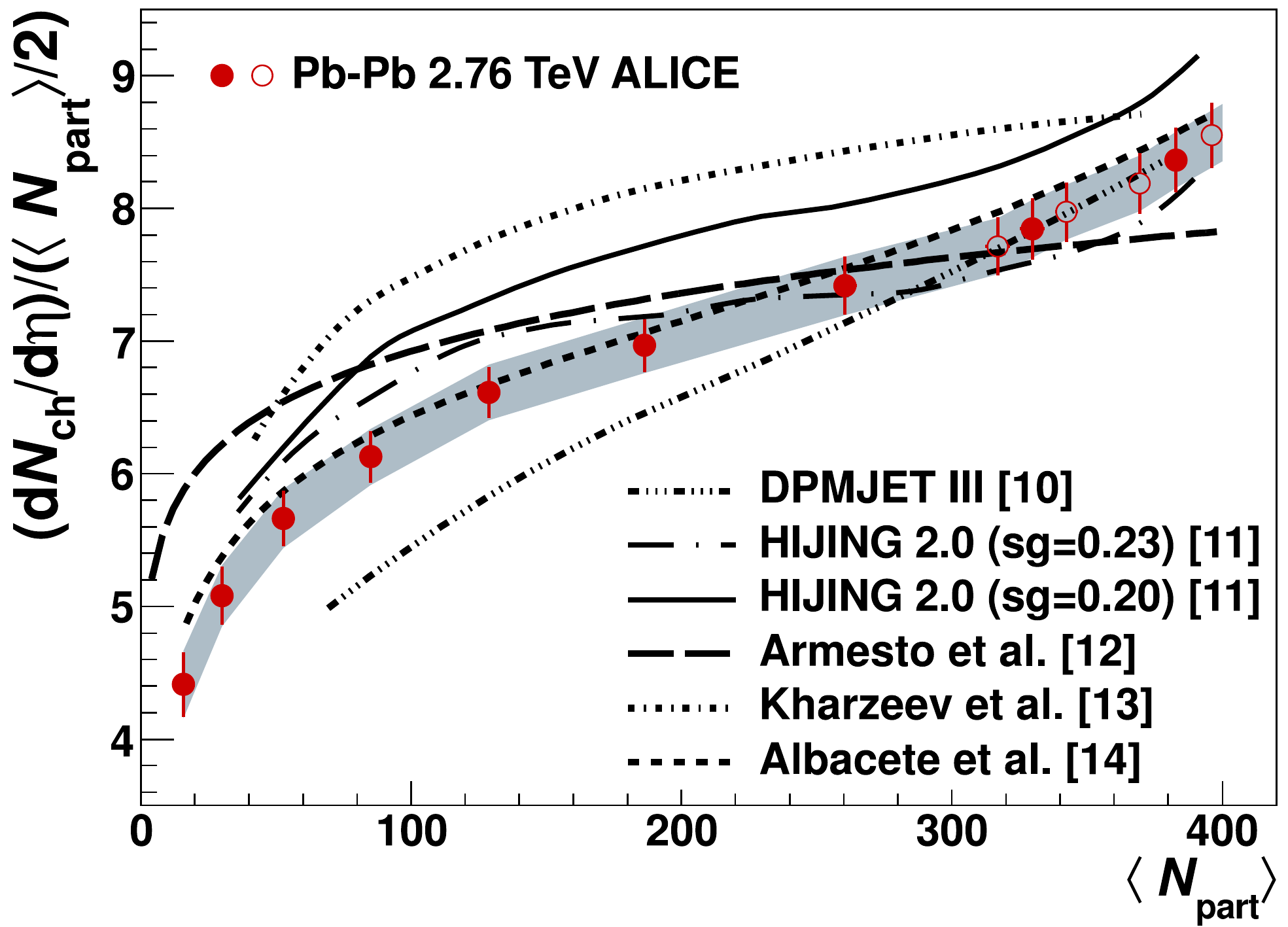}}
\resizebox{0.48\textwidth}{!}
{\includegraphics*{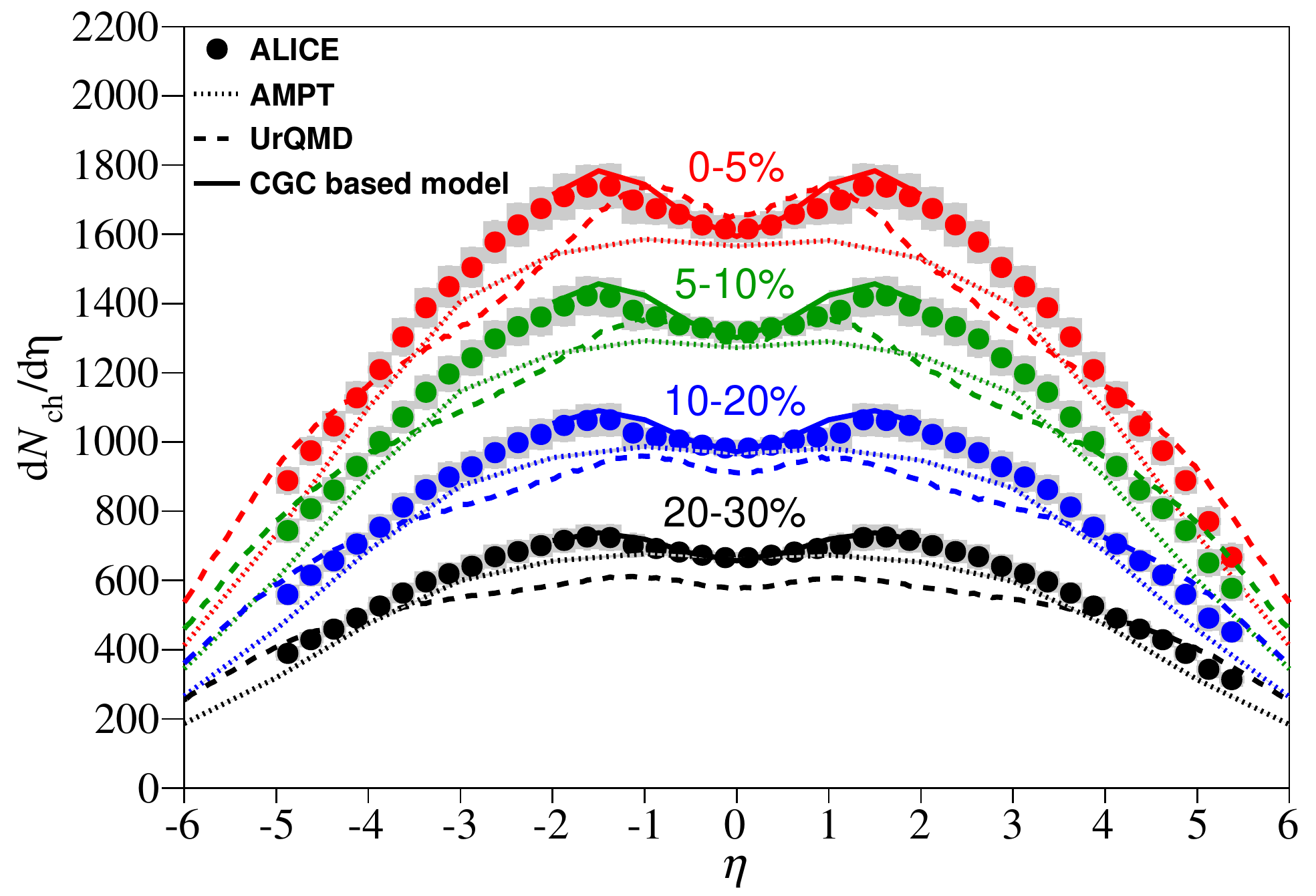}}
\caption{Left: Comparison of $({\rm d}N_{\rm ch}/{\rm d}\eta)/(\langle N_{\rm part}\rangle/2)$ with model calculations for PbPb at $\sqrt{s_{\rm NN}}=2.76$~TeV. Uncorrelated uncertainties are indicated by the error bars,
while correlated uncertainties are shown as the grey band.
Statistical errors are negligible. The HIJING 2.0 curve is shown for two values of the gluon shadowing~($s_{g}$) parameter. Right: ${\rm d}N_{\rm ch}/{\rm d}\eta$ in different centrality classes compared to model predictions~\cite{ALbacete:2010ad,Mitrovski:2008hb,Lin:2004en,Xu:2011fi}.} 
\label{fig:othermult}
\end{figure}

The multiplicity per colliding nucleon pair was found, at RHIC energies, to have a mild increase with the collision centrality~\cite{Adler:2004zn,Bearden:2001xw,Bearden:2001qq,Back:2002uc,Adams:2004cb}. This situation persists at LHC energy~\cite{Aamodt:2010cz,ATLAS:2011ag,Chatrchyan:2011pb}, where only a moderate variation, by less than a factor 2 from peripheral (75-80\%) to central (0-2\%) events, is observed, as can be seen in Fig.~\ref{fig:chmultET}(left, bottom). Remarkably, when the corresponding RHIC results are scaled in such a way to match the observed values for central collisions~\cite{Aamodt:2010cz}, a good agreement is found over all the centrality range. A similar behaviour was already found in the past when comparing results at RHIC and SPS energy~\cite{Back:2004dy}. Two main classes of calculations have been compared to the data, either two-component models based on a combination of perturbative QCD processes and soft interactions~\cite{Bopp:2007sa,Deng:2010xg}, or saturation models with various parametrisations for the energy and centrality dependence of the saturation scale~\cite{Armesto:2004ud,Kharzeev:2004if,ALbacete:2010ad}. As can be seen in Fig.~\ref{fig:othermult}(left), calculations from both classes of models were able to qualitatively predict the observed behaviour. 

Measurements of the centrality dependence of the pseudorapidity density distributions were performed for a wide $-5.0<\eta<5.5$ range~\cite{Abbas:2013bpa,Adam:2015kda}, allowing the investigation of the fragmentation region i.e. the region of phase space of particles where they have longitudinal momentum close to that of the original colliding nuclei. The main result is shown in Fig.~\ref{fig:othermult}(right), where a comparison with theoretical models~\cite{ALbacete:2010ad,Mitrovski:2008hb,Lin:2004en,Xu:2011fi} is also performed. The Color-Glass Condensate (CGC) model well describes the data in its (limited) domain of applicability, while no model is able to satisfactorily reproduce both  shape and absolute values of the yields over the whole $\eta$-range. A double gaussian fit, with a ratio of the widths independent of centrality, well reproduces the data. When compared to RHIC results~\cite{Bearden:2001qq,Alver:2010ck} as a function of the variable $\eta - y_{\rm beam}$, a scaling is observed in the large $\eta$-region (longitudinal scaling). Finally, the total integrated charged multiplicity can be obtained from these data. It reaches $N_{\rm ch}=17165\pm772$ for the 0-5\% most central collisions, and its evolution with centrality is identical to the one observed at RHIC~\cite{Alver:2010ck} once a 2.67 scaling factor is introduced.

\subsection{Azimuthal asymmetries}
\label{azim}

In heavy-ion collisions, the impact-parameter vector and the beam direction define the reaction plane. If the particle momenta do not only depend on 
the local conditions at their production point, but also on the overall event geometry, a correlation between the azimuthal angle of the produced particles 
and the reaction plane (anisotropic transverse flow) may be detected, and represents an unambiguous signature of a collective behaviour in the created medium.
In particular, in non-central heavy-ion collisions, the initial shape of the created fireball has a geometrical anisotropy, which induces different pressure gradients in the collectively expanding medium. As a consequence, an anisotropy of the azimuthal distributions of the produced particles arises, if interactions  in the medium are strong and consequently the mean free path small enough for relativistic hydrodynamics to be applicable \cite{Teaney:2009qa,Romatschke:2009im,Jeon:2015dfa} very early - an issue still pending a theoretical explanation \cite{Chesler:2015lsa,Gelis:2015gza}. The coefficients of a Fourier expansion of such distributions~\cite{Voloshin:1994mz}
can be used to characterize the observed asymmetry. Specifically, the second coefficient of the expansion (elliptic flow, $v_2$) is strongly related to the pressure
gradient difference, but also higher-order coefficients ($v_3$, $v_4$,...) provide valuable information on the evolution of the medium and on the geometric fluctuations 
of the initial state. At RHIC~\cite{Ackermann:2000tr}, the observation of $v_2$ values at low $p_{\rm T}$ in good agreement with ideal hydrodynamic calculations (i.e. zero viscosity) in 
semi-central and central collisions were at the basis of the claim for the production of a strongly interacting QGP phase behaving as an ideal liquid~\cite{Arsene:2004fa,Back:2004je,Adams:2005dq,Adcox:2004mh}.
In addition, a hierarchy of $v_2$ values was observed for identified particles, with higher values, at a given $p_{\rm T}$, for lighter particles~\cite{Adler:2001nb}. This characteristic 
ordering can be mainly understood in terms of a radial flow effect and models including a partonic phase, characterized by a harder equation of state (and consequently a higher speed of sound) with respect to the hadronic phase, showed a quantitative agreement with the 
results~\cite{Huovinen:2005gy}.

\begin{figure}[htbp]
\centering
\resizebox{0.5\textwidth}{!}
{\includegraphics*{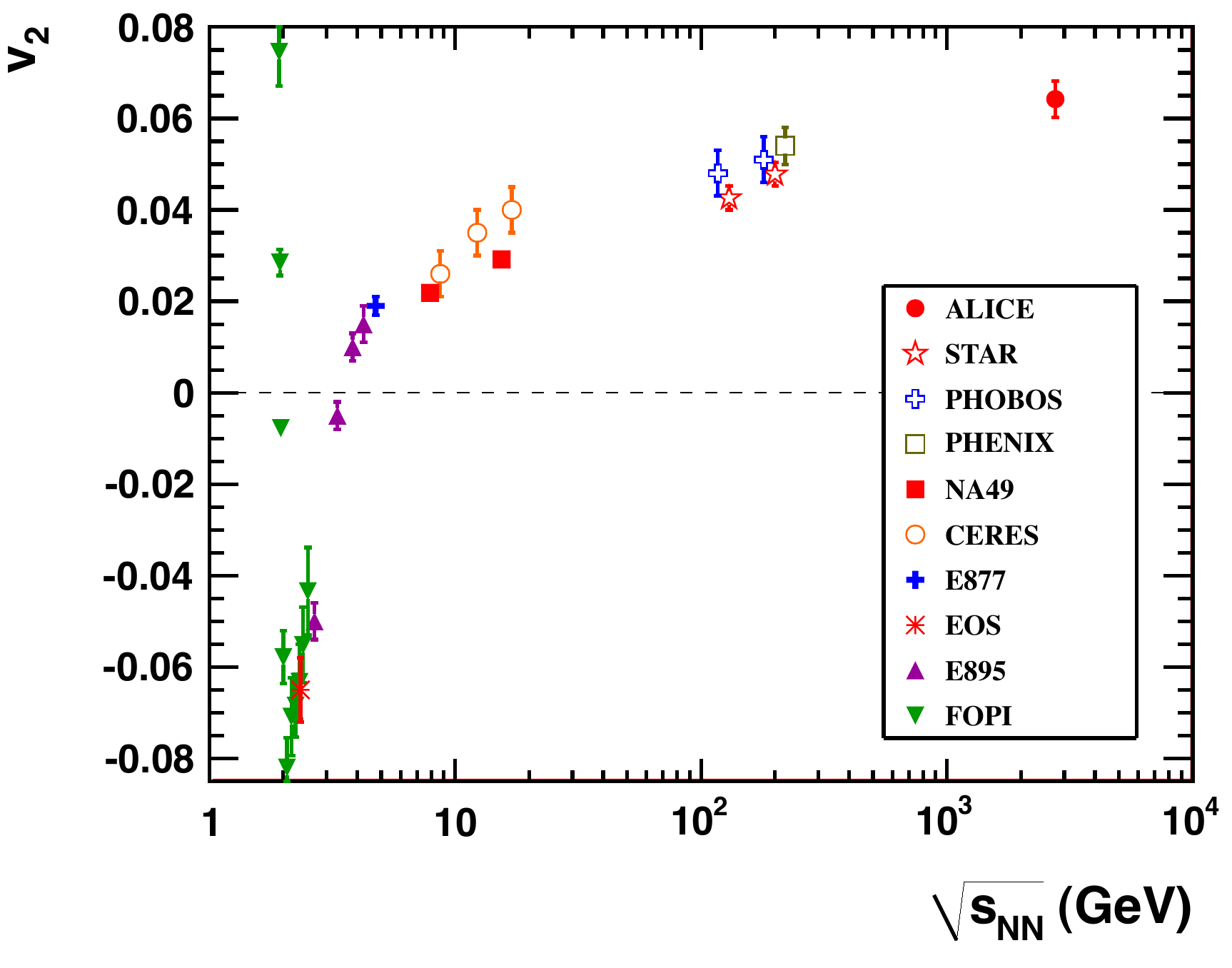}}
\resizebox{0.4\textwidth}{!}
{\includegraphics*{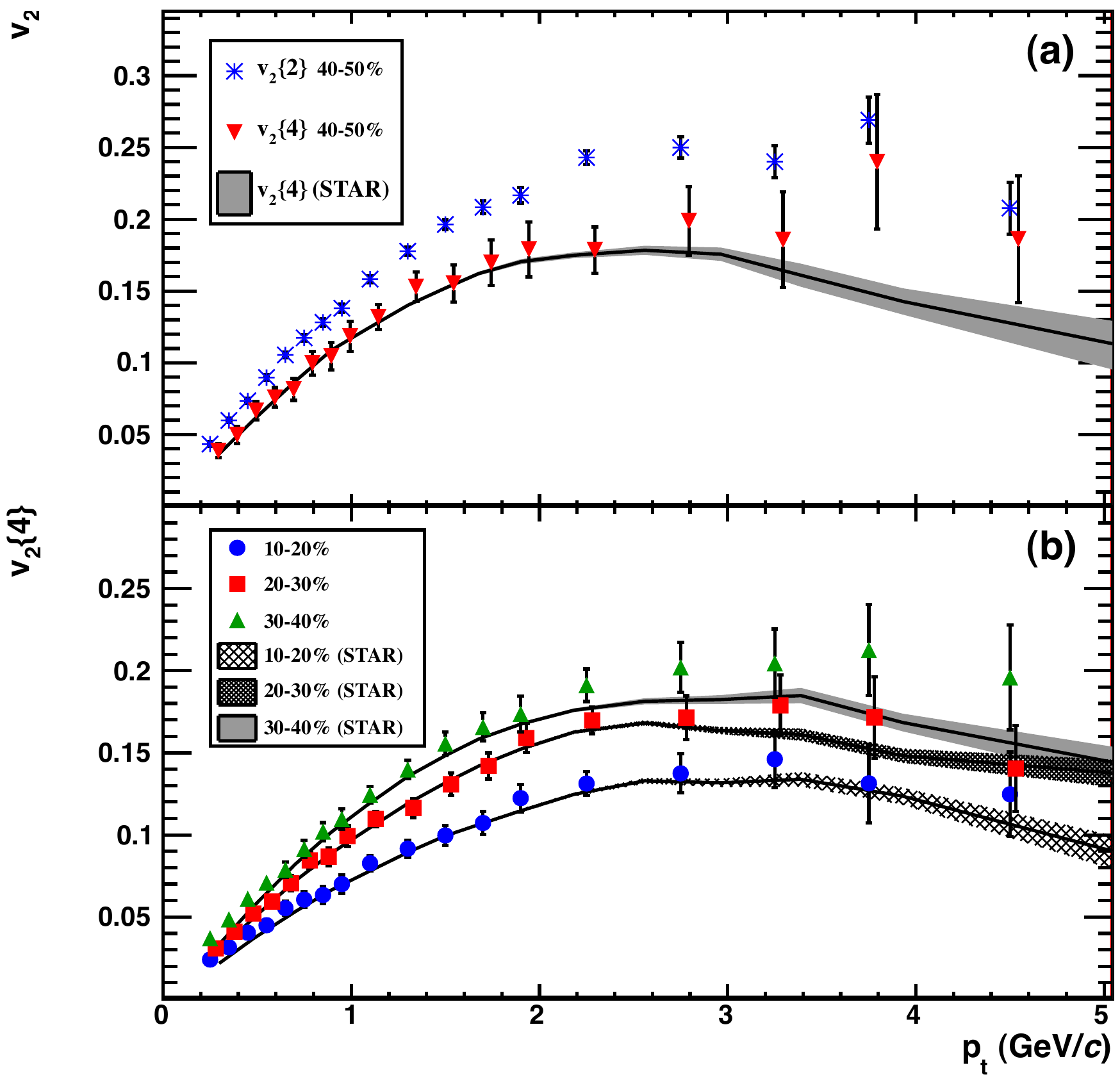}}
\caption{Left: The ALICE result on $p_{\rm T}$-integrated elliptic flow at 2.76~TeV in Pb--Pb 20--30\% centrality class~\cite{Aamodt:2010pa} compared with results from lower energies taken at similar  centralities~\cite{Voloshin:2008dg,Andronic:2004cp}. Right: a) $v_2$($p_{\rm T}$), measured by ALICE, for the centrality bin 40--50\% from the 2- and 4-particle cumulant methods and for Au--Au collisions at $\sqrt{s_{\rm NN}}=200$ GeV (STAR). b) $v_2$\{4\}($p_{\rm T}$) for various centralities compared to STAR measurements. } 
\label{fig:v2vsenergypt}
\end{figure}

At the LHC, anisotropic flow measurements have reached an unprecedent accuracy,  due to the larger charged multiplicities and to the large acceptance of the
experiments. Concerning $v_2$, it was observed (see Fig.~\ref{fig:v2vsenergypt}(left)) that the $p_{\rm T}$-integrated values increase by about 30\% with respect to RHIC, the increase being mainly due to 
the larger $\langle p_{\rm T}\rangle$ (as $v_2$ increases with $p_{\rm T}$ at relatively low $p_{\rm T}$)~\cite{Aamodt:2010pa}. At fixed $p_{\rm T}$, as shown in Fig.~\ref{fig:v2vsenergypt}(right), the observed values are similar  
at RHIC and LHC~\cite{Aamodt:2010pa,ATLAS:2011ah,Chatrchyan:2012xq}. It is interesting to note that such a similarity extends down to low RHIC energies~\cite{Adamczyk:2012ku}. 

\begin{figure}[htbp]
\centering
\resizebox{0.42\textwidth}{!}
{\includegraphics*{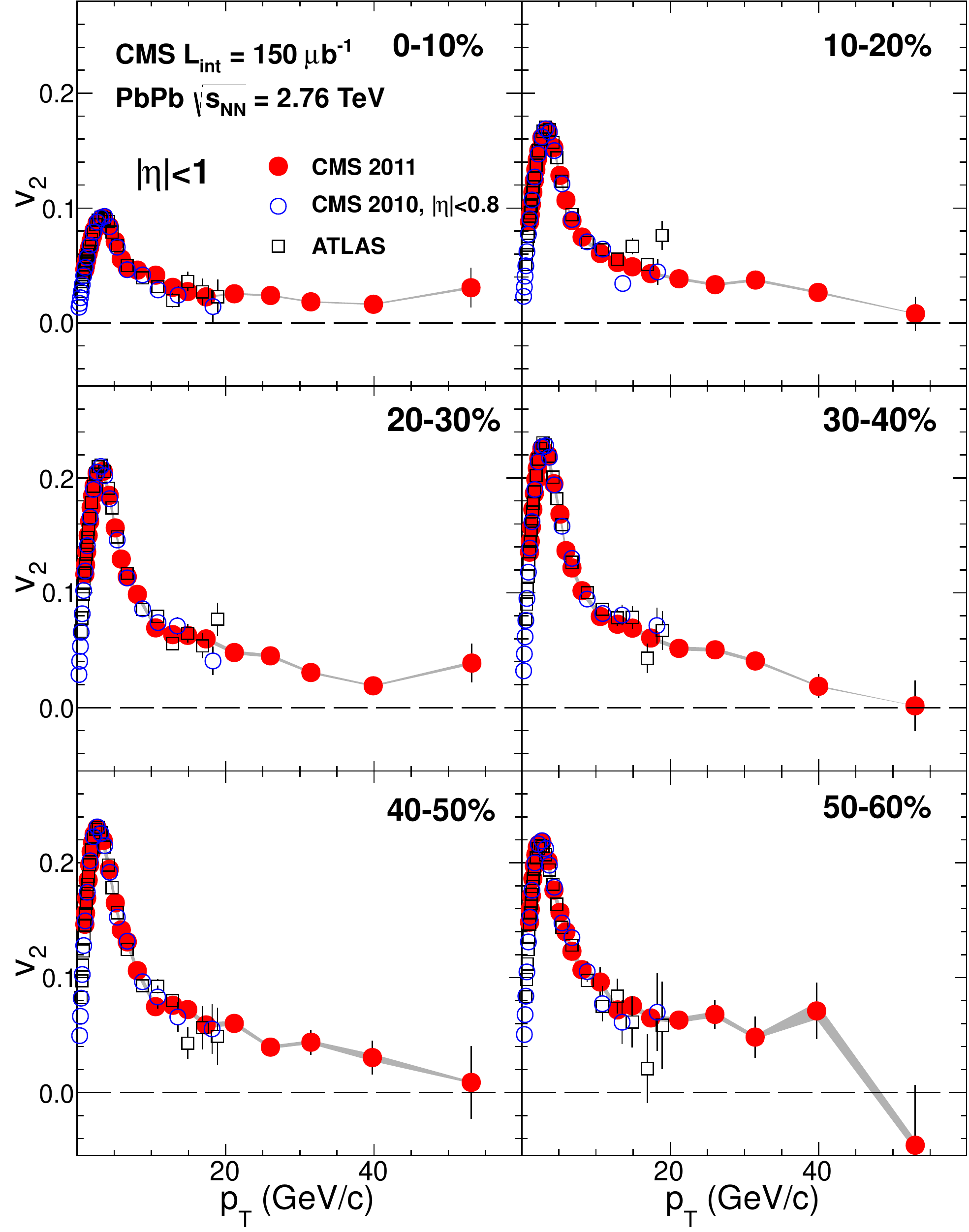}}
\resizebox{0.47\textwidth}{!}
{\includegraphics*{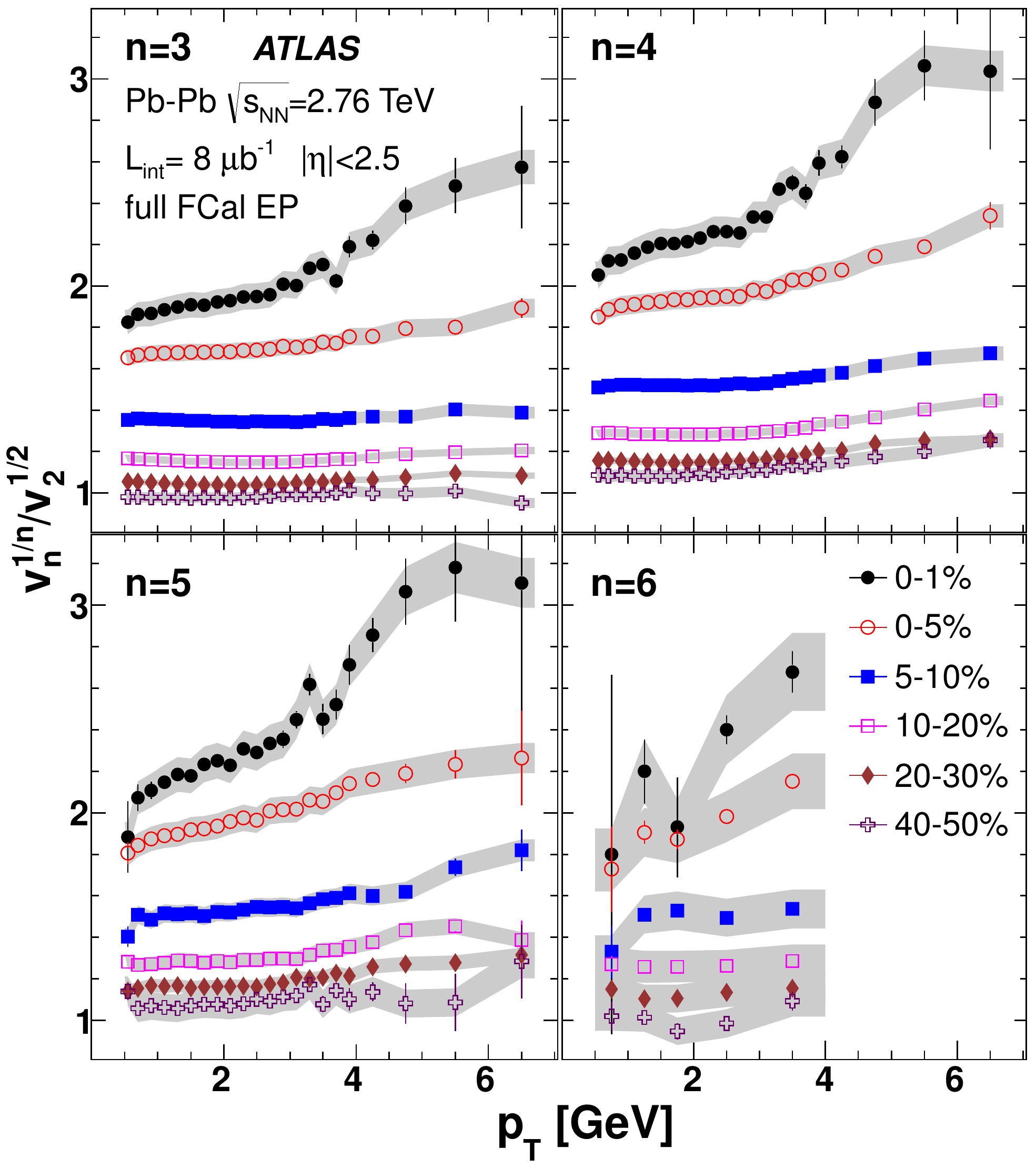}}
\caption{Left: The single-particle azimuthal anisotropy, $v_2$, as a function of the charged particle transverse momentum from 1 to 60 GeV/$c$ with $|\eta|<1$  
for six centrality ranges in Pb--Pb collisions at $\sqrt{s_{\rm NN}} = 2.76$ TeV, measured by the CMS experiment (solid markers)~\cite{Chatrchyan:2012xq}. Error bars denote the statistical uncertainties, while the grey bands correspond to the small systematic uncertainties. Comparison to results from the ATLAS (open squares) and CMS (open circles) experiments using data collected in 2010 is also shown. Right: $v_n^{1/n}/v_2^{1/2}$ vs. $p_{\rm T}$ for several centrality intervals~\cite{ATLAS:2012at}, measured by ATLAS. The shaded bands indicate the total systematic uncertainties.} 
\label{fig:v2highpt}
\end{figure}

The $p_{\rm T}$-reach of the elliptic flow measurements has been pushed to very-high $p_{\rm T}$ ($>50$ GeV/$c$)~\cite{Chatrchyan:2012xq}, as shown in Fig.~\ref{fig:v2highpt}(left). Here the interpretation in 
terms of collective effects does not hold any more and the observed non-zero $v_2$ is mainly related to the different pathlength in the medium for particles emitted in-plane and out-of-plane, which induces a different energy loss (see Section~\ref{largept}). Higher-order azimuthal asymmetries have also received considerable attention, with measurements extending up to $v_6$~\cite{ATLAS:2012at,ALICE:2011ab,Chatrchyan:2013kba}. Their study is particularly important to help constraining the initial conditions of the collision, which play a key role in the determination of the viscosity. The measured flow coefficients show 
little dependence on centrality, consistent with an anisotropy primarily associated with fluctuations in the initial geometry. The approximate scaling 
$v^{\rm 1/n}_{\rm n} (p_{\rm T}) \propto v^{\rm 1/2}_{\rm 2}(p_{\rm T})$, expected in a hydrodynamic scenario, has been observed for all but the most central events (see Fig.~\ref{fig:v2highpt}(right)~\cite{ATLAS:2012at}).
%with such a deviation being explained in terms of eccentricity fluctuations. 

\begin{figure}[htbp]
\centering
\includegraphics[width=0.73\textwidth]{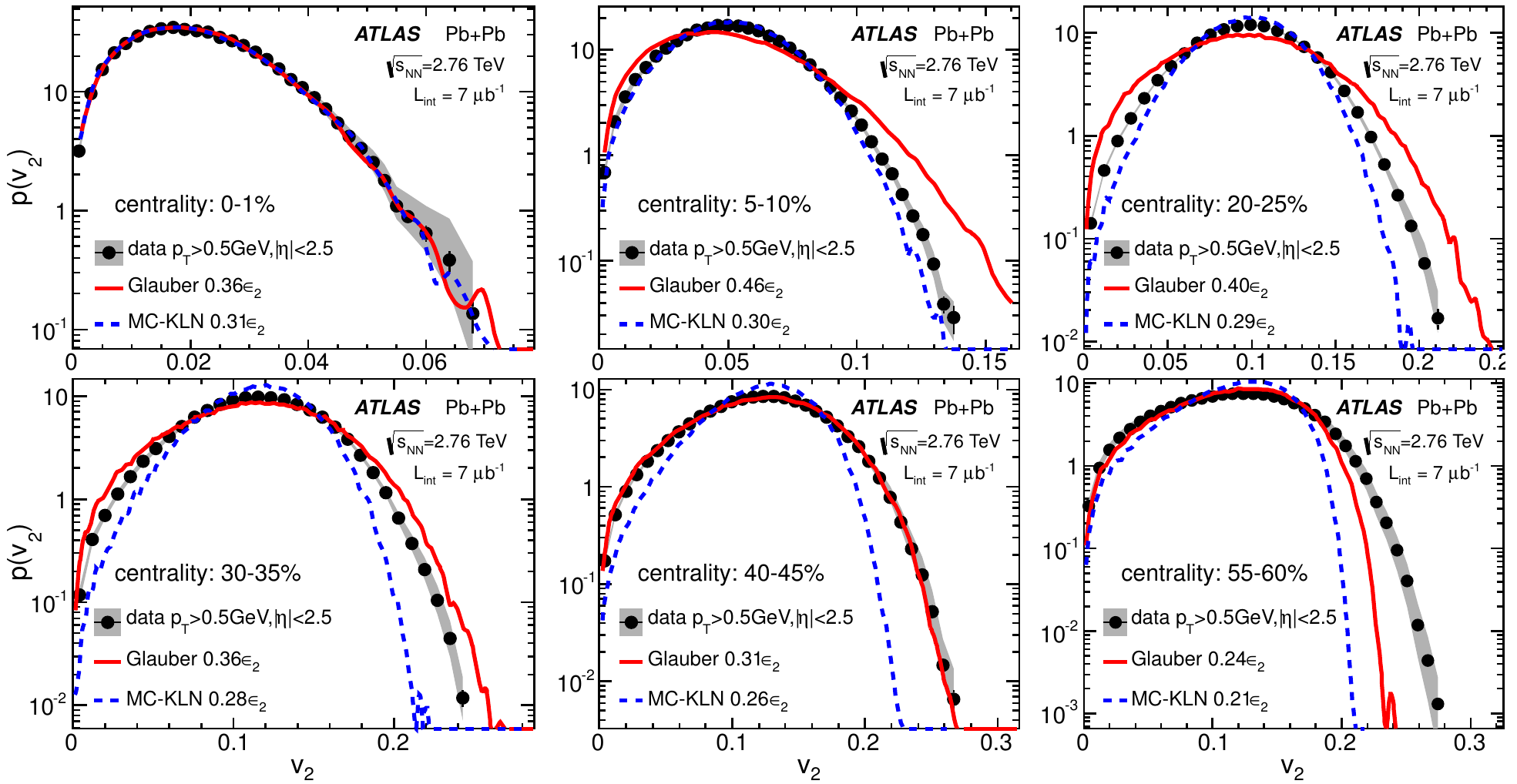}\\
%\hskip 0.5cm
\includegraphics[width=0.77\textwidth]{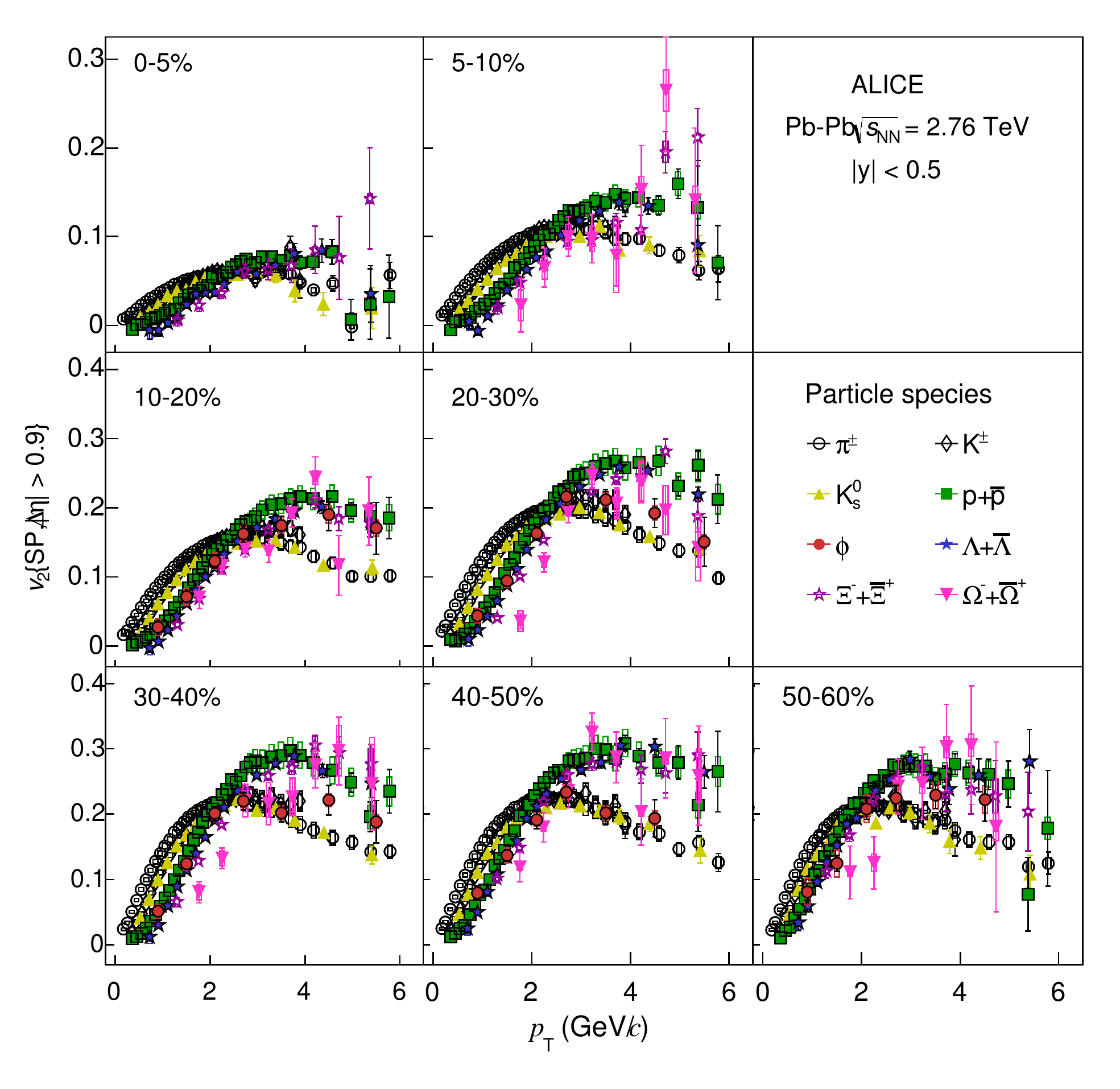}
%\centering
%\resizebox{0.57\textwidth}{!}
%{\includegraphics*{Figures/Flow/fig_18_ebevn}}
%\resizebox{0.49\textwidth}{!}
%{\includegraphics*{Figures/Flow/allspeciesSame}}
\caption{Top: The event-by-event $v_2$ distributions measured by ATLAS~\cite{Aad:2013xma} and compared with the $\epsilon_2$ eccentricity distributions from two initial geometry models: a Glauber model (solid lines) and the MC-KLN model (dashed lines). The $\epsilon_2$ distributions have been rescaled to the same mean values. The scale factors are indicated in the legends. Bottom: The $p_{\rm{T}}$-differential $v_2$ for different particle species grouped by centrality class in Pb--Pb collisions at $\sqrt{s_\mathrm{{NN}}} = 2.76$~TeV, from ALICE~\cite{Abelev:2014pua}.} 
\label{fig:ebevn}
\end{figure}

At LHC energies, first systematic studies of event-by-event flow coefficients have been carried out~\cite{Aad:2013xma}. They are particularly relevant for the assessment of fluctuations in the initial geometry, and were extensively compared to models. The results show that the $v_{\rm n}$ distributions broaden from central to peripheral collisions, and are in good agreement with a fluctuation-only scenario, except for $v_2$. Calculations based on eccentricity distributions predicted in the Glauber model~\cite{Miller:2007ri} or in the MC-KLN model (which includes saturation effects)~\cite{Drescher:2006pi} do not describe the data, as can be seen in Fig.~\ref{fig:ebevn}(top). Other approaches, such as  EKRT~\cite{Niemi:2015qia} and IP-Glasma~\cite{Schenke:2013aza}, were then shown to reach a better agreement with the ATLAS results.

Detailed studies of the elliptic flow of identified particles ($\pi^{\pm}$, $K^{\pm}$, $K_{\rm S}^0$, p+$\overline{\rm p}$, $\phi$, $\Lambda$+$\overline{\Lambda}$, $\Xi^-$+$\overline{\Xi}^+$, $\Omega^-$+$\overline{\Omega}^+$) have also been carried out~\cite{Abelev:2014pua}, see Fig.~\ref{fig:ebevn}(bottom). At low momentum, the mass ordering already seen at RHIC, and related to an interplay of radial flow (i.e. azimuthally symmetric)  and anisotropic flow, has been confirmed. At higher $p_{\rm T}$ values ($p_{\rm T}>3$ GeV/$c$) particles tend to group according to their type (mesons vs. baryons). However, systematic comparisons of $v_2/n_{\rm q}$, where $n_{\rm q}$ is the number of constituent quarks, as a function of $p_{\rm T}/n_{\rm q}$ (or $(m_{\rm T}-m_{\rm 0})/n_{\rm q}$, where $m_{\rm 0}$ is the mass of the particle),  show that this scaling is only approximate at LHC energies. Comparisons with hydrodynamical calculations coupling viscous fluid dynamics for the QGP with a hadronic cascade model (VISHNU~\cite{Song:2010mg}) show a similar mass ordering, indicating that a rather small value of the viscosity to entropy density ratio is favoured ($\eta/s=0.16$, about two times the $1/4\pi$ lower bound \cite{Kovtun:2004de}, is used at the beginning of the hydrodynamic evolution of the system).

\begin{figure}[htbp]
\centering
\resizebox{0.55\textwidth}{!}
{\includegraphics*{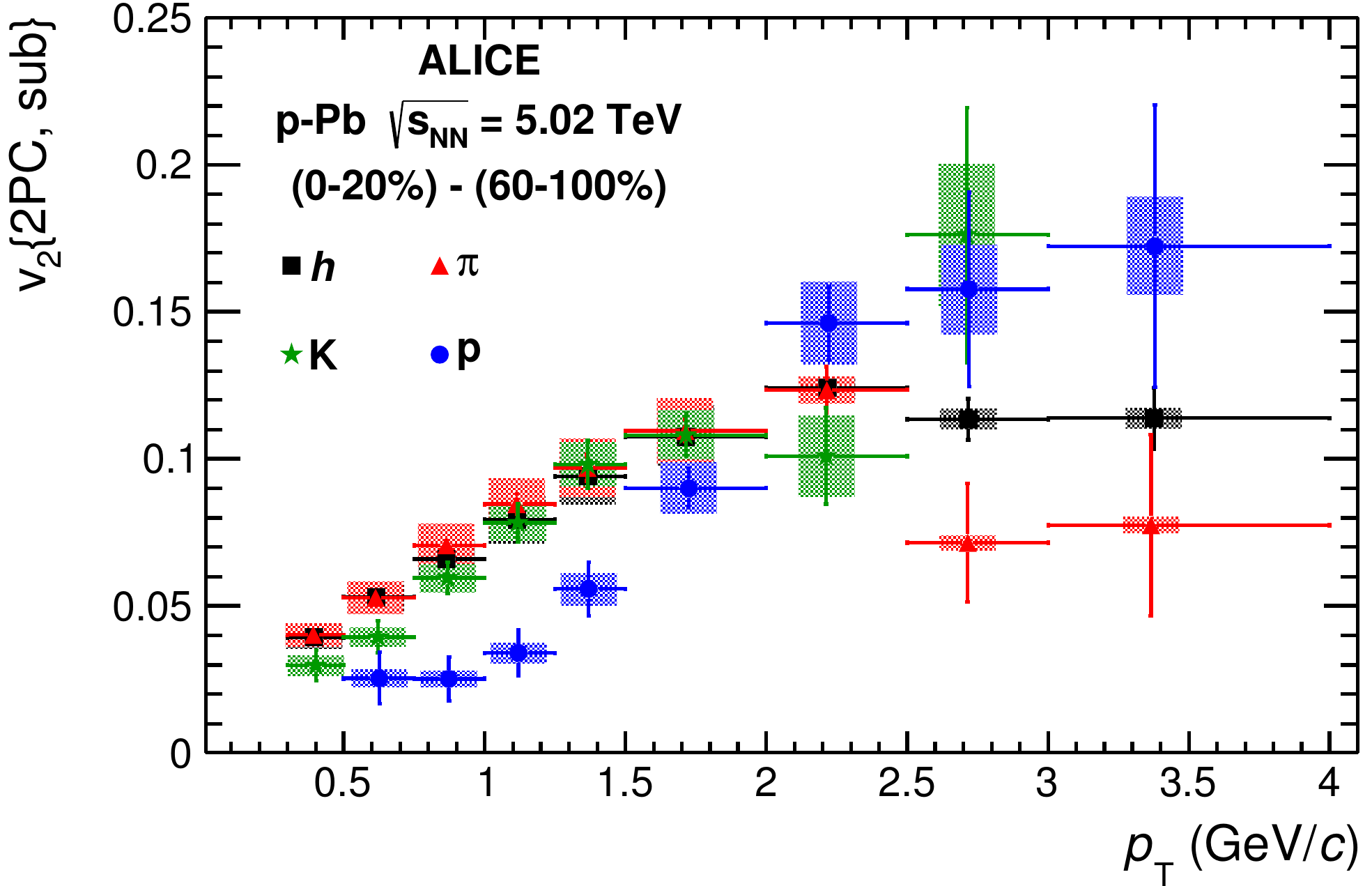}}
\resizebox{0.55\textwidth}{!}
{\includegraphics*{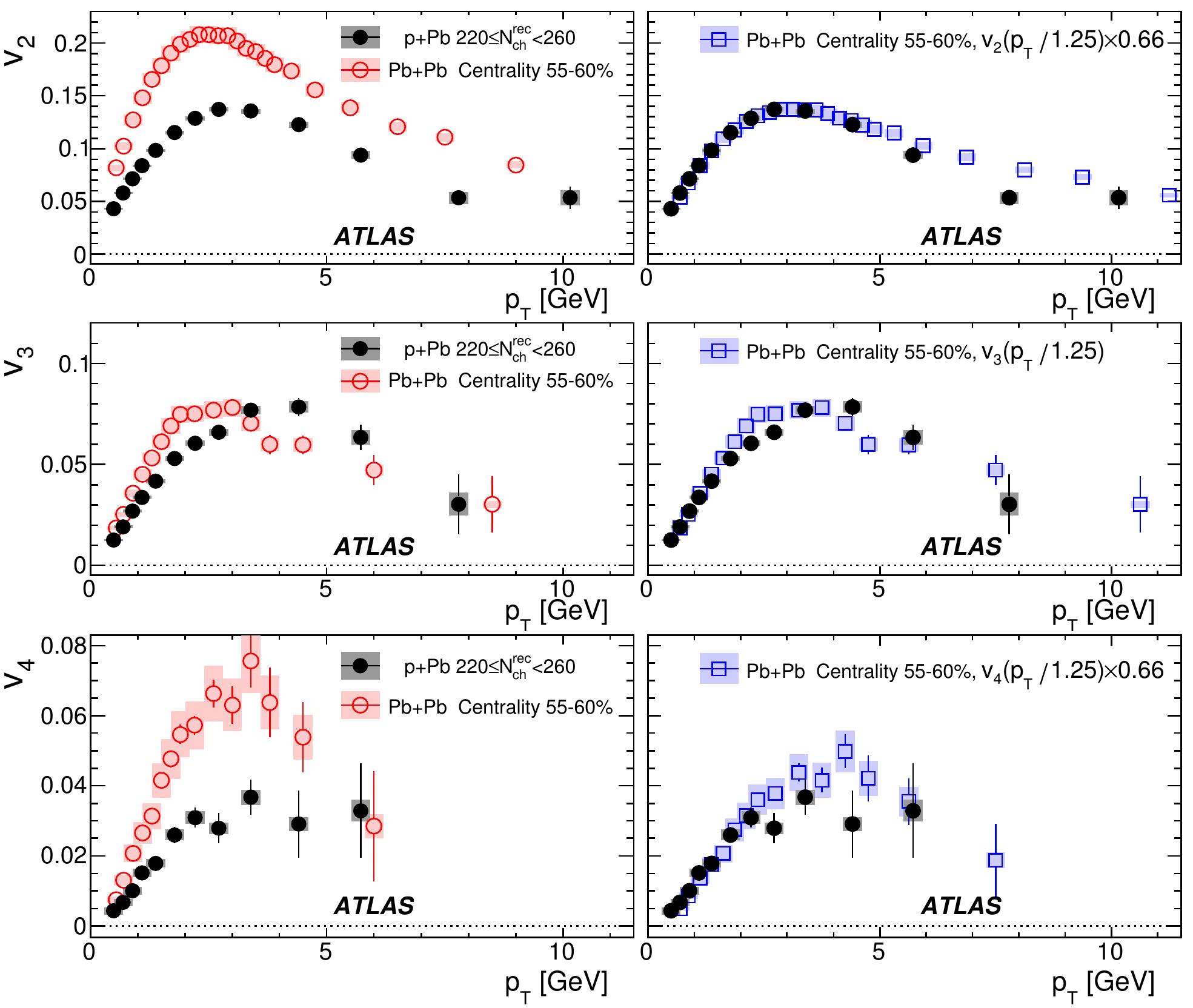}}
\caption{Top: The Fourier coefficient $v_2\{{\rm 2PC,sub}\}$ for hadrons (black squares), pions (red triangles), kaons (green
stars) and protons (blue circles) as a function of $p_{\rm T}$ from the correlation in the 0--20\%
multiplicity class after subtraction of the correlation from the 60--100\% multiplicity class, from ALICE~\cite{Abelev:2014mda}. Error bars show statistical
uncertainties while shaded areas denote systematic uncertainties. Bottom: The coefficients $v_2$ (top row), $v_3$ (middle row) and $v_4$ (bottom row) measured by ATLAS as a function of $p_{\rm T}$ compared between p--Pb  collisions with $220\leq N_{\rm ch}^{\rm rec}<260$~\cite{Aad:2014lta} and Pb--Pb collisions in 55--60\% centrality from Ref.~\cite{ATLAS:2012at}. The left column shows the original data with their statistical (error bars) and systematic uncertainties (shaded boxes). In the right column, the same Pb--Pb data are rescaled horizontally by a constant factor of 1.25, and the $v_2$ and $v_4$ are also down-scaled by an empirical factor of 0.66 to match the p--Pb  data.} 
\label{fig:v2pPb}
\end{figure}

Finally, important results have been obtained in the study of identified particle $v_2$ in \mbox{pPb} collisions, estimated from the study of two-particle correlations. Here, for high-multiplicity \mbox{pPb} collisions (0-20\%) one observes (see Fig.~\ref{fig:v2pPb}(top)~\cite{Abelev:2014mda}) a mass ordering similar to the one seen in \mbox{PbPb}, with $v_2^{\rm p}<v_2^{\pi}$ for $0.5<p_{\rm T}<1.5$ GeV/$c$ and larger beyond $p_{\rm T}\sim 2.5$ GeV/$c$. There is also a hint of $v_2^{\rm K}<v_2^{\pi}$ below 1 GeV/$c$. The striking similarity between \mbox{PbPb} and (central) \mbox{pPb} results, with the former being described by hydrodynamical calculations, suggests a collective behaviour also in \mbox{pPb} collisions at LHC energies. For unidentified charged particles, the coefficients from $v_2$ to $v_5$ have also been measured~\cite{Aad:2014lta,Chatrchyan:2013nka}. Comparing central \mbox{pPb} results with the corresponding peripheral \mbox{PbPb} collisions at a similar hadronic multiplicity, and after applying a scale factor to take into account the difference in the $\langle p_{\rm T}\rangle$ between the two systems~\cite{Basar:2013hea}, a good agreement of the $v_{\rm n}$ distributions is found (Fig.~\ref{fig:v2pPb}(bottom)), in particular in the low-$p_{\rm T}$ region where the statistical uncertainties are small. 

%
%event plane correlations \cite{Aad:2014fla}
%
%event plane fluctuations \cite{Khachatryan:2015oea}
%
%v2 for upto 6 particles in pPb \cite{Khachatryan:2015waa}
%
%flow in ultracentral PbPb \cite{CMS:2013bza}

\subsection{Correlations at low transverse momentum}
\label{ridge}

Correlations in azimuthal angle were treated in the previous Subsection, and those in momentum for particles with similar rapidities  will be treated in Subsection \ref{femto}. Here we review event-by-event fluctuations and correlations, charge-event plane correlations, and long-range rapidity correlations. The latter could be included in the previous Subsection but here we focus on their long-range nature in rapidity, while usually measurements of azimuthal asymmetries are performed without such consideration.

The analysis of event-by-event fluctuations was proposed as a probe of the matter
 generated in high-energy heavy-ion collisions, see the review \cite{Jeon:2003gk}. In the presence of either  a
second-order phase transition from the QGP to a hadron gas, or the existence of a critical point in the
phase diagram of strongly interacting matter, strong fluctuations of thermodynamic
quantities, such as temperature, or of global conserved charges, may occur. This could be reflected in dynamical event-by-event fluctuations of the
mean transverse momentum of final-state charged particles  or in net-charge
 fluctuations studied by ALICE in  \cite{Abelev:2014ckr} and \cite{Abelev:2012pv} respectively. The results indicate the existence of non-trivial collective effects in the collisions but are not yet conclusive about the signal of a well-defined, finite order phase transition\footnote{Lattice calculations at high temperatures \cite{Karsch:2001cy} indicate that the transition from confined to deconfined matter is a cross-over.}.

On the other hand, charge separation with respect to the event plane has been suggested as a signal of the Chiral Magnetic Effect that would point to the existence of domains in the created medium in which discrete symmetries like parity are violated \cite{Fukushima:2008xe}. The creation of strong magnetic fields  in heavy-ion  collisions, together with the existence of an imbalance in the populations of different helicity states, would produce an electric current and thus charge separation. ALICE has measured such separation \cite{Abelev:2012pa} with a magnitude similar to that found at RHIC. Whether this is a  signal of Chiral Magnetic Effect or not is still an open issue.

Besides, the charge balance functions (describing  the conditional probability that a particle in one momentum bin will be accompanied by a particle of opposite charge in another momentum bin) have been suggested to be sensitive
probes of the properties of the system, providing valuable insight into the charge creation mechanism
and into fundamental questions concerning hadronisation in heavy-ion
collisions \cite{Bass:2000az}. These balance functions are sensitive to the time of creation of the pairs of opposite-charge partons or particles. The  charge correlations measurements by ALICE \cite{Abelev:2013csa,Adam:2015gda} indicate that charge generation gets a sizeable contribution from pre-hadronic stages.

\begin{figure}[htbp]
\begin{center}
\includegraphics[width=0.54\textwidth]{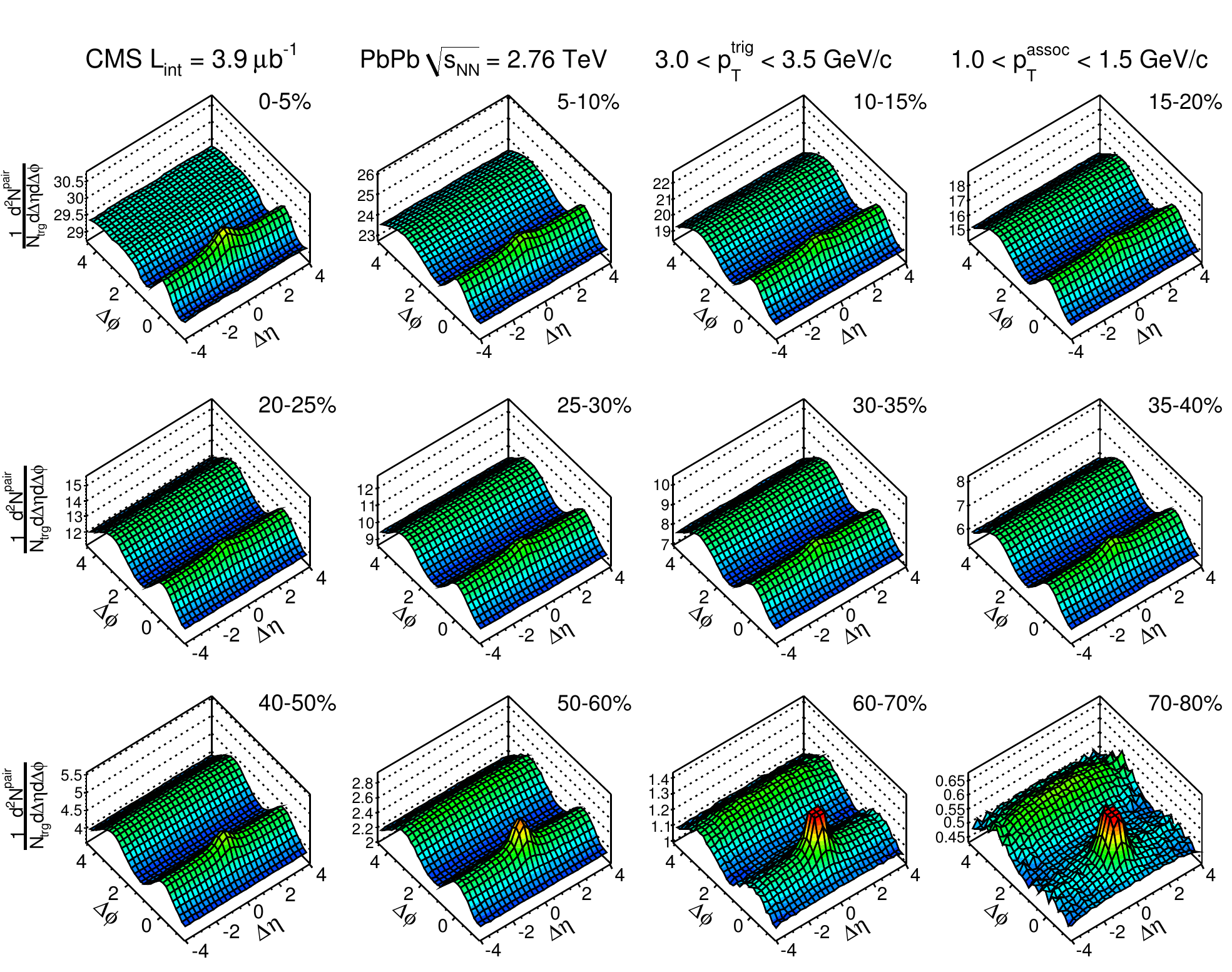}\hskip 0.3cm\includegraphics[width=0.42\textwidth]{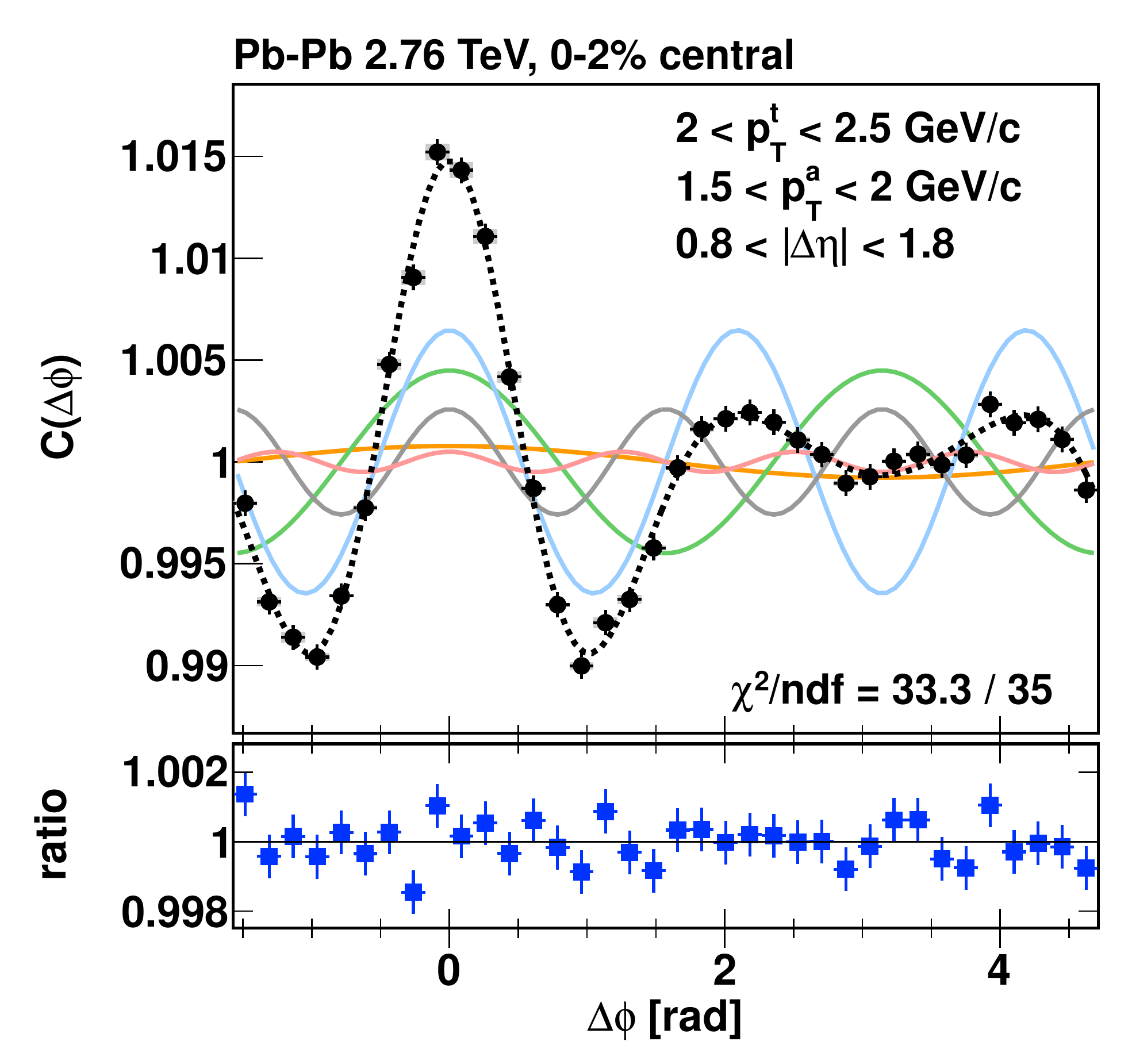}
\end{center}
\caption{Left:  Two-dimensional (2D) per-trigger-particle associated yield of charged hadrons as a
function of $|\Delta \eta|$ and $|\Delta \phi|$ for $3 < p_{\rm T}^{trig}
< 3.5$ GeV/c and $1 < p_{\rm T}^{assoc}
 < 1.5$ GeV/c, for twelve
centrality ranges of PbPb collisions at 2.76 TeV/nucleon [the near-side peak is truncated in the
two most peripheral distributions to better display the surrounding structure], from CMS.
Right: correlation for particle pairs at $|\Delta \eta| > 0.8$, with the Fourier harmonics for $n=1$ to 5
superimposed in colour, their sum shown as the dashed curve and the ratio of data to the $n \le 5$ sum 
shown in the lower panel, from ALICE.
Taken from  \cite{Chatrchyan:2012wg} and \cite{Aamodt:2011by}.} 
\label{fig:lowptcorr1}
\end{figure}

Now we turn to long-range correlations in rapidity. Simple causal arguments indicate that such correlations must have their origin in the initial stages of the collisions \cite{Dumitru:2008wn}. In AuAu collisions at RHIC \cite{Alver:2009id,Abelev:2009af}, a large correlation in the difference in rapidity and azimuthal angle for pairs of particles, extended along several units of rapidity and collimated in azimuth at 0 (near side) and 180 (away side) degrees (thus named ``ridge"), was observed for triggered and untriggered analysis, with a peak at (0,0) in $(\Delta \eta, \Delta \phi )$. Later, a strikingly similar structure was found in high-multiplicity pp collisions at the LHC by CMS \cite{Khachatryan:2010gv,Khachatryan:2015lva} and ATLAS \cite{Aad:2015gqa}, and then in PbPb collisions \cite{Aamodt:2011by,Chatrchyan:2011eka,Chatrchyan:2012wg}. The correlation becomes stronger with increasing centrality. As shown by ALICE, its structure in azimuth can be  characterised by a  Fourier decomposition even for pairs with large rapidity separations, see Fig. \ref{fig:lowptcorr1}, which provided a new view on this type of correlations.

\begin{figure}[htbp]
\begin{center}
\includegraphics[width=0.38\textwidth]{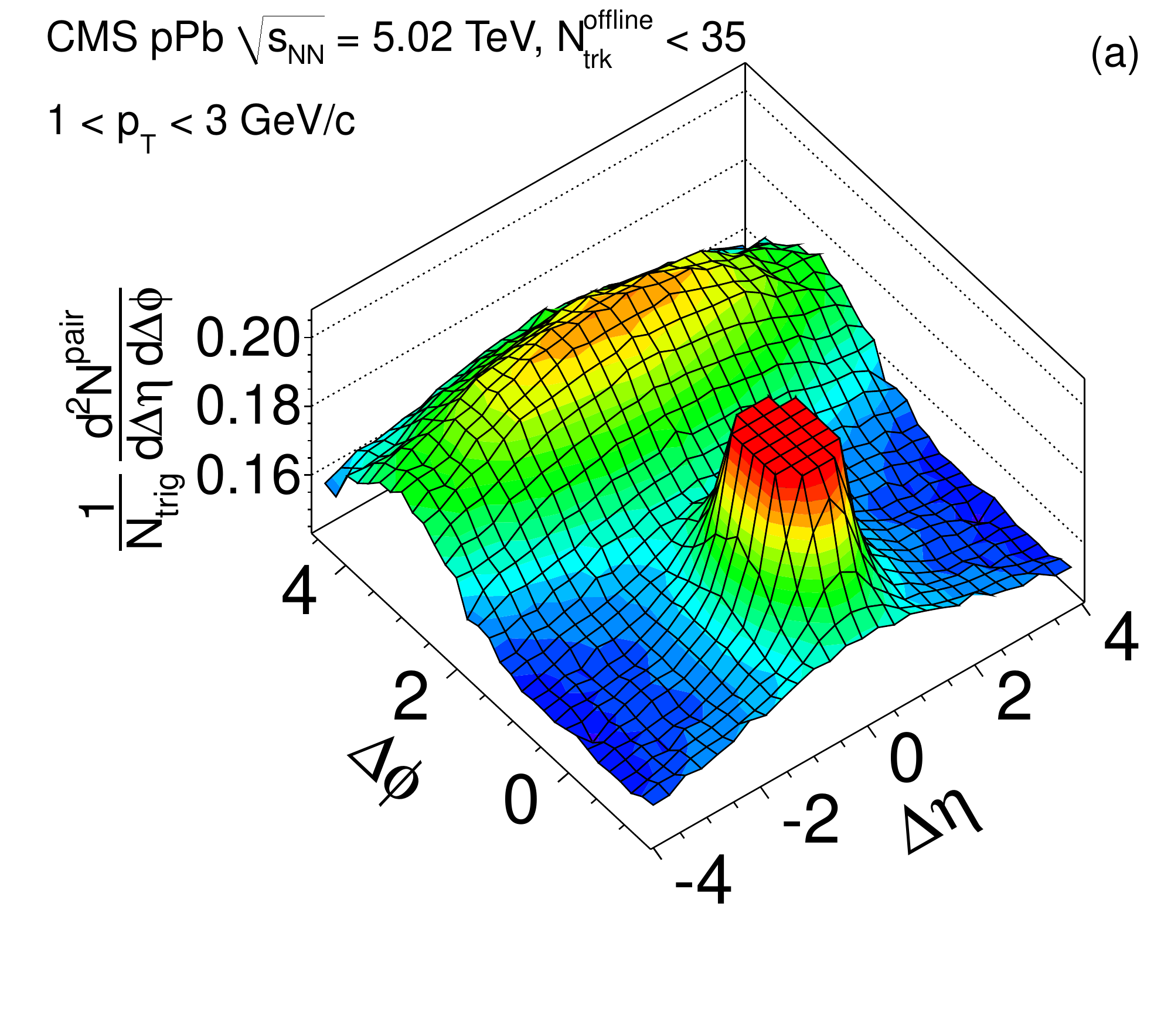}\hskip 0.3cm\includegraphics[width=0.38\textwidth]{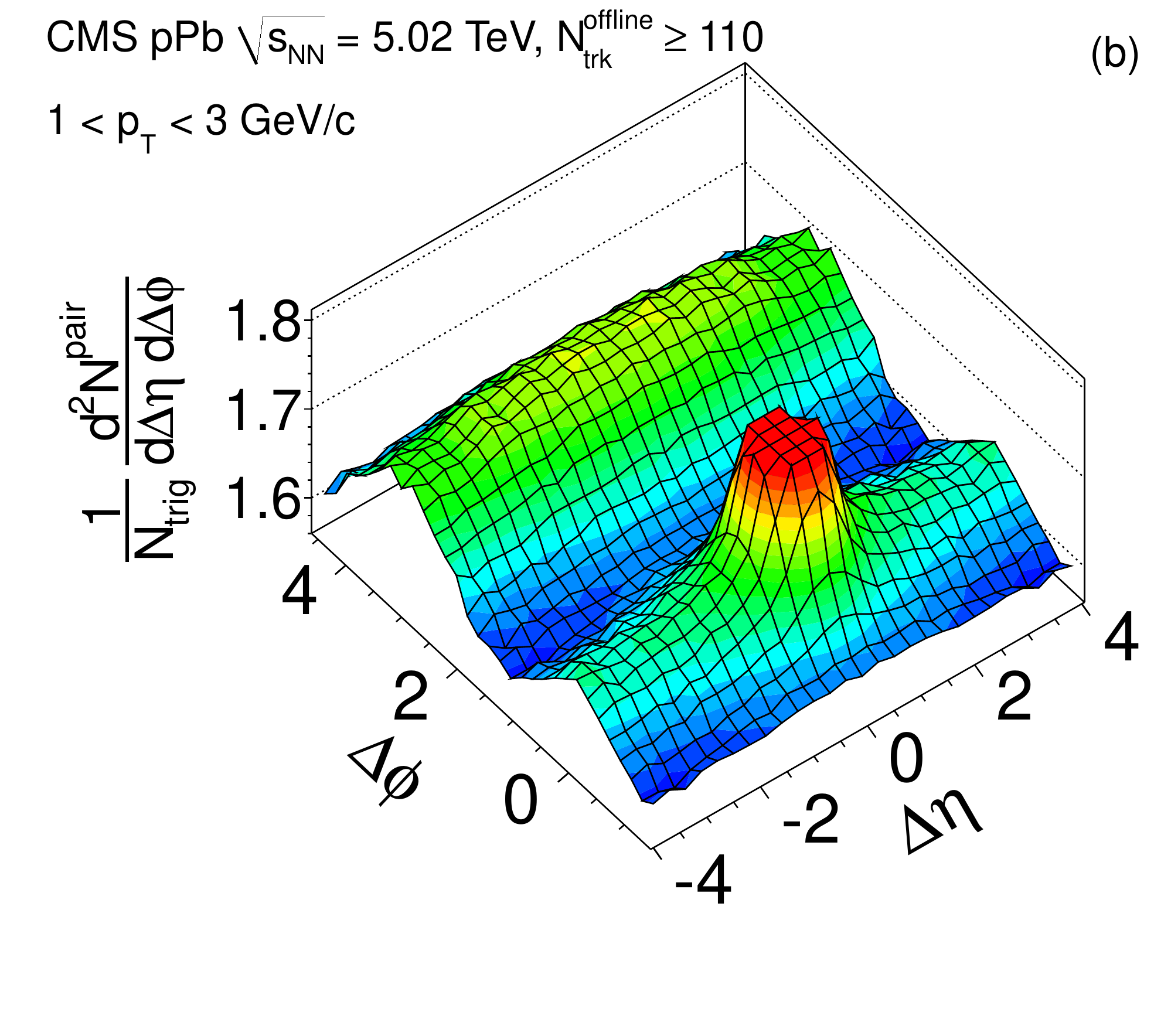}
\includegraphics[width=0.60\textwidth]{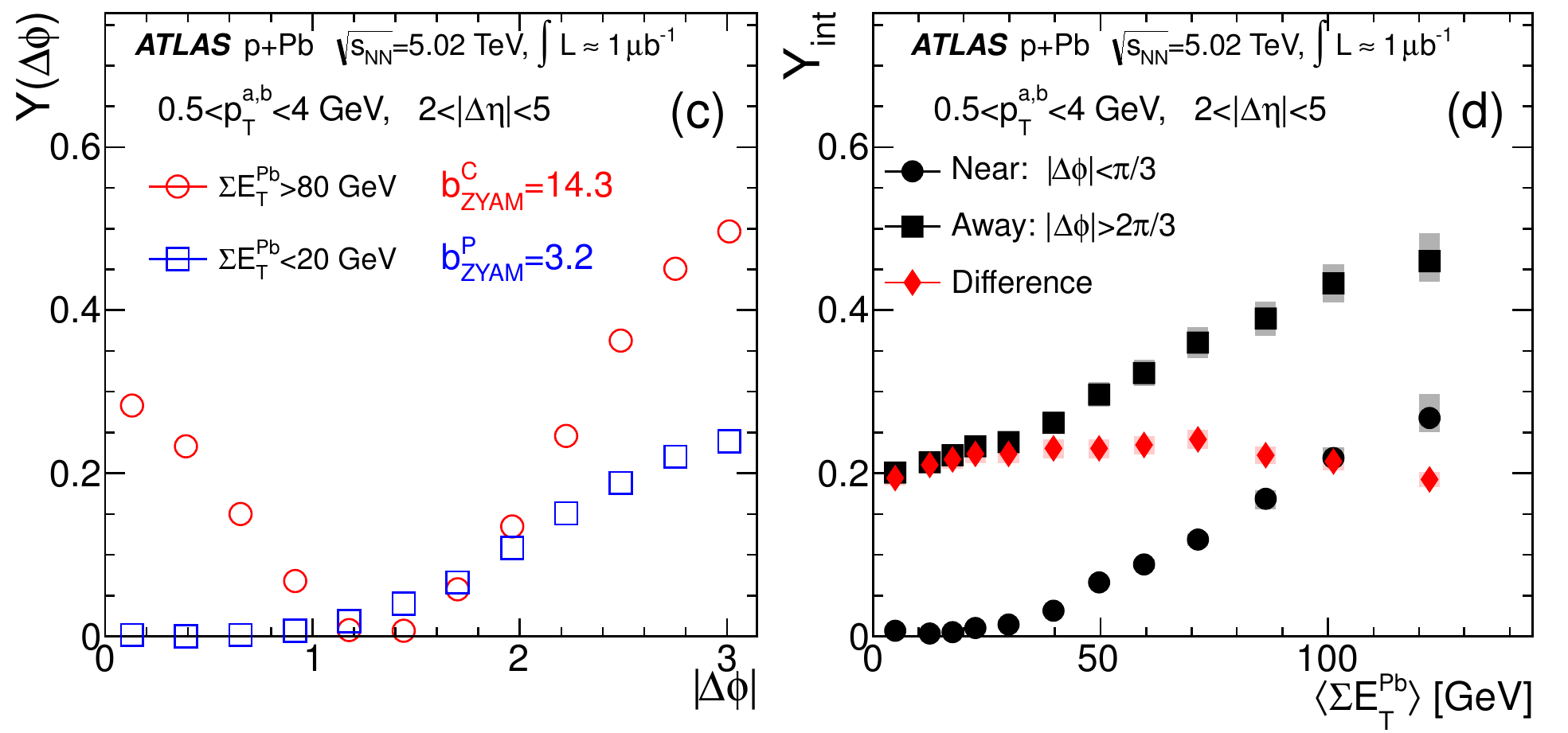}
\end{center}
\caption{Top:  2-D two-particle correlation functions for 5.02 TeV/nucleon pPb collisions for pairs of charged
particles with $1 < p_{\rm T} < 3$ GeV/c; results are shown (a) for low-multiplicity events ($N^{offline}_{trk} <
35$) and (b) for a high-multiplicity selection ($N^{offline}_{trk} \ge 110$), and the sharp near-side peaks from jet
correlations have been truncated to better illustrate the structure outside that region. Bottom: (c) the per-trigger yield $\Delta \phi$ distribution together with
pedestal levels for peripheral ($b^P_{ZYAM}$) and central ($b^C_{ZYAM}$) events, and (d) integrated per-trigger yield as function of
the sum of the transverse energy on the Pb side for pairs in $2 < |\Delta \eta | < 5$.
Taken from  CMS \cite{CMS:2012qk} and ATLAS \cite{Aad:2012gla}.} 
\label{fig:lowptcorr2}
\end{figure}

\begin{figure}[htbp]
\begin{center}
\includegraphics[width=0.48\textwidth]{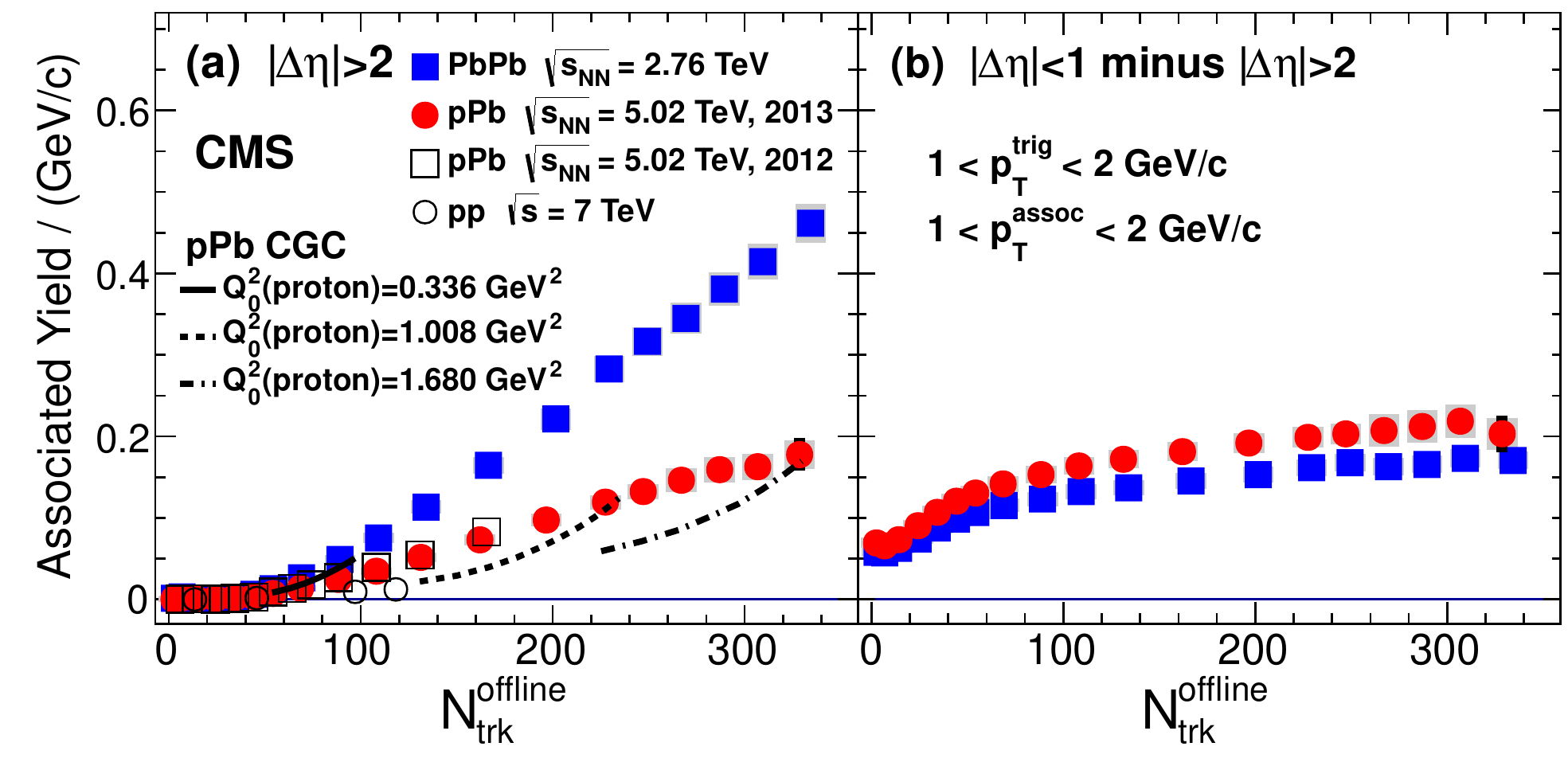}\hskip 0.3cm\includegraphics[width=0.48\textwidth]{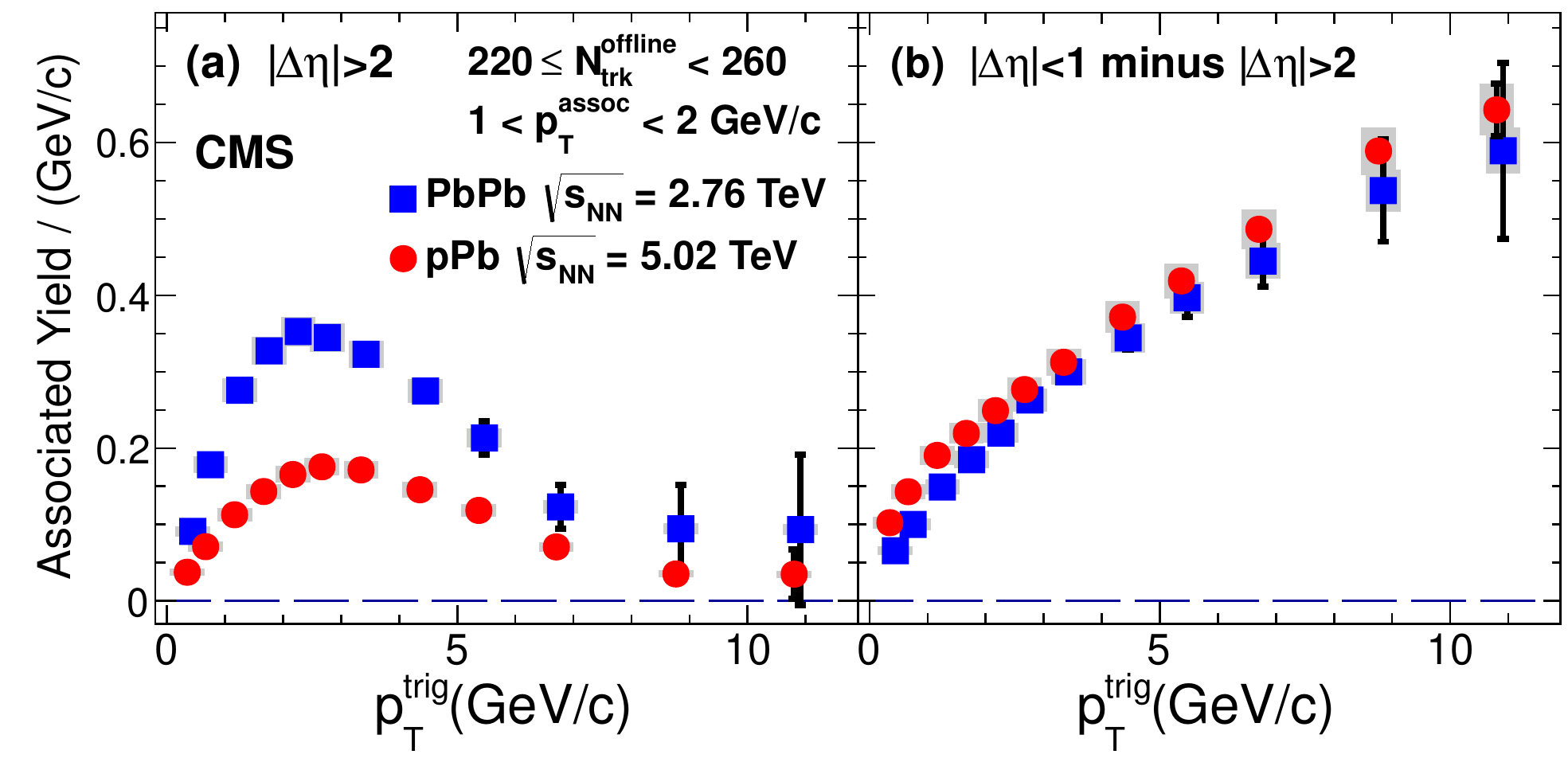}
\end{center}
\caption{Left: Associated event-normalized yields for the near-side correlation function as a function
of multiplicity $N^{offline}_{trk}$ for $1 < p^{trig}_T < 2$ GeV/c and $1 < p^{assoc}_T < 2$ GeV/c averaged over the (a) long-range ($|\Delta \eta| > 2$) region and (b) short-range
($|\Delta \eta| <1$) region, from which the event-normalized yield of the long-range region is
subtracted; the results for 7 TeV pp collisions (open circles) and 5.02 TeV/nucleon pPb
collisions from the 2012 run (open squares), as well as calculations from the Color Glass Condensate
(CGC) theory (curves), are also shown.
Right: Associated event-normalized yield for the near-side correlation function integrated
over the region $|\Delta \phi | < 1.2$, averaged over the (a) long-range ($|\Delta \eta| > 2$) region and (b) short-range
($|\Delta \eta| <1$) region, from which the event-normalized yield of the long-range region is
subtracted; the results are shown as a function of $p^{trig}_T$
at $1 < p^{assoc}_T < 2$ GeV/c for events with $220 \le N^{offline}_{trk} < 260$ for 5.02/nucleon TeV pPb collisions (filled circles) and 2.76 TeV/nucleon PbPb collisions
(filled squares).
Taken from  \cite{Chatrchyan:2013nka}.} 
\label{fig:lowptcorr3}
\end{figure}

Both the rapidity and azimuth structures have also been observed in pPb collisions \cite{CMS:2012qk,Abelev:2012ola,Aad:2012gla,Aad:2014lta,Adam:2015bka}, with again a clear transition from the absence of a near side ridge to a large signal with increasing centrality, see Fig. \ref{fig:lowptcorr2}, while the jet signal remains roughly constant with centrality \cite{Abelev:2014mva}. Finally, the Fourier azimuthal coefficients of identified particles extracted  in pPb \cite{ABELEV:2013wsa} showed the mass ordering expected if such azimuthal correlations would come from a collective flow driven by relativistic hydrodynamics. In addition, correlations between strange particles and charged hadrons in pPb and PbPb were also measured \cite{Khachatryan:2014jra}. A comparison between the yields  in pPb and PbPb at the same multiplicity \cite{Chatrchyan:2013nka}, see Fig. \ref{fig:lowptcorr3}, shows different features and magnitudes  with increasing multiplicity and transverse momenta for correlations of short- and long-range in rapidity, but a very similar size for both colliding systems. All these observations, together with the measurement of up to 8-particle correlations in pPb collisions \cite{Khachatryan:2015waa}, have  triggered an intense debate on the origin of this phenomenon and on the possibility of large degrees of collectivity in very high-multiplicity pp and  pPb collisions at LHC energies.

Many theoretical explanations have been proposed for the ridge. While not the only ones, see others in \cite{Hwa:2008um,Bjorken:2013boa,Shuryak:2013sra,Andres:2014bia}, two main lines are currently under debate. One is the pure hydrodynamical explanation that considers that, provided an initial condition that contains long-range correlations in rapidity, the coupling to an expanding medium automatically produces the azimuthal correlation \cite{Bozek:2012gr,Shuryak:2013ke,Bzdak:2013zma,Werner:2010ss,Gavin:2008ev}. The other one, based on ideas of the CGC that indeed provides a long-range rapidity correlation, considers that the azimuthal asymmetry originates from the initial correlations in the wave function of the colliding hadrons \cite{Kovner:2010xk,Levin:2011fb,Dumitru:2010iy,Dusling:2012iga,Dusling:2012wy,Altinoluk:2015uaa} e.g. Bose enhancement for gluon and the existence of domains in the hadron with oriented chromoelectric fields or density gradients. While explanations based on CGC ideas have been successful in reproducing quantitatively the ridge in pp, flow is required to reproduce its magnitude in PbPb and the situation in pPb is still unclear. This intense activity is linked with the attempts to understand the origin of the fluctuations in the initial conditions for relativistic hydrodynamics, see e.g. \cite{Dumitru:2014yza}. In any case, the apparent collectivity suggested by pp and pPb data is among the largest surprises from Run 1 at the LHC, although such possibility had already been put forward \cite{Cunqueiro:2008uu,CasalderreySolana:2009uk}.

\subsection{Hadrochemistry}
\label{hadroch}

The study of the chemical composition of the particles produced in heavy-ion collisions has been a key feature of the experimental programmes and theoretical discussions for decades \cite{Koch:1986ud}. The success of the description of hadrochemistry assuming an equilibrated system ruled  by the grand canonical ensemble, the statistical model of hadronisation \cite{BraunMunzinger:2003zd}, has been taken as evidence of both the approximate equilibration and  the partonic nature of the produced medium, as well as of the simple statistical nature of the process.

\begin{figure}[htbp]
\begin{center}
\includegraphics[width=0.38\textwidth]{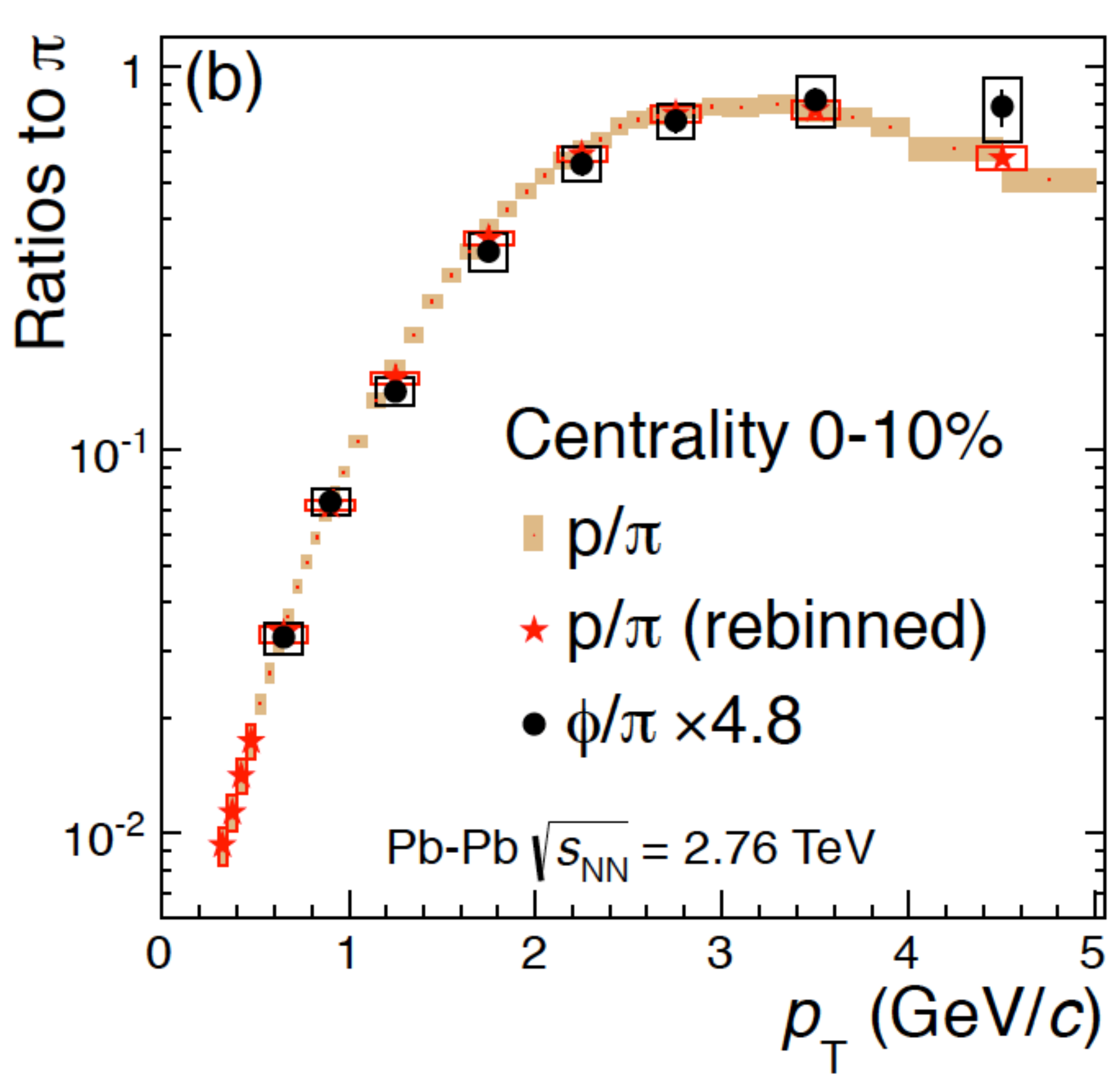}\hskip 2cm\includegraphics[width=0.38\textwidth]{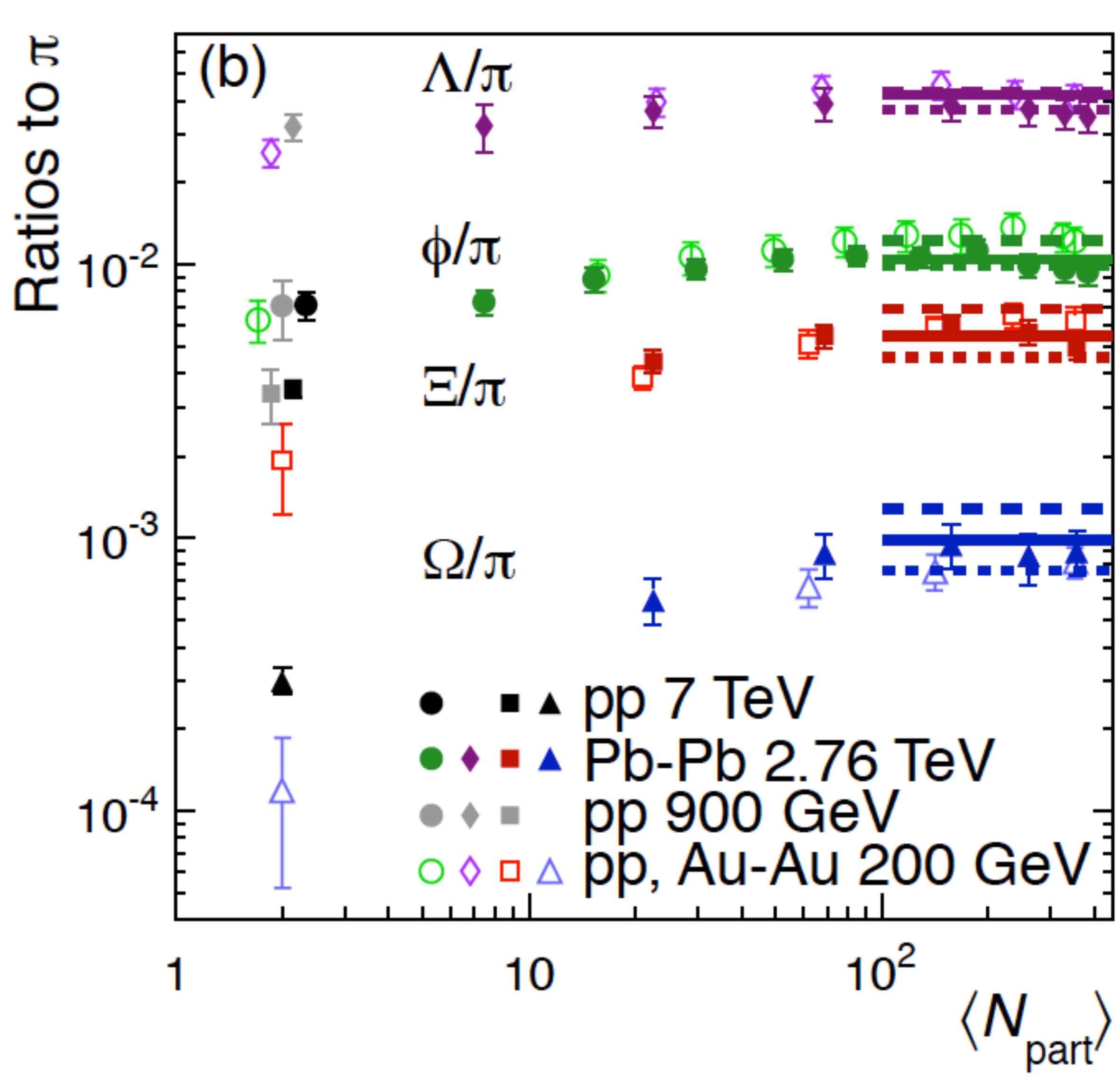}
\end{center}
\caption{Left: ratios of p and $\phi$ yields to charged pions as a function of $p_{\rm T}$ for central PbPb collisions
at 2.76 TeV/nucleon; the p$/\pi$ ratio is presented using two $p_{\rm T}$ binning schemes: the ratio with its original
measured bins is shown along with a recalculated version that uses the same bins as the $\phi$ meson $p_{\rm T}$ distribution
for $0.5 < p_{\rm T} < 5$ GeV/c.
Right: ratios of
particle yields to charged pion yields for PbÐPb collisions at 2.76 TeV/nucleon, AuÐAu collisions
at 200 GeV/nucleon, and pp collisions at 200 GeV, 900 GeV
and 7 TeV; the lines show ratios given by grand-canonical thermal models with temperatures of
170 MeV 
\cite{Cleymans:2006xj}
 (upper dashed lines), 164 MeV 
 \cite{Andronic:2008gu}
  (solid lines), and 156 MeV
  \cite{Stachel:2013zma}
   (lower dashed lines); the total
uncertainties (including centrality-uncorrelated and centrality-correlated components) are shown as bars. Some
of the measurements at $N_{part} = 2$ have been shifted horizontally for visibility; the two most central $\phi/\pi$ values
for AuÐAu collisions are for overlapping centrality intervals (0-5\% and 0-10\%).
Taken from ALICE \cite{Abelev:2014uua}.} 
\label{fig:hadrochem1}
\end{figure}

In PbPb collisions at the LHC, ALICE has made a wealth of 
$p_{\rm T}$-differential analyses of identified particle production. In \cite{ABELEV:2013zaa} and \cite{Abelev:2014uua}, the relative yields of multistrange hadrons and of $\phi$ mesons have been measured. As shown in Fig. \ref{fig:hadrochem1}, the $\phi$ mesons exhibit, with respect to pions, a relative enhancement with increasing $p_{\rm T}$ that closely follows that of protons, suggesting that a mass effect is present, as resulting from hydrodynamic calculations where the radial flow is linked to the mass of the particles (but other options exist, see e.g. \cite{Ortiz:2013yxa}).
The relative ratios of multistrange particles and $\Lambda$'s with respect to pions increase gradually with centrality, and agree with the expectations of the statistical models for mid-central and central collisions. The same trends have been observed in the $\Lambda/K_S^0$ ratios  \cite{Abelev:2013xaa}, or for the transverse momentum and centrality dependence of pions, kaons and protons \cite{Abelev:2012wca,Abelev:2013vea} that show a behaviour compatible with relativistic hydrodynamics.

\begin{figure}[htbp]
\begin{center}
\includegraphics[width=0.4\textwidth]{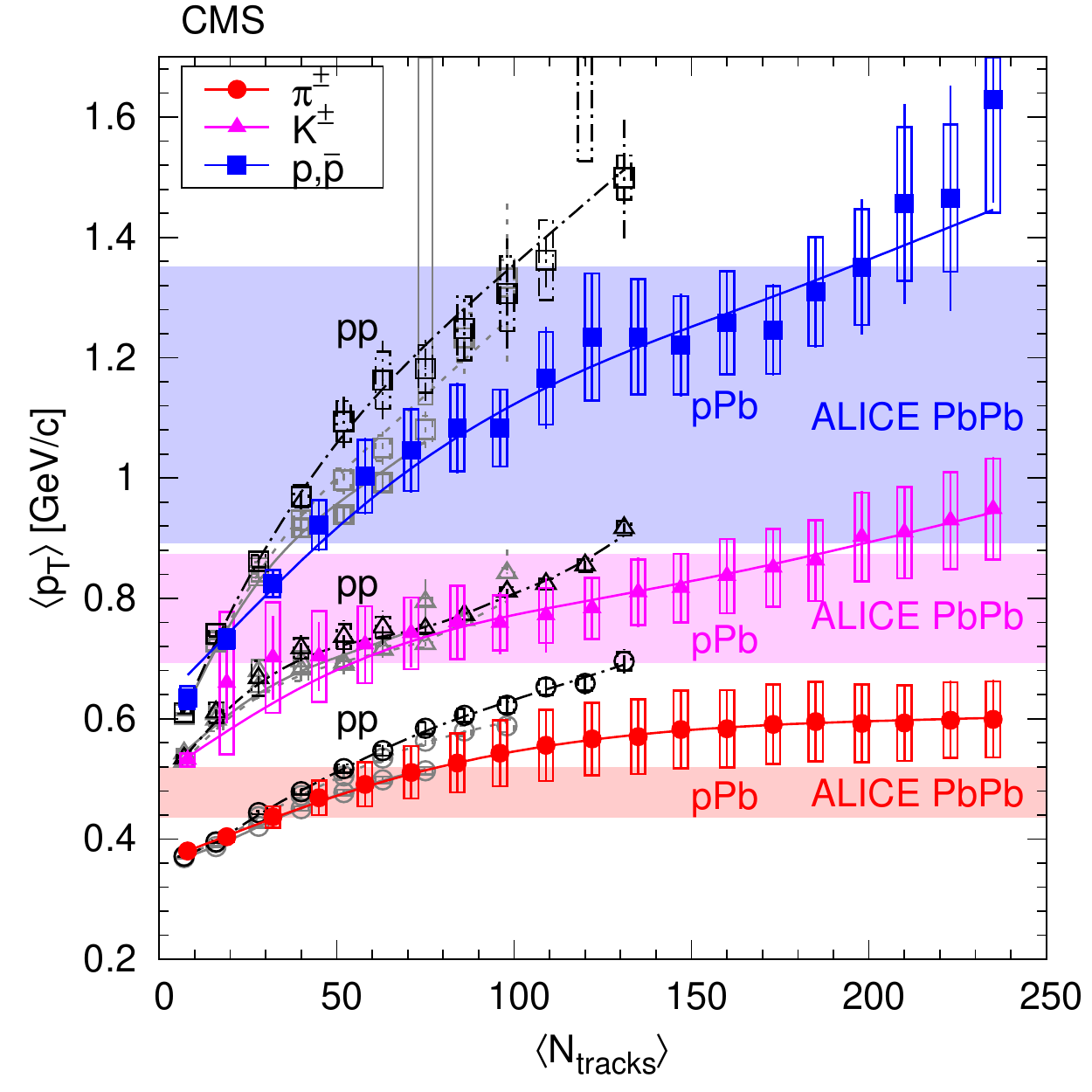}\hskip 2cm\includegraphics[width=0.4\textwidth]{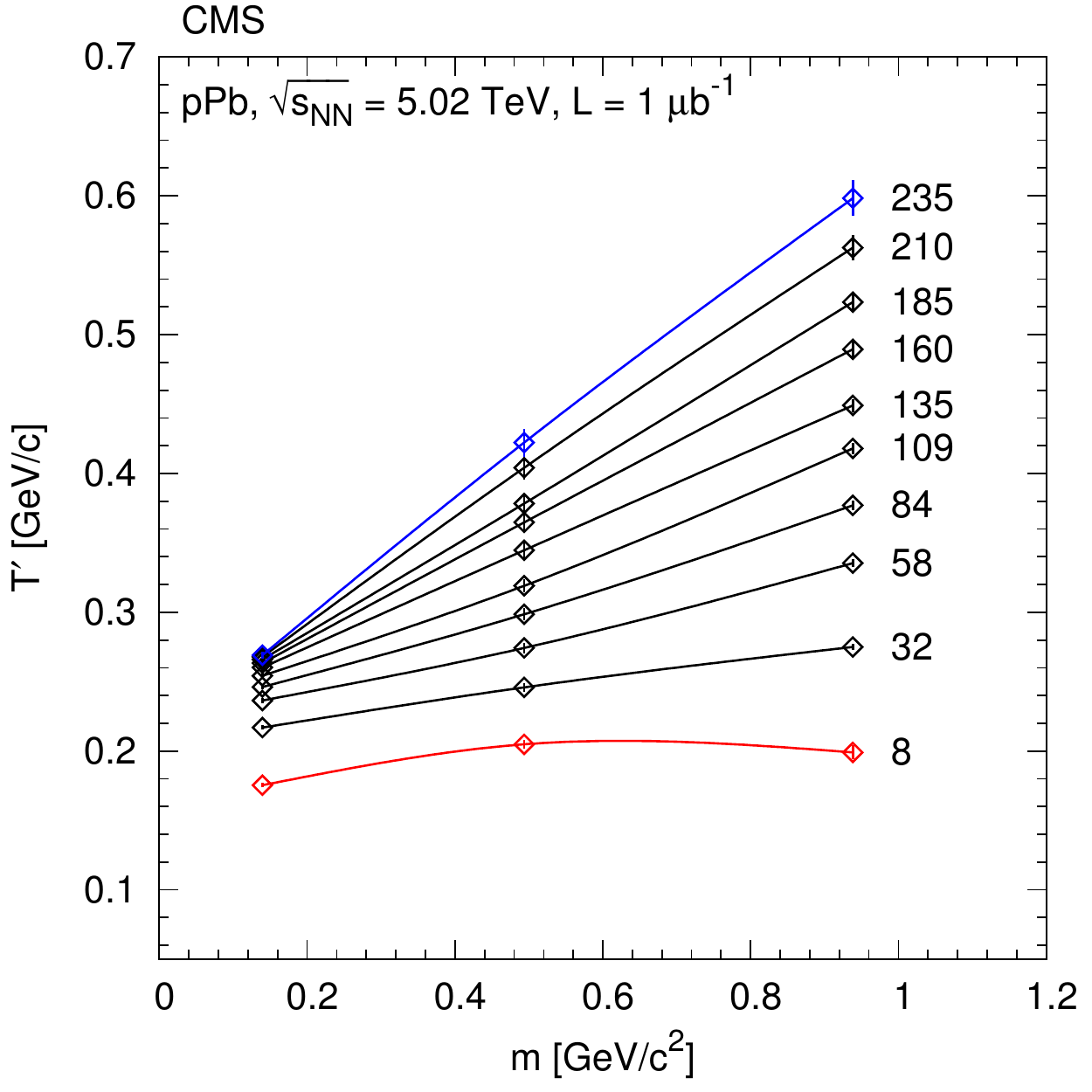}
\end{center}
\caption{Left: average transverse momentum of identified charged hadrons (pions, kaons, protons)  in the range $|y| < 1$ as a function of the corrected
track multiplicity for $|\eta | < 2.4$, for pp collisions (open symbols) at several energies
\cite{Chatrchyan:2012qb},
and for pPb collisions (filled symbols) at 5.02 TeV/nucleon; both $\langle p_{\rm T}\rangle $ and yield ratios were
computed assuming a Tsallis-Pareto distribution in the unmeasured range; lines are drawn to guide the eye (gray solid Ð pp 0.9 TeV, gray dotted Ð pp 2.76 TeV, black
dash-dotted Ð pp 7 TeV, colored solid Ð pPb 5.02 TeV/nucleon); the ranges of $\langle p_{\rm T}\rangle $ values
measured by ALICE in various centrality PbPb collisions at 2.76 TeV/nucleon \cite{Abelev:2013vea} are
indicated with horizontal bands.
Right:  inverse slope parameters $T^\prime$
from fits of pion, kaon, and proton spectra (both charges)
with a form proportional to $p_{\rm T} \exp(-m_{\rm T}/T^\prime)$; results for a selection of multiplicity classes,
with different $\langle N_{tracks}\rangle$ as indicated, are plotted for pPb data; the curves are drawn to guide the eye.
Taken from  \cite{Chatrchyan:2013eya}.} 
\label{fig:hadrochem3}
\end{figure}

In pPb collisions, see Fig. \ref{fig:hadrochem3}, both ALICE \cite{Abelev:2013haa}  and CMS \cite{Chatrchyan:2013eya} have measured a hardening of the transverse momentum spectra of different particle species, larger for particles with larger masses, and smoothly increasing with centrality.
This hardening with increasing hadron multiplicity is present in both pp, pPb and PbPb
and is more pronounced for smaller systems, a feature also observed for unidentified charged particle production.
Noticeably, the observed mean transverse momentum or the extracted effective temperature show mass ordering like in PbPb. Besides, ALICE has measured the spectrum of $\phi$ mesons \cite{Adam:2015jca}, that shows a large enhancement (compared to scaled pp) for intermediate transverse momentum ($\sim 3-4$ GeV/c) in the backward (Pb-going) region and suppression at smaller transverse momentum in the forward (p-going) region.

\begin{figure}[htbp]
\begin{center}
\includegraphics[width=0.48\textwidth]{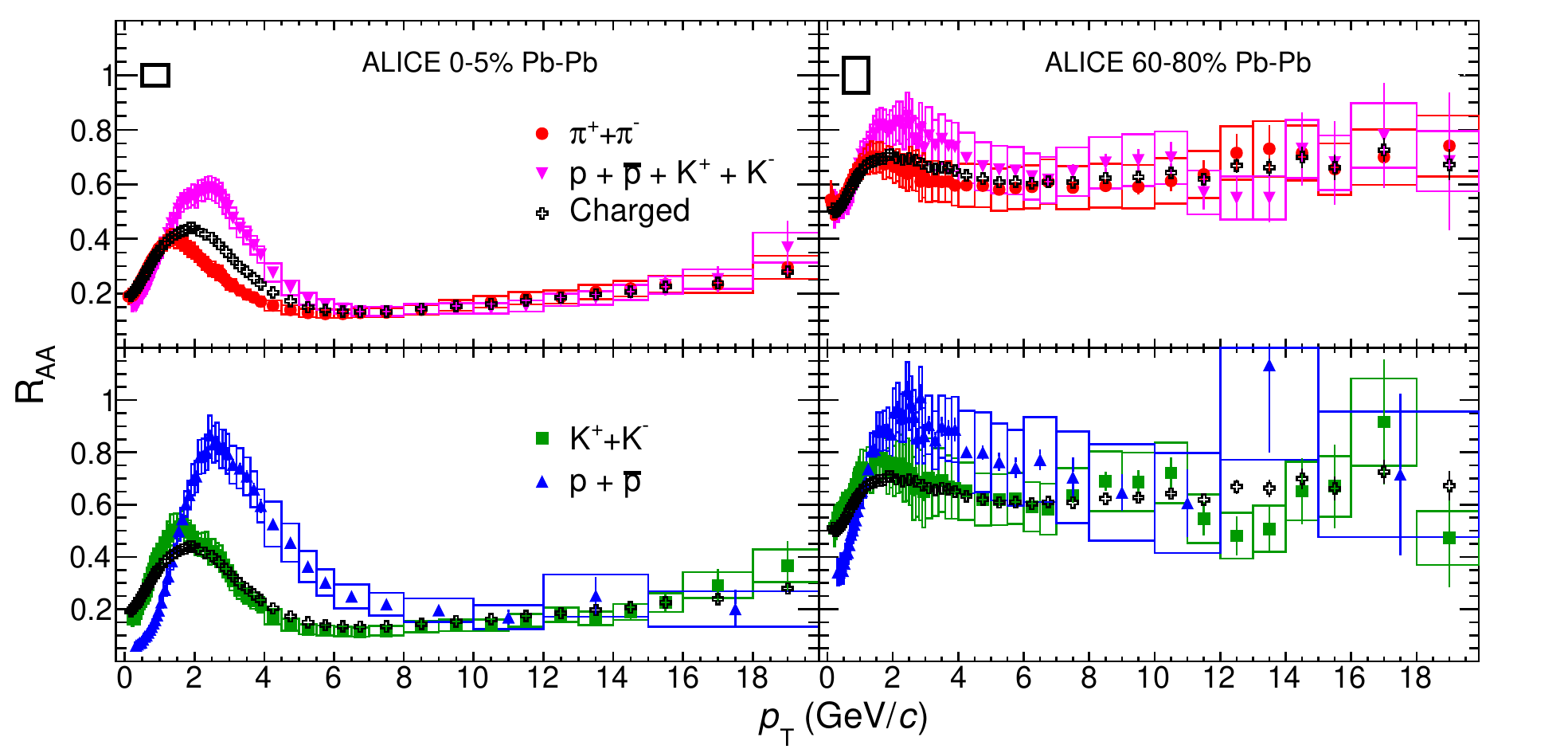}\hskip 0.3cm \includegraphics[width=0.48\textwidth]{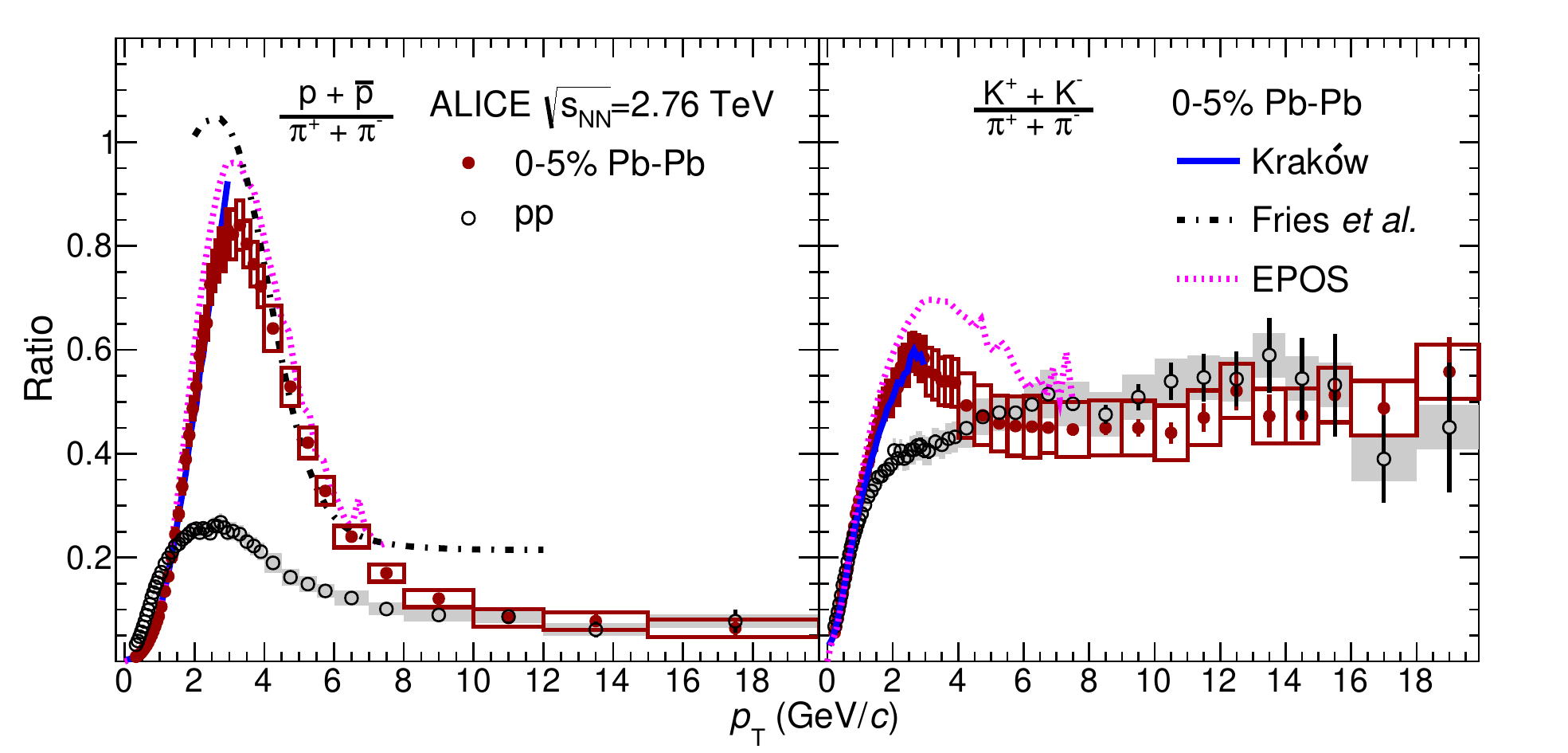}
\end{center}
\caption{Left: nuclear modification factor $R_{\rm AA}$ as a function of $p_{\rm T}$ for different particle species; results for 0-5\%
(left) and 60-80\% (right) collision centralities are shown.
Right: particle ratios as a function of $p_{\rm T}$ measured in pp and the most central, 0-5\%, PbÐPb collisions; the theoretical predictions
refer to PbÐPb collisions.
Taken from  \cite{Abelev:2014laa}.} 
\label{fig:hadrochem2}
\end{figure}

The spectra of $\pi^0$ at mid-rapidity has been measured for different centralities in PbPb collisions 
by ALICE \cite{Abelev:2014ypa}. Their nuclear modification factor (see the definition in Section \ref{hard}) shows a suppression slightly larger than that found at RHIC and compatible with the one of charged hadrons that we will discuss in more length in Subsection \ref{largept}. Interestingly, see Fig. \ref{fig:hadrochem2}, ALICE \cite{Abelev:2014laa,Adam:2015kca} has also studied the nuclear modification factor for pions, kaons and protons for different centralities and up to $p_{\rm T} \sim 20$ GeV/c. For all centralities, the nuclear modification factors for all species, and the relative chemical composition in pp and PbPb, become equal above $p_{\rm T} \sim 8$ GeV/c, indicating that hadronisation of such particles is not affected by the medium and that the origin of the suppression is partonic - as already observed at RHIC but for smaller transverse momenta \cite{Adare:2008qa,Adare:2010dc,Agakishiev:2011dc}. This poses strong constrains on ideas that consider non-perturbative \cite{Fries:2003kq}
 or perturbative modifications of hadronisation inside the medium  \cite{Sapeta:2007ad}.

Finally, ALICE has also presented results \cite{Adam:2015nca,Adam:2015yta,Adam:2015vda} on weakly decaying bound states of $\Lambda$, $\overline \Lambda$ and (anti)nucleons, and on hypertritons, tritons, deuterons and their corresponding antinuclei.
%Such data constrain the hadronisation mechanism in thermal  or coalescence models \cite{Andronic:2010qu,Csernai:1986qf}. 
Furthermore, studies of the mass-over-charge differences between light nuclei and antinuclei confirm CPT invariance with improved precision in the nuclear domain \cite{Adam:2015pna}.

\subsection{Femtoscopy}
\label{femto}
The size of the particle-emitting region at freeze-out can be extracted via two-particle correlations
of identical particles, usually pions (referred to as Bose-Einstein, or Hanbury-Brown--Twiss ``HBT'', correlations).
This technique, also known as femtoscopy, was first used in particle physics in the '60s~\cite{Goldhaber:1960sf} and measures the apparent width of the distribution of relative separation of
emission points, which is conventionally called the ''radius parameter''. The radius can be determined independently
for three directions: $R_{\rm long}$ along the beam axis, $R_{\rm out}$ along the pair transverse momentum, and
$R_{\rm side}$, perpendicular to the other two. Femtoscopic analyses were extensively carried out already at AGS, SPS
and RHIC energies, providing valuable information~\cite{Lisa:2005dd}. Among the main findings in nucleus-nucleus collisions, 
a linear increase of the radius parameters was measured as a function of $\langle{\rm d}N_{\rm ch}/{\rm d}\eta\rangle^{1/3}$. 
In addition, a significant decrease of the radii with the momentum of the pair was also established. This is a characteristic 
feature of expanding particle sources since the HBT radii describe the homogeneity length~\cite{Akkelin:1995gh} rather than the overall size 
of the particle-emitting system. The homogeneity length is defined as the size of the region that
contributes to the pion spectrum at a particular three-momentum $\vec{p}$. The decrease of the size with
$k_{\rm T}$, defined as half of the modulus of the vector sum of the two transverse momenta, is observed in experimental data from heavy-ion collisions at all centralities, various collision energies
and colliding system types, and is well described quantitatively in hydrodynamic models~\cite{Broniowski:2008vp,Karpenko:2012yf}.

\begin{figure}[htbp]
\begin{center}
\includegraphics[width=0.27\textwidth]{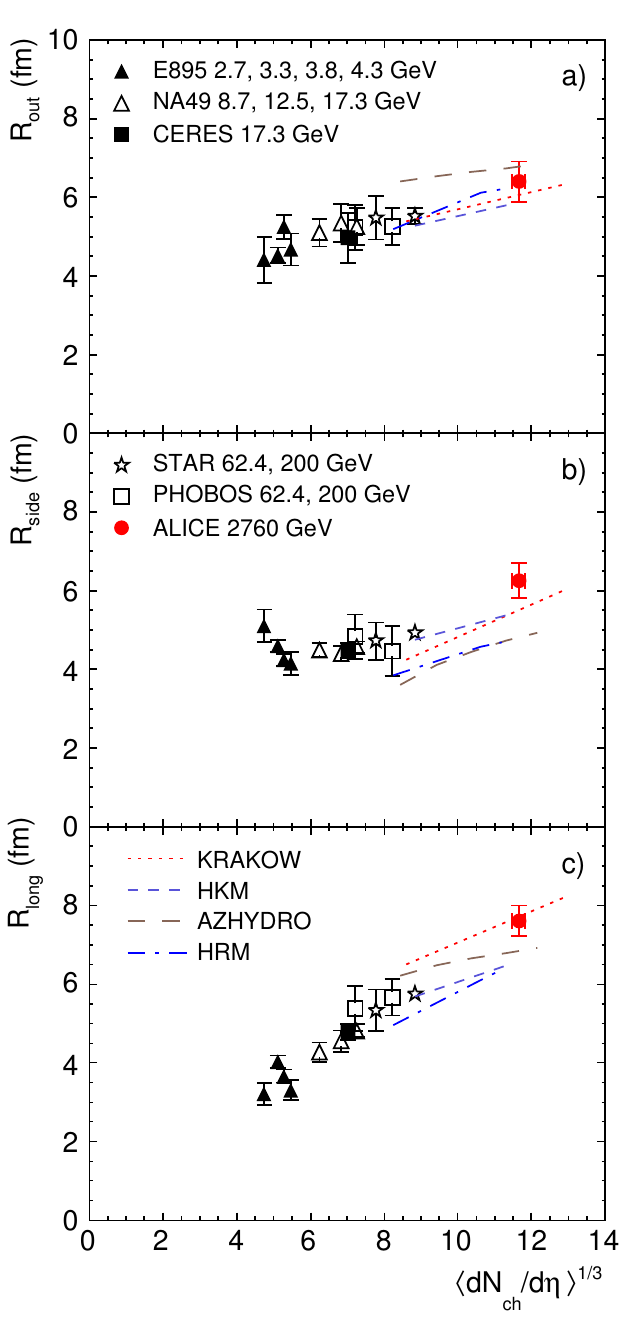} \includegraphics[width=0.55\textwidth]{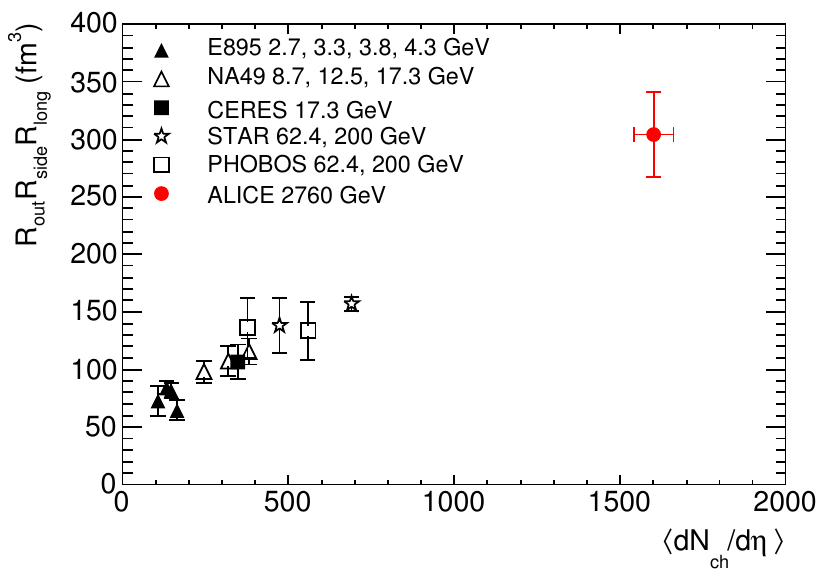}
\end{center}
\caption{Left: Pion HBT radii at $k_{\rm T}=0.3$ GeV/$c$ for the 5\% most central PbPb at $\sqrt{s_{\rm NN}}=2.76$ TeV 
(red filled dot) and the radii obtained for central gold and lead collisions 
at lower energies at the 
AGS~\cite{Lisa:2000hw}, SPS~\cite{Alt:2007uj,Afanasiev:2002mx,Adamova:2002wi},   and RHIC~\cite{Abelev:2009tp,Back:2004ug,Back:2005hs,Back:2002wb,Adams:2004yc,Abelev:2008ab}. Right: product of the three pion HBT radii at $k_{\rm T}=0.3$ GeV/$c$. The ALICE result (red filled dot) is compared to those obtained 
for central gold and lead collisions at lower energies.} 
\label{fig:femtocentral}
\end{figure}

At the LHC, the ALICE experiment has performed an extensive set of femtoscopic measurements~\cite{Aamodt:2011mr,Abelev:2013pqa,Abelev:2014pja,Adam:2015pya,Adam:2015vja,Adam:2015vna}. In Fig.~\ref{fig:femtocentral}(left)
the results for 0--5\% Pb--Pb collisions at $\sqrt{s_{\rm NN}}=2.76$ TeV~\cite{Aamodt:2011mr} are compared, separately for the three radius
parameters, to lower energy experiment results. The LHC values are higher than the top RHIC energy values
by 10-35\%, and the scaling with $\langle{\rm d}N_{\rm ch}/{\rm d}\eta\rangle^{1/3}$ is seen to hold up to LHC energies.
The product of the radii is connected to the volume of the homogeneity region and is shown in Fig.~\ref{fig:femtocentral}(right)
for $k_{\rm T}=0.3$ GeV/$c$~\cite{Aamodt:2011mr}. Compared with previous results it shows a linear dependence on the charged-particle
pseudorapidity density and is two times larger at the LHC than at RHIC. Estimates of the decoupling time for pions, based on 
$R_{\rm long}$ measurements, exceeds 10 fm/$c$, and are 40\% larger than at RHIC.

\begin{figure}[htbp]
\begin{center}
\includegraphics[width=0.35\textwidth]{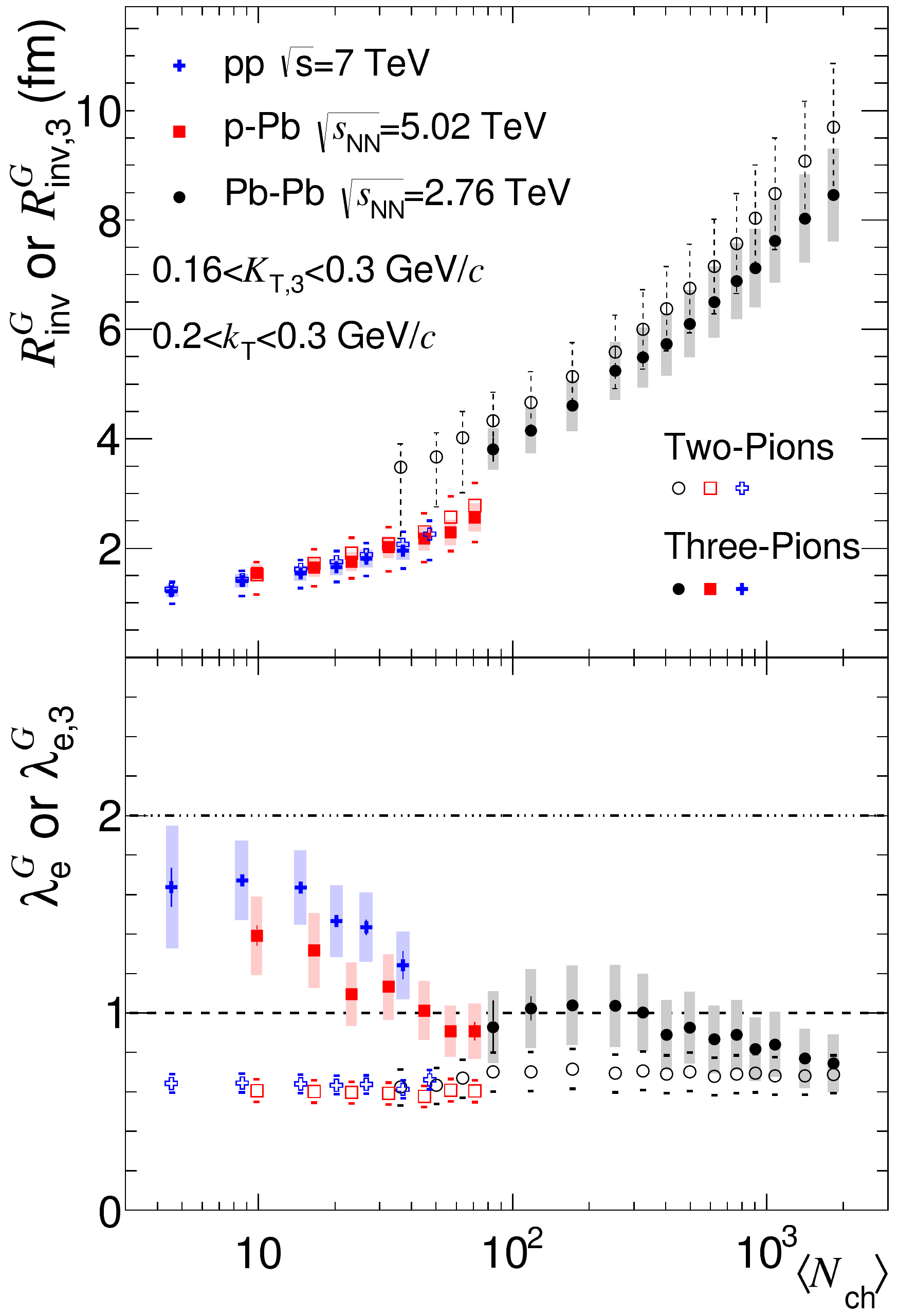} \includegraphics[width=0.5\textwidth]{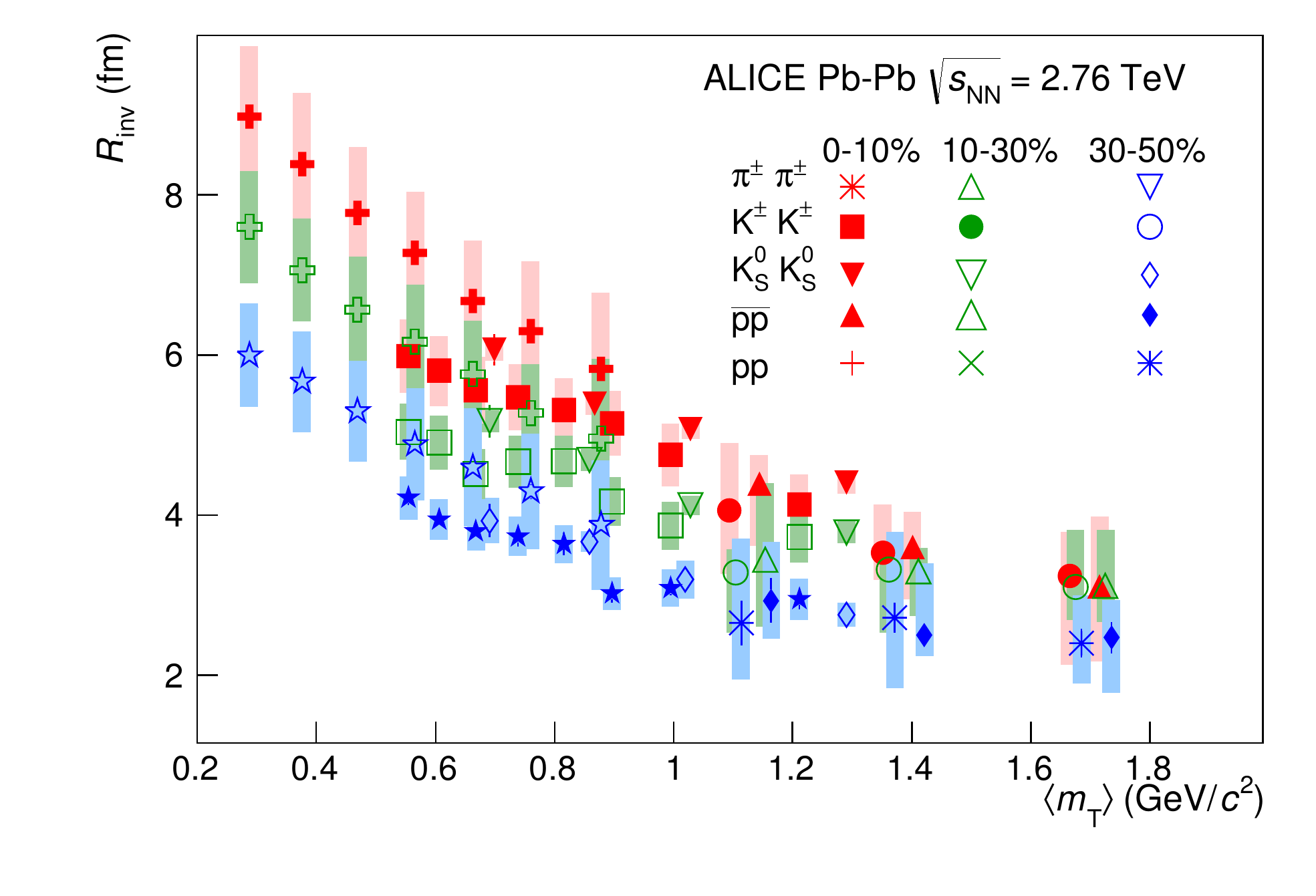}
\end{center}
\caption{Two- and three-pion Gaussian fit parameters versus 
$\langle N_{\rm ch}\rangle$ in pp, pPb and PbPb collision systems for low average values of the two- and three-pion average transverse momenta, $k_{\rm T}$ and $K_{\rm T,3}$. The top panel shows the Gaussian radii $R^{\rm G}_{\rm inv}$ and $R^{\rm G}_{\rm inv,3}$. The bottom panel (not discussed in the text) shows the effective Gaussian intercept parameters $\lambda^{\rm G}_{\rm e}$ and $\lambda^{\rm G}_{\rm e,3}$ (see~\cite{Abelev:2014pja} for details).
Right: $R_{\rm inv}$ parameters vs. $m_{\rm T}$ for the three centralities considered for $\pi^{\pm}\pi^{\pm}$, $K^{\pm}K^{\pm}$, $K_{\rm S}^0 K_{\rm S}^0$, pp, and ${\overline {\rm p}}{\overline {\rm p}}$~\cite{Adam:2015vja}. Statistical (thin lines) and systematic (boxes) uncertainties are shown.} 
\label{fig:femtosystems}
\end{figure}

As a further study, two- and three-pion correlations of same and mixed-charge were measured at low relative momentum to estimate 
the coherent fraction of charged pions in PbPb collisions~\cite{Abelev:2013pqa}. The presence of coherent production reduces the maximum of Bose-Einstein
correlations and the estimated fraction was found to be 23$\pm8$\% for low triplet momentum.
Comparisons of HBT parameters were also carried out between pp, pPb and PbPb (see Fig.~\ref{fig:femtosystems}(left), where the radii obtained from a Gaussian fit of the two- and three-pions correlation functions are reported)~\cite{Abelev:2014pja}. 
At similar multiplicity, a modest increase of
the parameters was found between pp and pPb (5-15\%), while the difference between pPb and PbPb is larger (35-55\%). The 
weak increase between pp and pPb favours models with a CGC initial phase~\cite{Bzdak:2013zma} with respect to those assuming
a hydrodynamic evolution of the system~\cite{Bozek:2013df}. Finally, femtoscopic analyses have also been carried out in Pb--Pb collisions 
for identified particles, including $\pi^\pm$, $K^\pm$, $K_S^0$, p and $\overline{\rm p}$~\cite{Adam:2015vja}. Common $m_{\rm T}$ scaling of the 
$\pi$ and $K$ parameters, with a decrease of the radii with increasing $m_{\rm T}$, were predicted to be a robust signature of
hydrodynamic flow~\cite{Makhlin:1987gm}. Furthermore, $K$ results are considered as a cleaner signal due to a smaller decay contribution, and 
(anti)proton results provide a possibility for checking if baryons
are included in the collective motion. In Fig.\ref{fig:femtosystems}(right), results on 
$R_{\rm inv}$, the invariant radius from two-particle correlations evaluated in the pair-rest frame are shown.
For overlapping $m_{\rm T}$, the radius parameters are mostly consistent with each other within uncertainties, though the pion radii are
generally larger than the kaon radii (the latter effect can be explained as a consequence of the increase of the Lorentz factor with decreasing particle mass~\cite{Kisiel:2014upa}). The radius parameters show an increase with  centrality as would be expected from a simple geometric picture of the collisions. They
also show a decreasing size with increasing $m_{\rm T}$, as would be expected in the presence of collective radial
flow. The results are found to be in good agreement with hydrodynamical models~\cite{Shapoval:2014wya}.

\section{Hard and electromagnetic probes}
\label{hard}

A powerful way to analyse the medium created in high-energy heavy-ion collisions comes from the study of probes\footnote{Due to the transient nature of the medium, they must be self-generated.} created in hard scatterings. Such probes, due to their large scale (mass, energy, transverse momentum,$\dots$, larger than any medium scale like average transverse momentum, temperature, saturation scale, $\Lambda_{QCD}$,$\dots$), 
are: (a) clearly separated from the medium that they analyse, (b) accurately computable in perturbative QCD so that the separation between production of the hard probe and propagation through the medium is under theoretical control, and (c) produced very early in the collision. In this way, it
is possible to establish their initial flux and to study the change in observed yields or cross sections due to the presence of the medium. Experimentally,
this is done by comparing results on pp and pA/AA collisions. The quantity usually employed to quantify such variation is the nuclear modification factor
\begin{equation}
R_{AA}(y,p_{\rm T})=\frac{\frac{d^2N^{\rm AA}}{dydp_{\rm T}}}{\langle N_{coll}\rangle \frac{d^2N^{\rm pp}}{dydp_{\rm T}}},
\end{equation}
with $\langle N_{coll}\rangle$ the number of collisions in a given centrality class, usually computed in the frame of the Glauber model. With this normalisation, the nuclear modification factor would be 1 if no nuclear effect exists, i.e., if an AA collision were a mere superposition of nucleon-nucleon ones. A similar quantity, called $R_{CP}$, can also be defined using a peripheral class for reference instead of pp. The numerator and denominator may be not yields but cross sections, in which case the normalisation factor is different.

These probes are generically known as {\it hard probes} and include high transverse momentum particles, jets, heavy flavours and quarkonium states. In addition, there are also other penetrating probes, so-called {\it electromagnetic} - those provided by particles that do not interact strongly with the medium: electro-weak bosons, and real and virtual photons at high and low transverse momentum.  Electro-weak bosons, and real and virtual photons at high transverse momentum, constitute control probes that check our understanding of perturbative QCD in nuclear collisions, in the absence of any medium, particularly for determining the nuclear modification of parton densities nPDFs\footnote{The study of hard probes in pA collisions is mandatory not only for the determination of nPDFs but also for the understanding of other CNM effects like absorption of produced bound states in normal nuclear matter and, in general, of factorisation in nuclear collisions.}. Low transverse momentum photons and dileptons are instead directly sensitive to the medium, because they can be thermally emitted by the medium itself. We refer the reader to the reviews \cite{qgp1,qgp2,qgp3,qgp4,Abreu:2007kv,Armesto:2009ug,Accardi:2004be,Accardi:2004gp,Bedjidian:2004gd,Arleo:2004gn,Majumder:2010qh,Mehtar-Tani:2013pia,Ghiglieri:2015zma,Blaizot:2015lma,jqqgp5} for extensive information.

\subsection{Particle production at large transverse momentum}
\label{largept}

The suppression of particles produced at large transverse momentum and the disappearance of back-to-back correlations was one of the golden measurements at RHIC that established the dense partonic nature of the medium produced in nucleus-nucleus collisions, see \cite{Adler:2003au,Adams:2003kv,Adams:2003im,Adler:2005ee,Adams:2005ph}. At the LHC, analogous results have been found for the nuclear modification factor \cite{Aamodt:2010jd,CMS:2012aa,Abelev:2012hxa,Aad:2015wga,Abelev:2014ypa,Abelev:2014laa,Adam:2015kca} and for the suppression of back-to-back correlations \cite{Aamodt:2011vg}.

\begin{figure}[htbp]
\begin{center}
\includegraphics[width=0.46\textwidth]{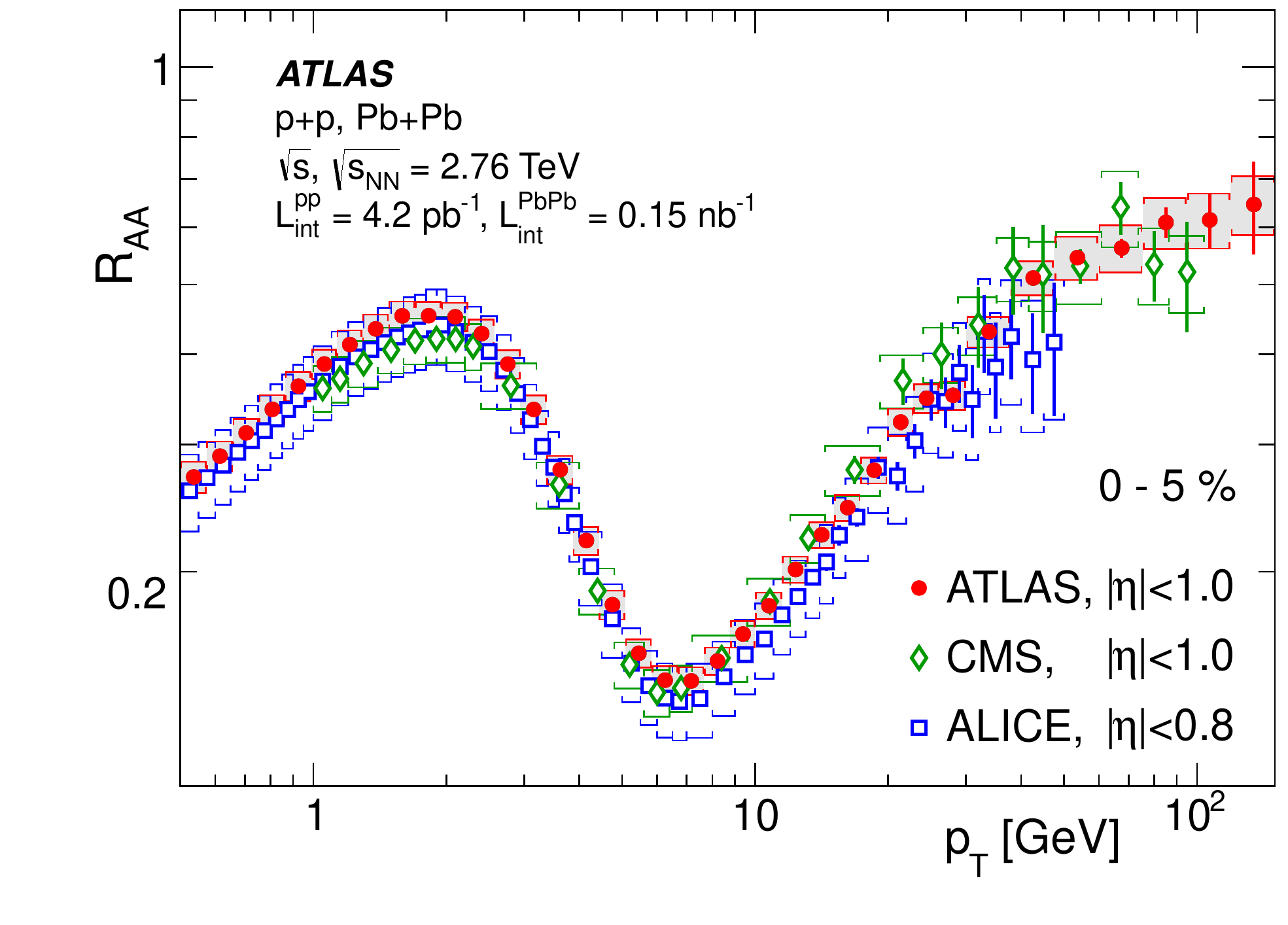}\hskip 1cm\includegraphics[width=0.47\textwidth]{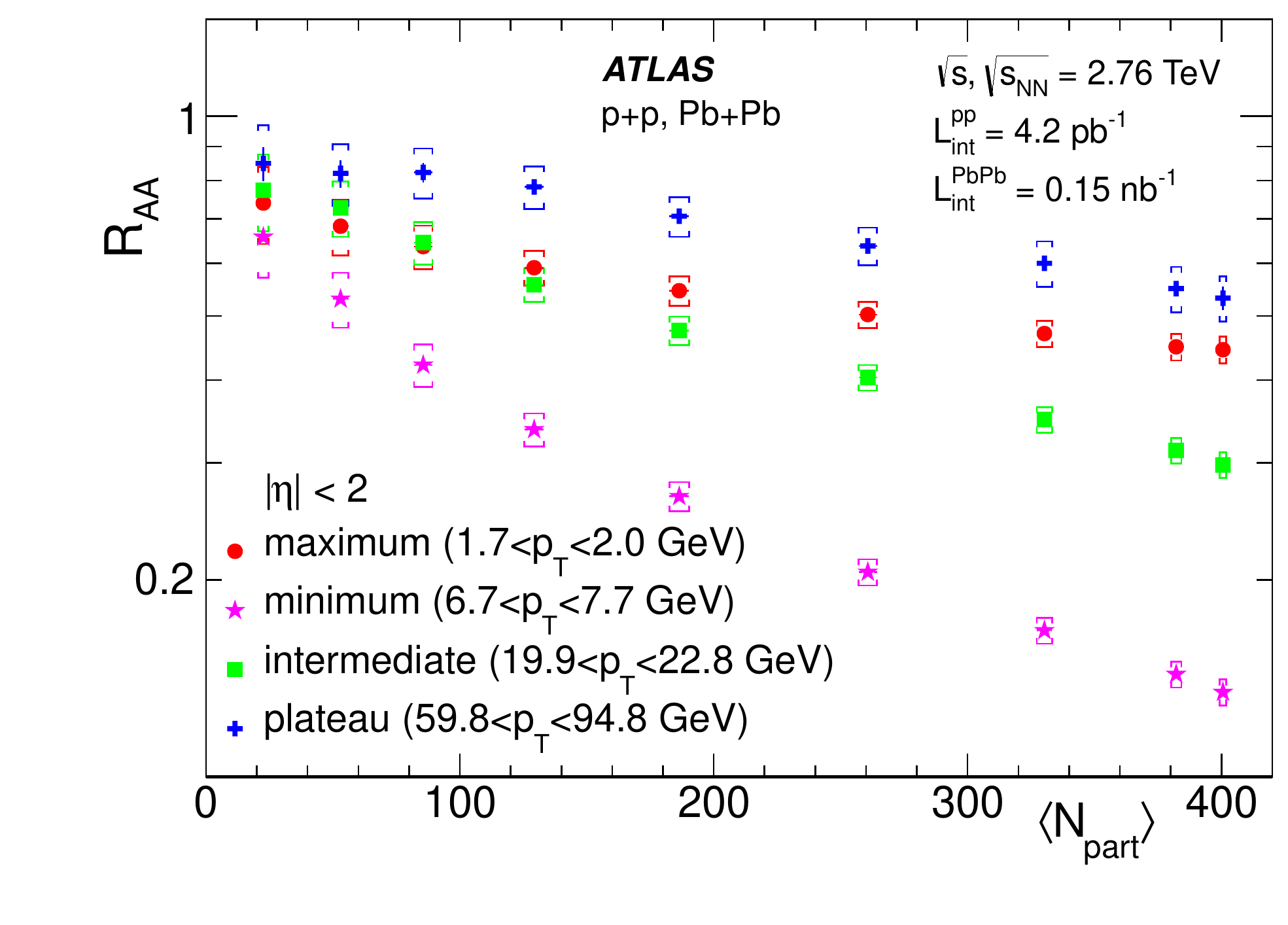}
\end{center}
\caption{The nuclear modification factor for charged particles measured by ATLAS \cite{Aad:2015wga}, versus transverse momentum (left, compared to CMS \cite{CMS:2012aa} and ALICE \cite{Abelev:2012hxa} results) and versus the number of participants (right). Taken from  \cite{Aad:2015wga}.} 
\label{fig:RAAparticles1}
\end{figure}

In Fig. \ref{fig:RAAparticles1}, results for the nuclear modification factor of charged particles versus transverse momentum and number of participants are shown, taken from \cite{Aad:2015wga}. The measurements extend to $p_{\rm T}>100$ GeV that corresponds to much higher jet transverse momentum. The agreement between different experiments when the results are plotted for similar centralities and rapidity windows is notable. For that, the use of a reference pp spectrum measured at the same centre-of-mass energy, 2.76 TeV, has been crucial (see the uncertainties due to interpolation of the reference in \cite{Aamodt:2010jd}). A steady decrease with increasing centrality is seen for all transverse momenta.

\begin{figure}[htbp]
\begin{center}
\includegraphics[width=0.35\textwidth]{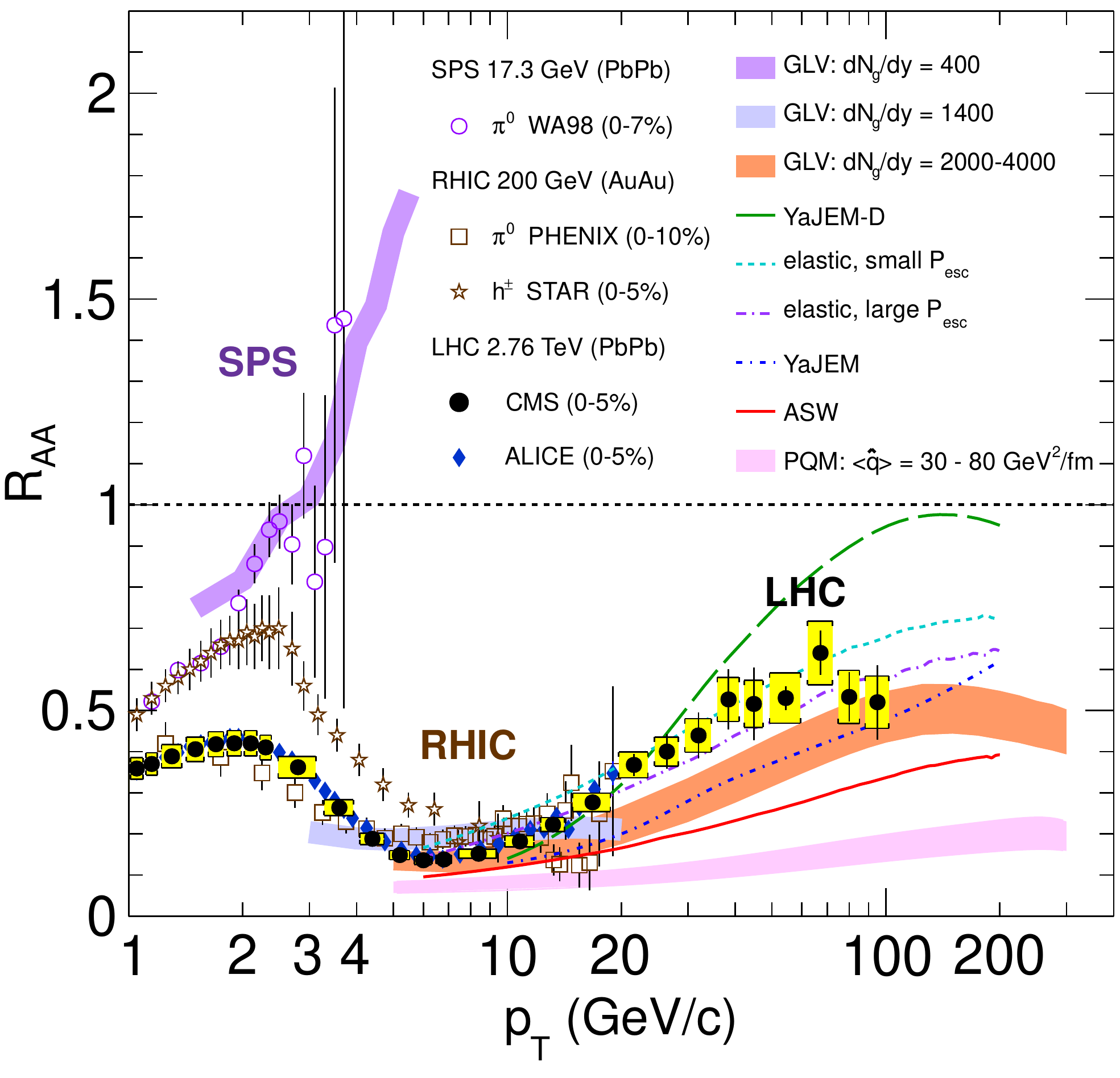}\hskip 1cm\includegraphics[width=0.42\textwidth]{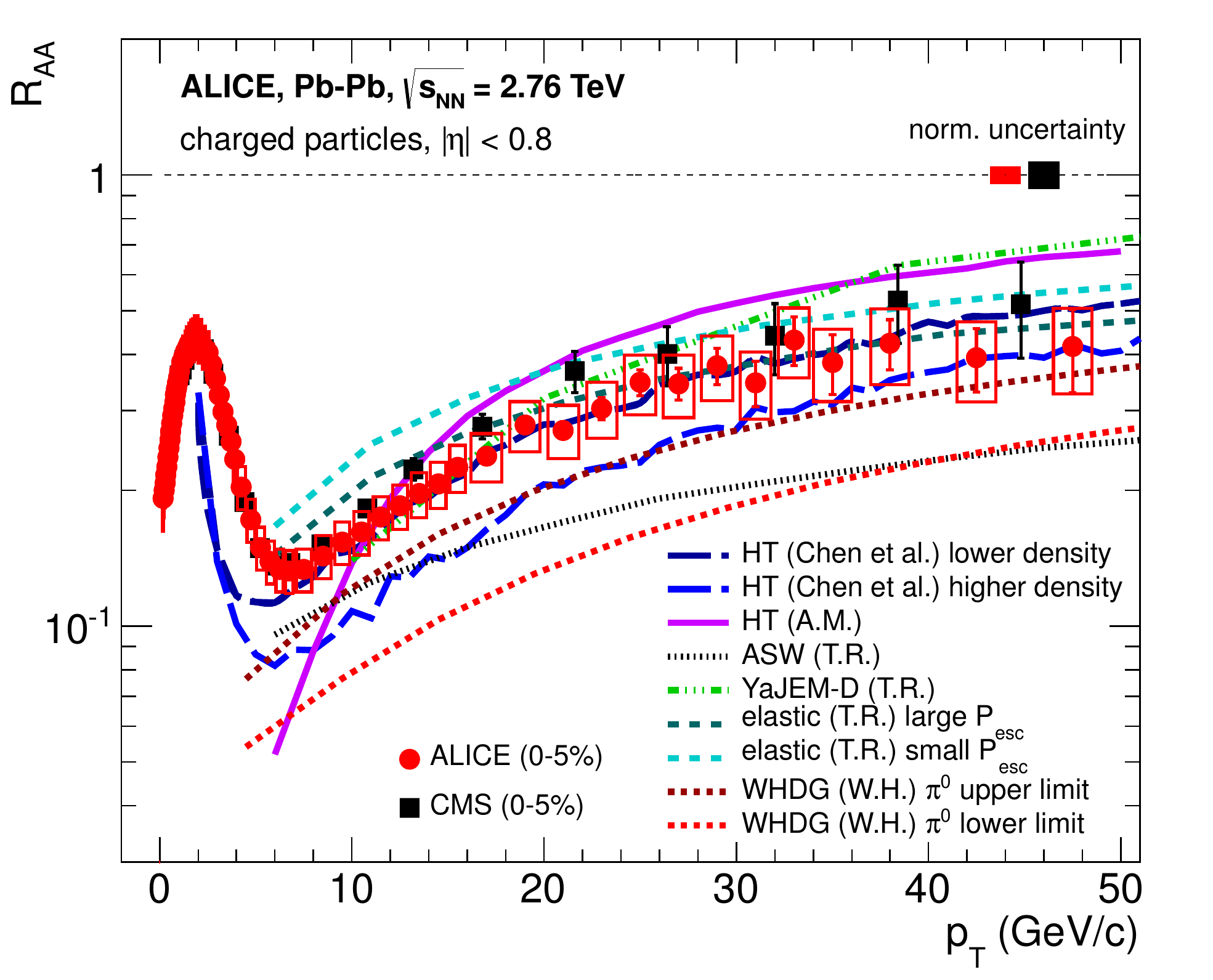}
\end{center}
\caption{The nuclear modification factor for charged particles compared to different models, for CMS \cite{CMS:2012aa} (left) and ALICE \cite{Abelev:2012hxa} (right) results. Taken from  \cite{CMS:2012aa} and \cite{Abelev:2012hxa}.} 
\label{fig:RAAparticles2}
\end{figure}

In Fig. \ref{fig:RAAparticles2}, experimental results of CMS \cite{CMS:2012aa} and ALICE \cite{Abelev:2012hxa} for the nuclear modification factor of charged particles are compared with different models that take into account energy loss of the parton traversing the medium whose hadronisation leads to the observed charged particle. This energy loss is considered to be dominantly radiative (medium-induced gluon radiation) at high energies of the parton, but elastic energy loss is included in some models, see \cite{Majumder:2010qh,Mehtar-Tani:2013pia,Blaizot:2015lma,jqqgp5} and references therein. In all these calculations, the medium is characterised by its length and dynamical behaviour, and by the strength of the interaction of the parton with the medium constituents, usually given by the transport coefficient $\hat q$ (the average transverse momentum squared transferred from the medium to the parton per mean free path). Such interaction leads to an enhancement of radiation off the parton that implies a longitudinal medium-induced energy degradation. Several models where the energy loss mechanism is embedded in a realistic dynamical model for the medium were compared 
to experimental data. In this way, values of the transport coefficient lying between those expected for weak and strong coupling 
pictures of the medium have been extracted \cite{Burke:2013yra,Liu:2015vna}.
However, a good agreement with the data can be found by various models that mainly differ in the inclusion or not of elastic energy loss, and in the way in which energy and momentum conservation is imposed on top of the theoretical calculation \cite{Armesto:2011ht}, done in the high-energy limit but applied to finite energies.
This is due to the fact that the nuclear modification factor is subject to several biases related to the steeply falling partonic spectrum, the different production points in the medium (usually called surface bias) and the hardness of the fragmentation functions. Besides, models assume that hadronisation takes place outside the medium. Energy loss is therefore a purely partonic process, an assumption that may not be fully true up to sizeable transverse momenta that at the LHC may be as large as 10 GeV/c, see Subsection \ref{hadroch}. Therefore, more differential measurements like those of jets are required to disentangle the various mechanisms of energy loss including full jet-medium interactions, and the weak or strong coupling nature of the medium.

To conclude this Subsection, we comment on pPb collisions. The nuclear modification factor for neutral pions measured at RHIC was used to provide information about the nuclear modification of parton densities \cite{Eskola:2009uj,deFlorian:2011fp}. Charged particle $R_{pPb}$ has been also measured at the LHC. While both ALICE  \cite{ALICE:2012mj,Abelev:2014dsa} and CMS   \cite{Khachatryan:2015xaa} observe a nuclear modification factor clearly compatible with unity for transverse momenta from roughly 3 to 30 GeV/c, CMS data\footnote{Preliminary ATLAS  data \cite{ATLAS-CONF-2014-029} also indicate an enhancement.} show an enhancement as large as 1.4 for larger transverse momenta, while ALICE data - that extend to smaller $p_{\rm T}$ than those of CMS - seem to be compatible with unity. Discussions are undergoing on the possible origin of the apparent discrepancy, like the different estimates of the reference pp spectra (based on interpolation procedures) used by the experiments. In such a situation, a pp run at $\sqrt{s}=5.02$ TeV can be considered as mandatory.

\subsection{Jets}
\label{jets}

Jets - sprays of collimated hadrons - are the closest objects to partons that in QCD can be properly defined, see \cite{Salam:2009jx}. The study of their modification inside a coloured medium was proposed long ago as a sensitive proof of the mechanisms of energy loss, of the interaction strength with the medium and of its density \cite{Baier:1999ds,Salgado:2003rv}.

While jets have been measured in heavy-ion collisions at RHIC \cite{Perepelitsa:2012gf,Adamczyk:2013jei}, their measurements there suffer from limitations due to the accessible energy range and to the limited detector acceptances. On the other hand, at the LHC \cite{Aad:2010bu,Aad:2012vca,Chatrchyan:2011sx,Chatrchyan:2012nia,Chatrchyan:2012gt,Chatrchyan:2012gw,Chatrchyan:2013kwa,Aad:2013sla,Abelev:2013kqa,Adam:2015ewa,Chatrchyan:2014ava,Aad:2014wha,Aad:2014bxa,Adam:2015doa,Aad:2015bsa,Adam:2015mda,Khachatryan:2015lha} the much higher centre-of-mass energy provides jets with large transverse energy that can be reconstructed above the fluctuations of energy contributions from the background event that is produced in heavy-ion collisions.  Furthermore, the large calorimeters of the experiments allow full jet reconstruction and detailed differential studies. In this respect, ATLAS and CMS have traditional large acceptance electromagnetic and hadronic calorimeters, while ALICE reconstructs jets with electromagnetic energy and tracks.

\begin{figure}[htbp]
\begin{center}
\includegraphics[width=0.48\textwidth]{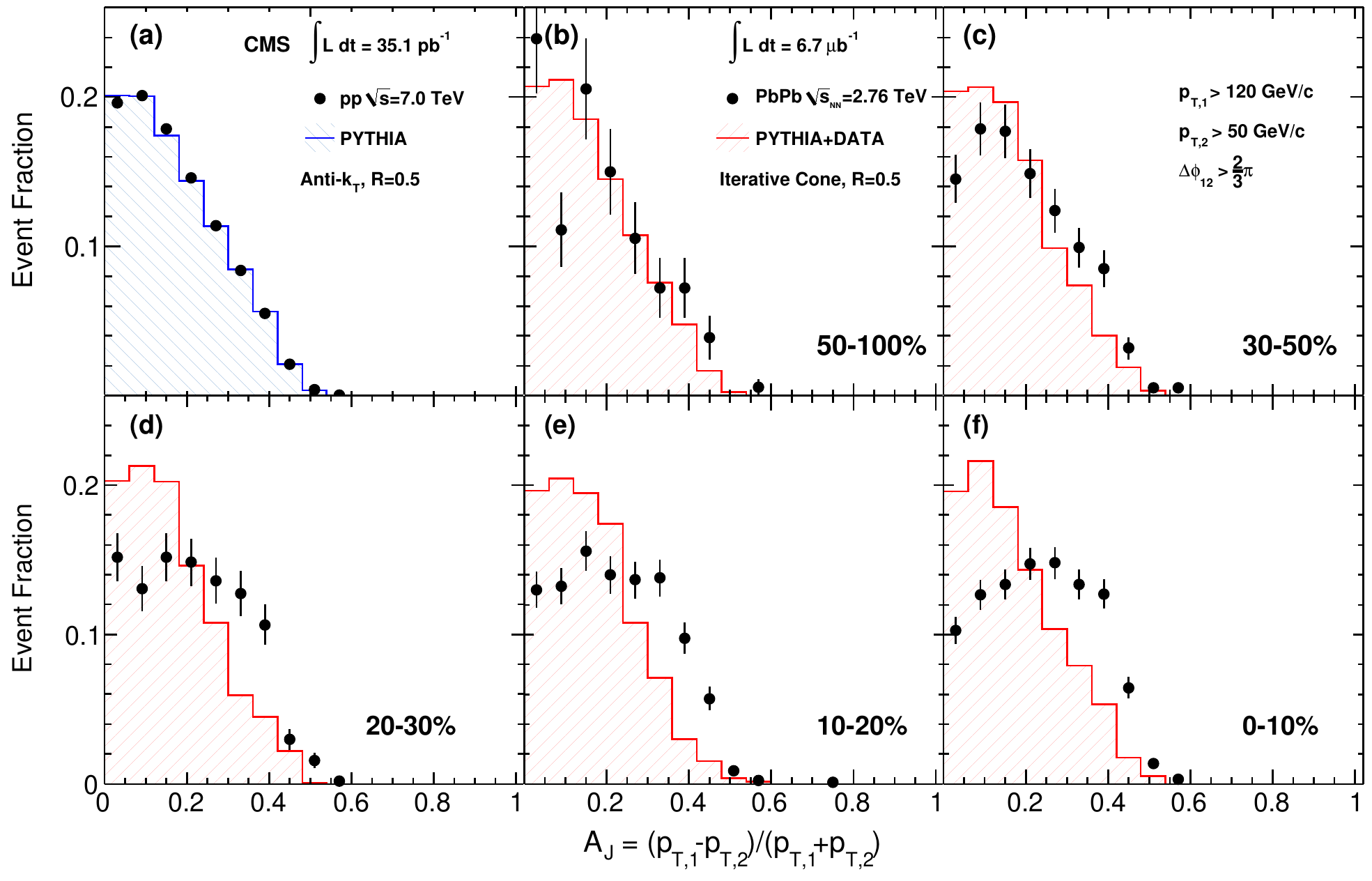}\hskip 0.3cm\includegraphics[width=0.48\textwidth]{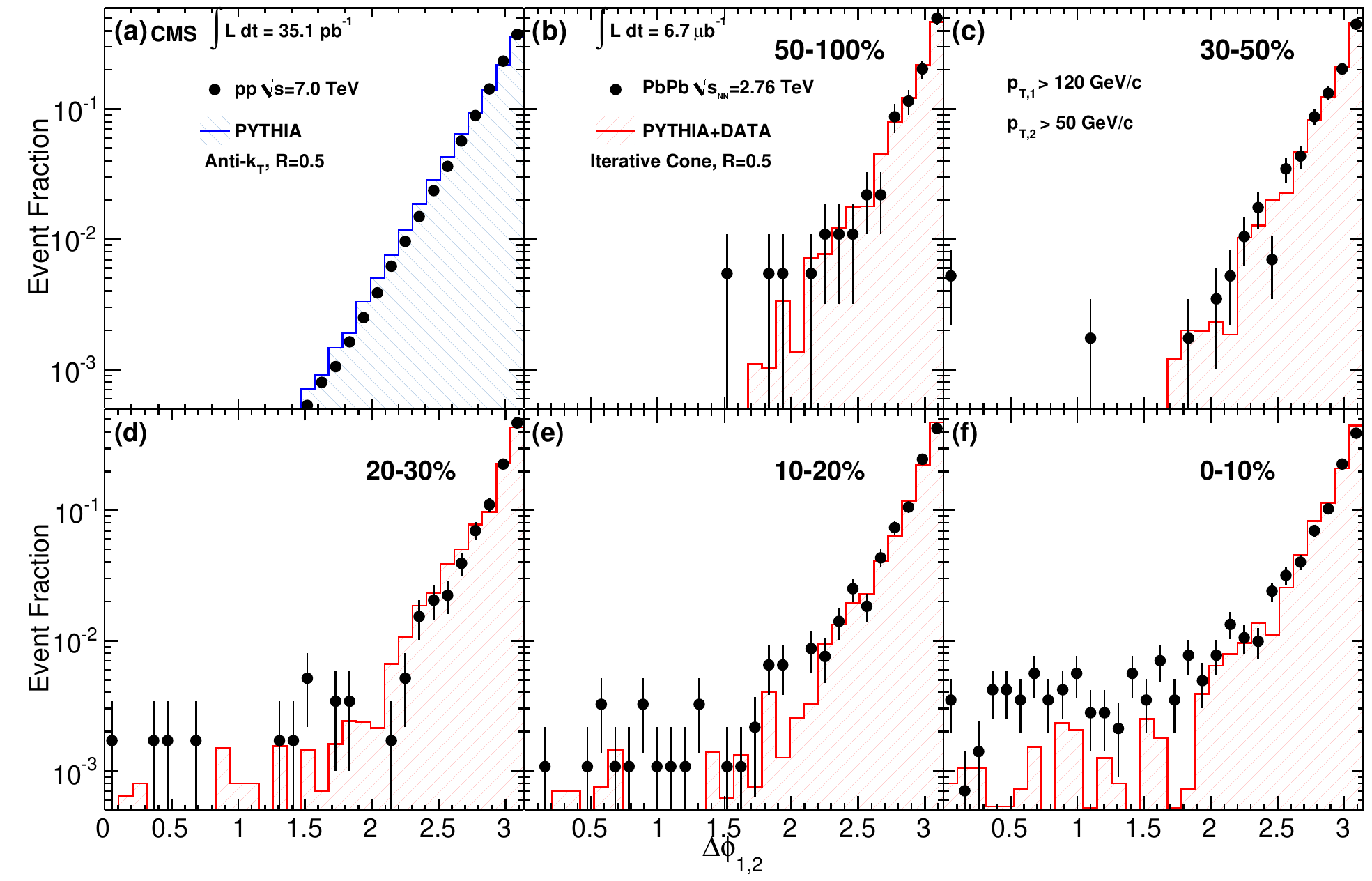}
\end{center}
\caption{Distributions in the energy asymmetry (left) and in the azimuthal angle (right) of the two hardest jets (with $p_{T,1}>120$ GeV/c and $p_{T,2}>50$ GeV/c) in pp collisions at 7 TeV, and PbPb collisions at 2.76 TeV/nucleon for different centralities, see the legends on the plots. Taken from  \cite{Chatrchyan:2011sx}.} 
\label{fig:jets1}
\end{figure}

In Fig. \ref{fig:jets1} we show the energy asymmetry $A_J=(p_{T,1}-p_{T,2})/(p_{T,1}+p_{T,2})$ and the azimuthal distribution for dijets 
 first studied by ATLAS \cite{Aad:2010bu} and then by CMS \cite{Chatrchyan:2011sx,Chatrchyan:2012nia}. A clear increase in the dijet asymmetry without sizeable azimuthal decorrelation can be observed. The trend of the data was subsequently verified for $\gamma$+jet final states  \cite{Chatrchyan:2012gt}. Note that the fraction of energy loss due to medium effects $\langle p_{T,2}/p_{T,1}\rangle$ is of the order of 10 \%. The missing transverse momentum is recovered in the form of relatively soft particles at large angles \cite{Chatrchyan:2011sx,Khachatryan:2015lha}.

\begin{figure}[htbp]
\begin{center}
\includegraphics[width=0.48\textwidth]{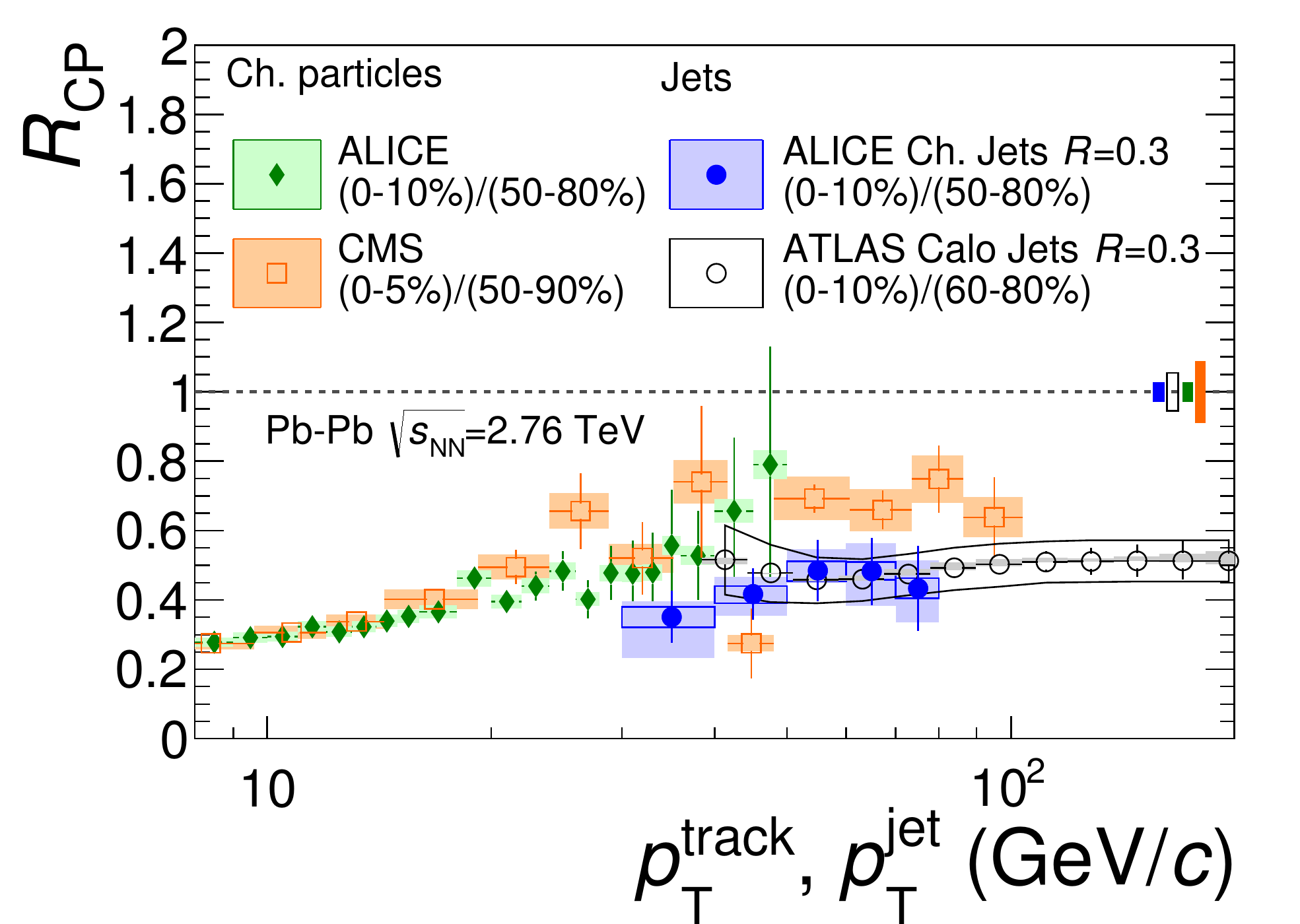}\hskip 0.3cm\includegraphics[width=0.48\textwidth]{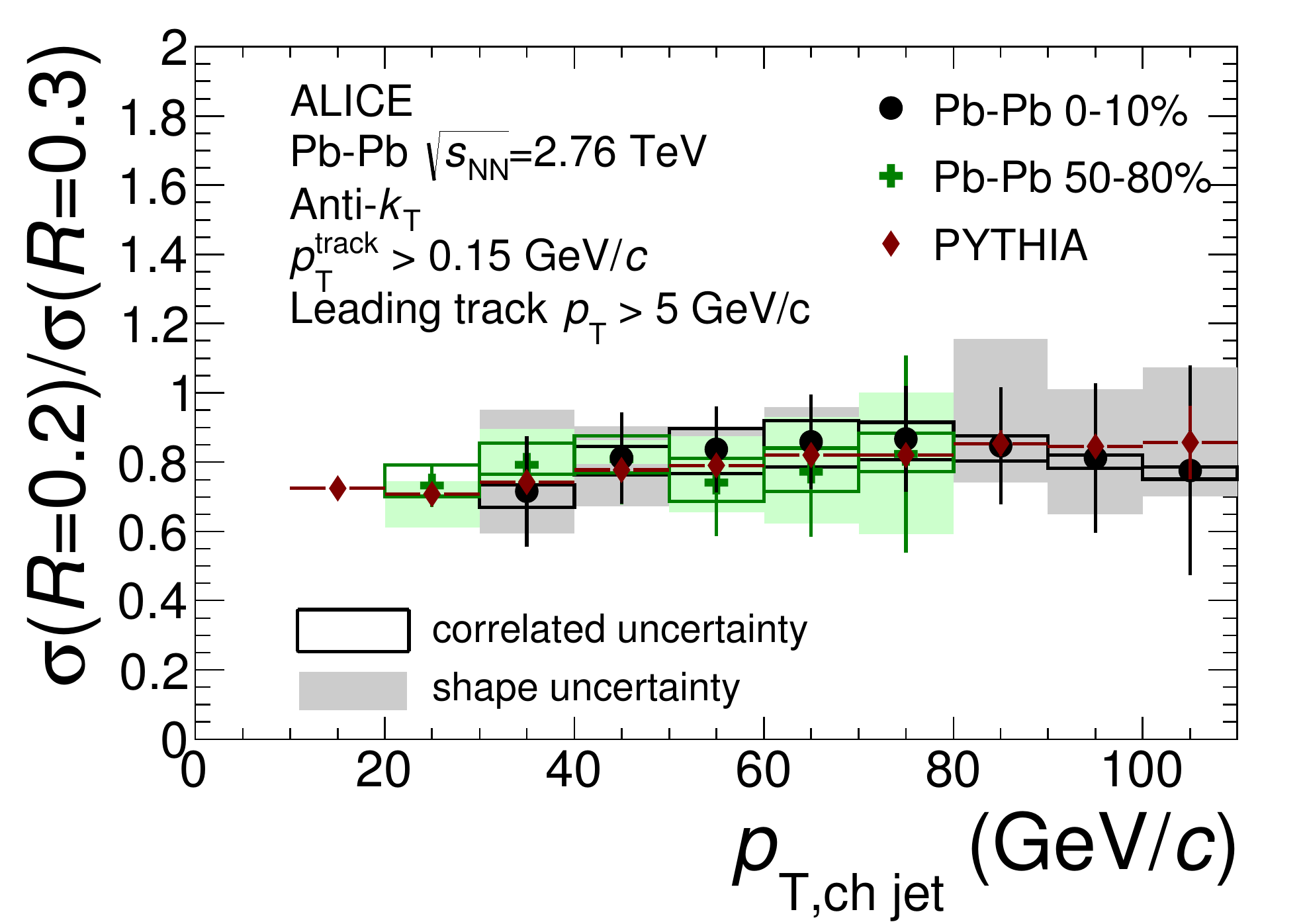}
\end{center}
\caption{Left: nuclear modification factor of central collisions compared to peripheral ones, for charged jets, fully reconstructed jets and charged hadrons, versus the corresponding transverse momentum of the jet or hadron. Right: ratio of spectra of charged jets defined with reconstruction parameter $R=0.2$ and $R=0.3$, versus charged jet transverse momentum. Taken from \cite{Abelev:2013kqa}.} 
\label{fig:jets2}
\end{figure}

In Fig.  \ref{fig:jets2} the nuclear modification factor for jets (both charged \cite{Abelev:2013kqa} and fully calorimetric \cite{Aad:2012vca} jets) shows a suppression similar to that found for charged particles (with due caution on the comparison of the different transverse momentum scales), and no sizeable difference for different jet reconstruction parameter in the algorithm.
% at least for the relative small values used in heavy-ion collisions.
The suppression shows a significant dependence with the azimuthal angle with respect to the reaction plane \cite{Aad:2013sla,Adam:2015mda}, as expected from energy loss processes.

\begin{figure}[htbp]
\begin{center}
\includegraphics[width=0.44\textwidth]{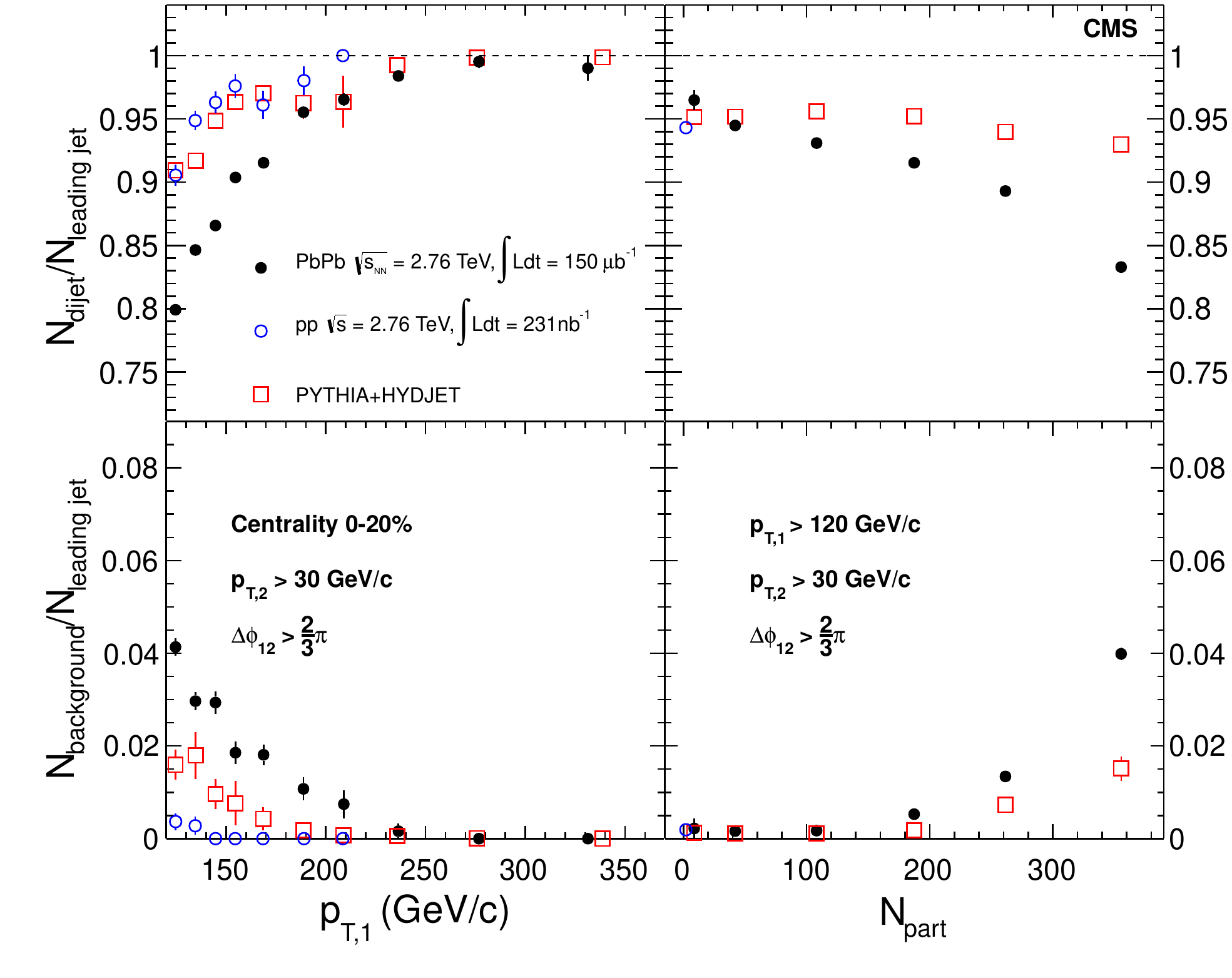}\hskip 0.3cm\includegraphics[width=0.54\textwidth]{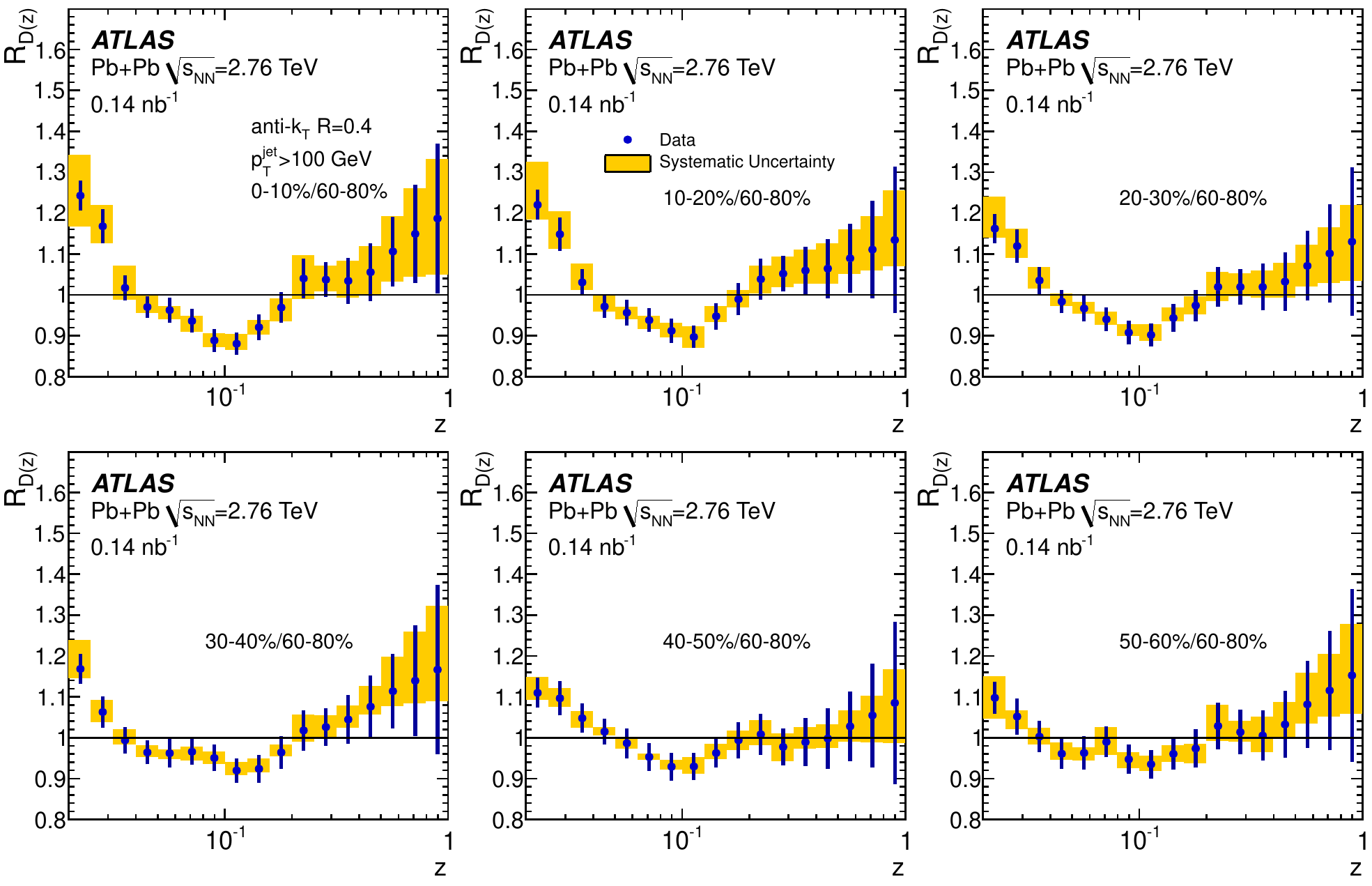}
\end{center}
\caption{Left top: ratio of events with a second jet over the total number of leading jets, for pp and PbPb collisions versus the transverse momentum of the leading jet and versus the number of participants. Left bottom: id. for the number of background jets over the number of leading jets. Right: ratio of jet fragmentation functions in PbPb and pp, versus the momentum fraction of the energy of the jet carried by the measured hadron. Taken from \cite{Chatrchyan:2012nia} and \cite{Aad:2014wha}.} 
\label{fig:jets3}
\end{figure}

On the other hand, the reduction in the number of events in which the subleading jet has disappeared (either by degradation below the cut or by migration outside the acceptance) compared to pp collisions is moderate, of the order of 10 \% for most central collisions, and the contamination from background jets is small, see Fig. \ref{fig:jets3} left \cite{Chatrchyan:2012nia}.  All features behave smoothly  and tend to become less and less pronounced with increasing jet energy. The fragmentation function \cite{Chatrchyan:2012gw,Chatrchyan:2013kwa,Chatrchyan:2014ava,Aad:2014wha} shows an enhancement of soft particles, a suppression of particles with intermediate momentum fractions, and little modification of hard ones (the small enhancement is usually interpreted in terms of the modification of the denominator - the jet energy - due to some energy loss), see Fig. \ref{fig:jets3} right.

Note that jet measurements in heavy-ion collisions require a procedure for background subtraction (see studies for characterising the background in \cite{Abelev:2012ej}) that, for correlation observables, must be carefully considered as it can mimic medium effects \cite{Cacciari:2011tm}.

From a theoretical point of view, 
the observation of: (i) a sizeable increase, with respect to the proton-proton case, of the energy asymmetry in dijet 
systems without modification of their azimuthal correlation, (ii) the recovery of missing energy at large 
angles away from the jet axis in the form of soft particles, and (iii) the lack of strong modification of the hard part of the 
jet fragmentation function, pose serious challenges for the standard explanation of jet quenching 
in terms  of medium-induced gluon radiation in which energy loss and broadening are intrinsically connected and 
the emitted radiation is semihard. While several phenomenological explanations, see \cite{CasalderreySolana:2010eh,Qin:2010mn,He:2011pd,Young:2011qx,Lokhtin:2011qq,Renk:2012cx,Apolinario:2012cg,Zapp:2012ak,CasalderreySolana:2012ef,Mehtar-Tani:2014yea}, have  been put forward (that go from the collimation by scattering-away of the soft components of the jet in the medium, to  the important role of colour coherence between emitters and its medium disruption), none provides a complete quantitative description of the data nor proper theoretical justification for some of the assumptions.

In pPb collisions \cite{Chatrchyan:2014hqa,ATLAS:2014cpa,Adam:2015xea,Adam:2015hoa}, the nuclear modification 
for dijet \cite{Chatrchyan:2014hqa} and single jet distributions \cite{ATLAS:2014cpa,Adam:2015hoa} in minimum 
bias events is very mild and in agreement with the expectations from the nuclear modification of parton densities 
(actually these data can be used to constrain them \cite{Paukkunen:2014pha}). Negligible nuclear effects are seen
on the dijet acoplanarity \cite{Adam:2015xea}. On the other hand, studies on the centrality and pseudorapidity 
dependence of these observables \cite{Chatrchyan:2014hqa,ATLAS:2014cpa} point to the existence of non-trivial 
effects coupling the hard process to the centrality estimator, 
%see \cite{Martinez-Garcia:2014ada,Bzdak:2014rca,Alvioli:2014eda,Perepelitsa:2014yta,Armesto:2015kwa} for several interpretation 
see \cite{Bzdak:2014rca,Alvioli:2014eda,Perepelitsa:2014yta,Armesto:2015kwa} for several interpretation attempts.

\subsection{Heavy Flavours}
\label{HQ}

Heavy quarks are an important probe of the medium created in nuclear collisions.
Due to their mass, much larger than the temperature of the system, they are not created or destroyed in the medium, but rather produced by hard scattering in the early stage of the collision process. Therefore, they are particularly sensitive to the QGP phase. 
The study of the modification of their yields and of their angular distribution can give information on the degree of thermalization of the medium and on its transport coefficient.
%~\cite{Baier:1996kr}.
In particular, heavy quarks can lose energy and participate in the collective motion of the created system, the mainly related observables being their nuclear modification factor and their elliptic flow. A definite prediction of the radiative energy loss models is a substantial dependence of the specific energy loss on the mass of the quark~\cite{Dokshitzer:2001zm,Armesto:2005iq}, with the heavy quarks losing less energy than the light ones (often called ``the dead cone effect''). Systematic studies of heavy quark production were started at RHIC, where the measurements were essentially based on the detection of the non-photonic electrons (muons), a name related to the main background source (photon conversions) for leptons from heavy-flavour decays. A strong suppression of non-photonic electrons, of the same order of magnitude as the one seen for light hadrons, was observed by PHENIX~\cite{Adare:2006nq} and STAR~\cite{Abelev:2006db}, although these measurements could not separate the contributions from bottom and charm quarks. 
More recently, direct measurements of the hadronic decay of $D$ mesons by STAR~\cite{Adamczyk:2014uip}, opened the way towards a more detailed understanding of heavy-quark energy loss. Finally, thanks to the recent experiment upgrades (Si vertex tracking systems), PHENIX is now able to 
separate charm/bottom contributions in the lepton decay channel~\cite{Adare:2015hla}.

At the LHC, thanks to the much increased heavy-quark production cross section, the indirect measurements have been extended towards higher $p_{\rm T}$ (ATLAS~\cite{Perepelitsa:2012cza}, ALICE~\cite{Abelev:2012qh,Adam:2015pga,Adam:2015qda}), a detailed study of hadronic decays for various charmed mesons has taken place (ALICE~\cite{Adam:2015nna,ALICE:2012ab,Abelev:2013lca,Abelev:2014ipa,Adam:2015sza,Adam:2015jda}), and results on beauty production via either detection of J/$\psi$ from the decay $B\rightarrow {\rm J}/\psi + X$ (ALICE~\cite{Adam:2015rba}, CMS~\cite{Chatrchyan:2012np}) or hadronic decays (CMS~\cite{Khachatryan:2015uja}) are now available.
In addition, the study of jets originated from a b-quark has now been extended for the first time from pp to heavy-ion collisions by CMS~\cite{Chatrchyan:2013exa,Khachatryan:2015sva}.  

\begin{figure}[htbp]
\centering
\resizebox{0.4\textwidth}{!}
{\includegraphics*{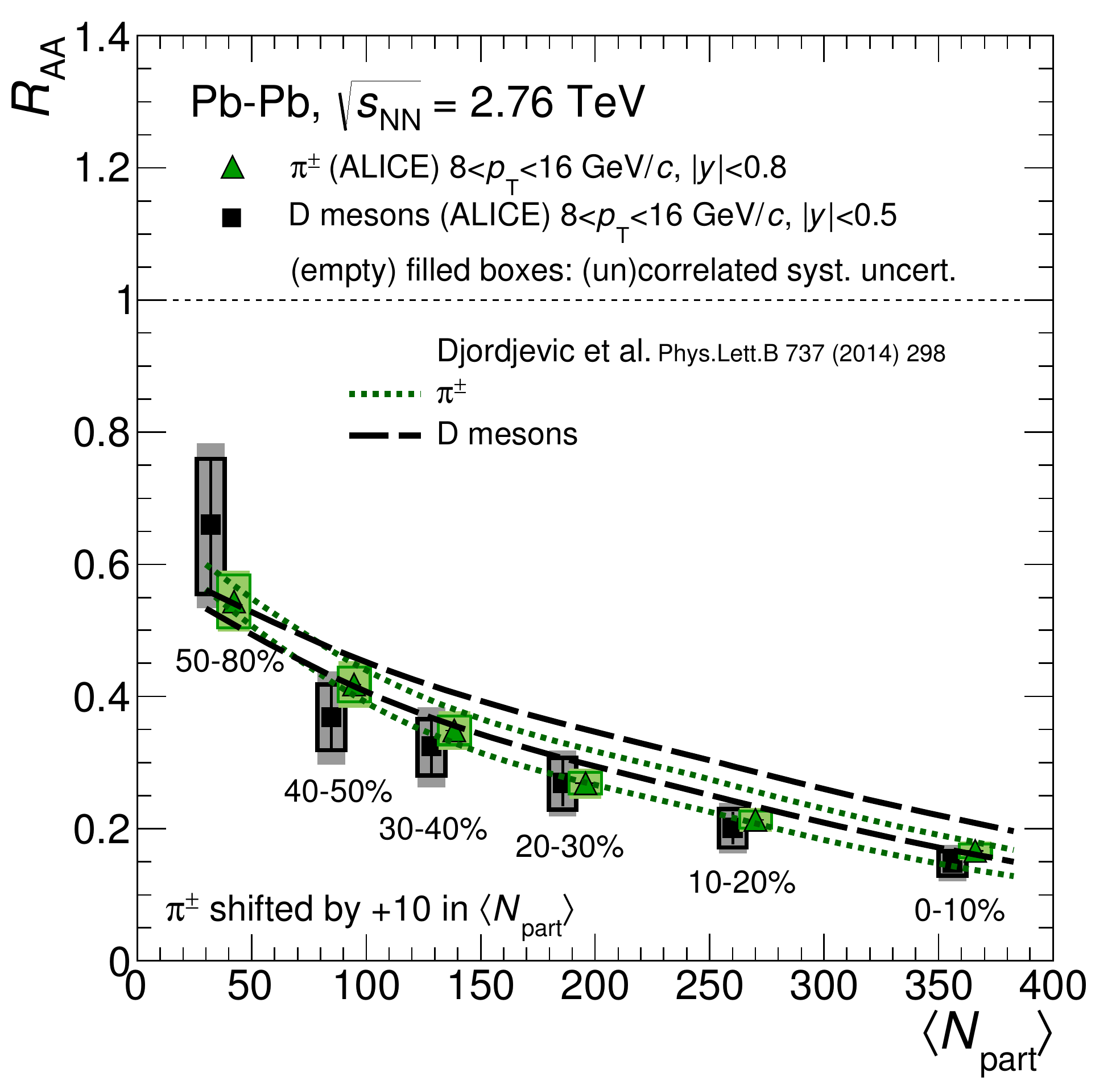}}
\resizebox{0.4\textwidth}{!}
{\includegraphics*{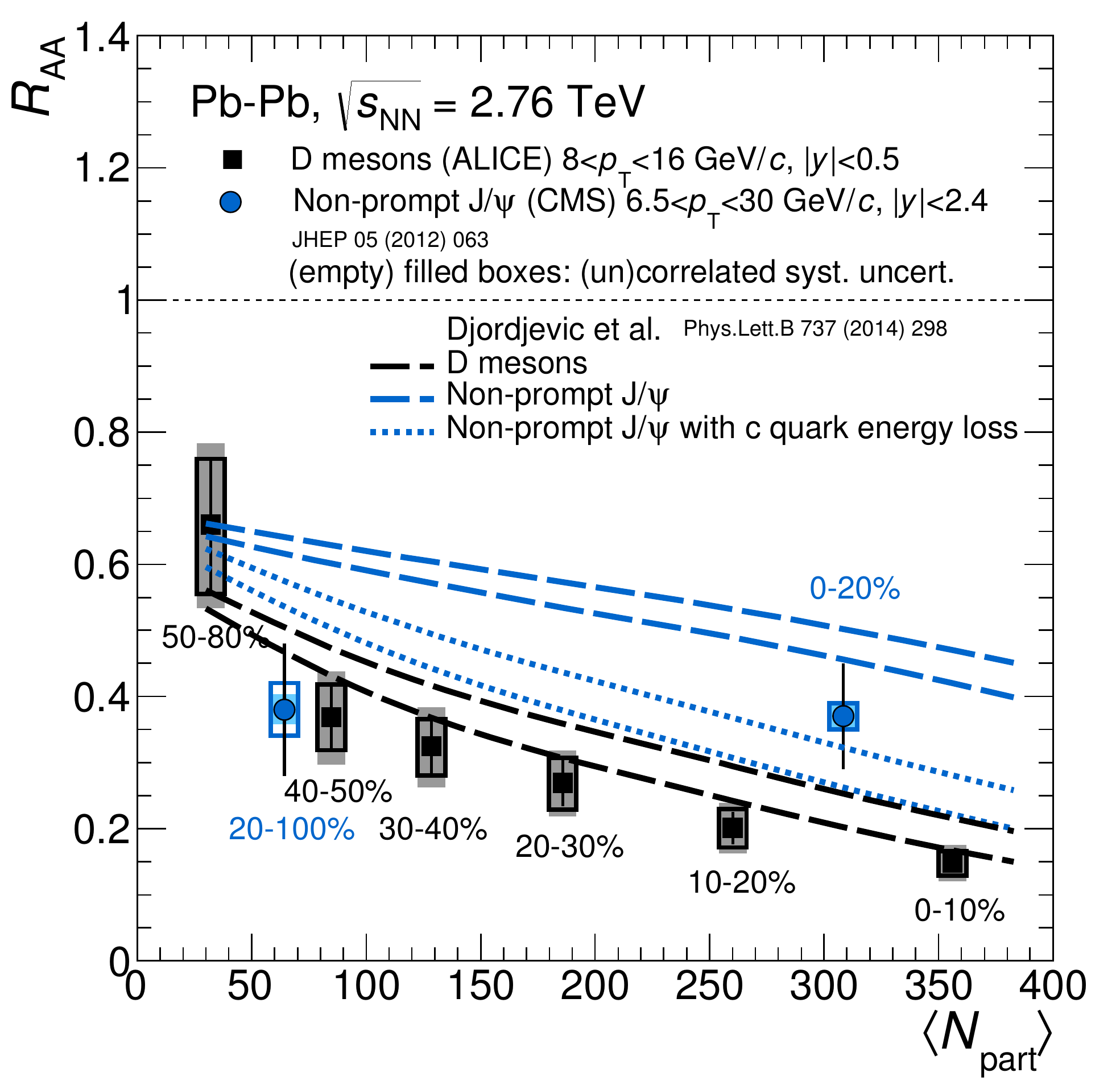}}
\caption{Left: comparison of $D$-meson and charged pion $R_{\rm AA}$, measured by ALICE in PbPb collisions at $\sqrt{s_{\rm NN}}=2.76$ TeV~\cite{Adam:2015nna}. Right: same $D$-meson results, compared to non-prompt J/$\psi$ $R_{\rm AA}$ measured by CMS~\cite{Chatrchyan:2012np}.
 The results are compared to the theoretical model of Ref.~\cite{Djordjevic:2014tka}.} 
\label{fig:DRAA}
\end{figure}

In Fig.~\ref{fig:DRAA} (left) we show the R$_{\rm AA}$ for $D$ mesons measured in PbPb collisions at $\sqrt{s_{\rm NN}}$= 2.76 TeV by ALICE~\cite{Adam:2015nna}, in the region $8<p_{\rm T}<16$ GeV/$c$, as a function of the centrality. The result is obtained as the average of $D^0$, $D^+$ and $D^{*,+}$ $R_{\rm AA}$ values. It is compared with the corresponding quantity for charged pions and to a calculation implementing energy loss for gluons, light and heavy quarks, including both radiative and collisional processes~\cite{Djordjevic:2014tka}. In Fig.~\ref{fig:DRAA} (right) the result for $D$-mesons is compared to CMS data for J/$\psi$ from $b$-decays~\cite{Chatrchyan:2012np}, in the $p_{\rm T}$ range $6.5<p_{\rm T}<30$ GeV/$c$, and to the same theoretical calculation. It can be seen that the $D$-meson suppression reaches a factor 5 for central (0-10\%) \mbox{PbPb} events, and is compatible with the corresponding result for charged pions. The expected smaller suppression for particles resulting from the fragmentation of heavy quarks is not seen neither in the data nor in the model calculations, and is interpreted (mainly) as a consequence of the different fragmentation function and $p_{\rm T}$ spectrum of gluons, light and $c$-quarks~\cite{Armesto:2005iq,Djordjevic:2013pba}. On the contrary, when comparing $b$- and $c$-quark suppression~\cite{Adam:2015nna}, a clear ordering is seen, with the heavier $b$-quark being less suppressed. It has to be noted that the effect is even more evident when the ALICE results of Fig.~\ref{fig:DRAA} are compared with preliminary higher-statistics results from CMS~\cite{CMS:2012vxa}.

\begin{figure}[htbp]
\centering
\resizebox{0.4\textwidth}{!}
{\includegraphics*{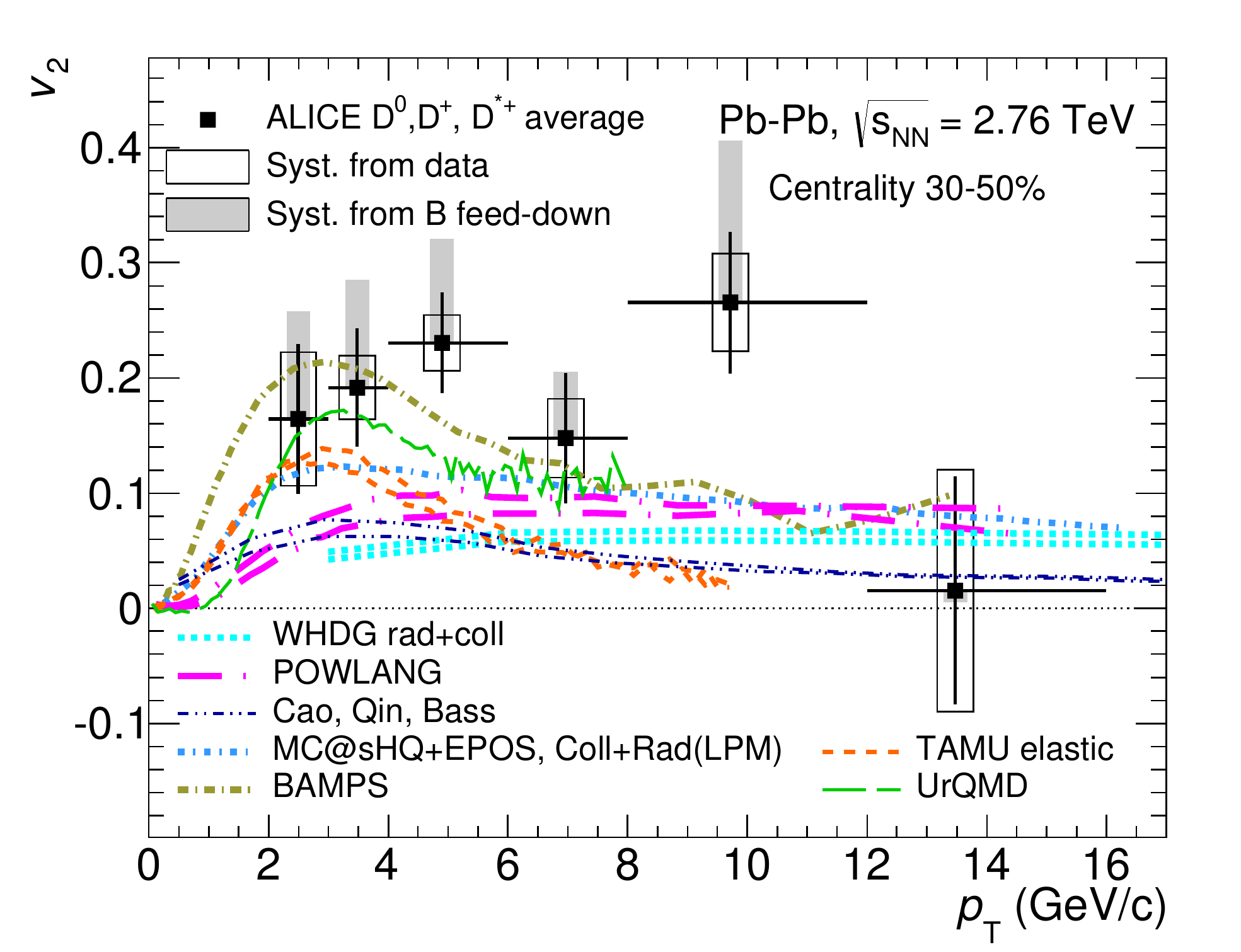}}
\resizebox{0.4\textwidth}{!}
{\includegraphics*{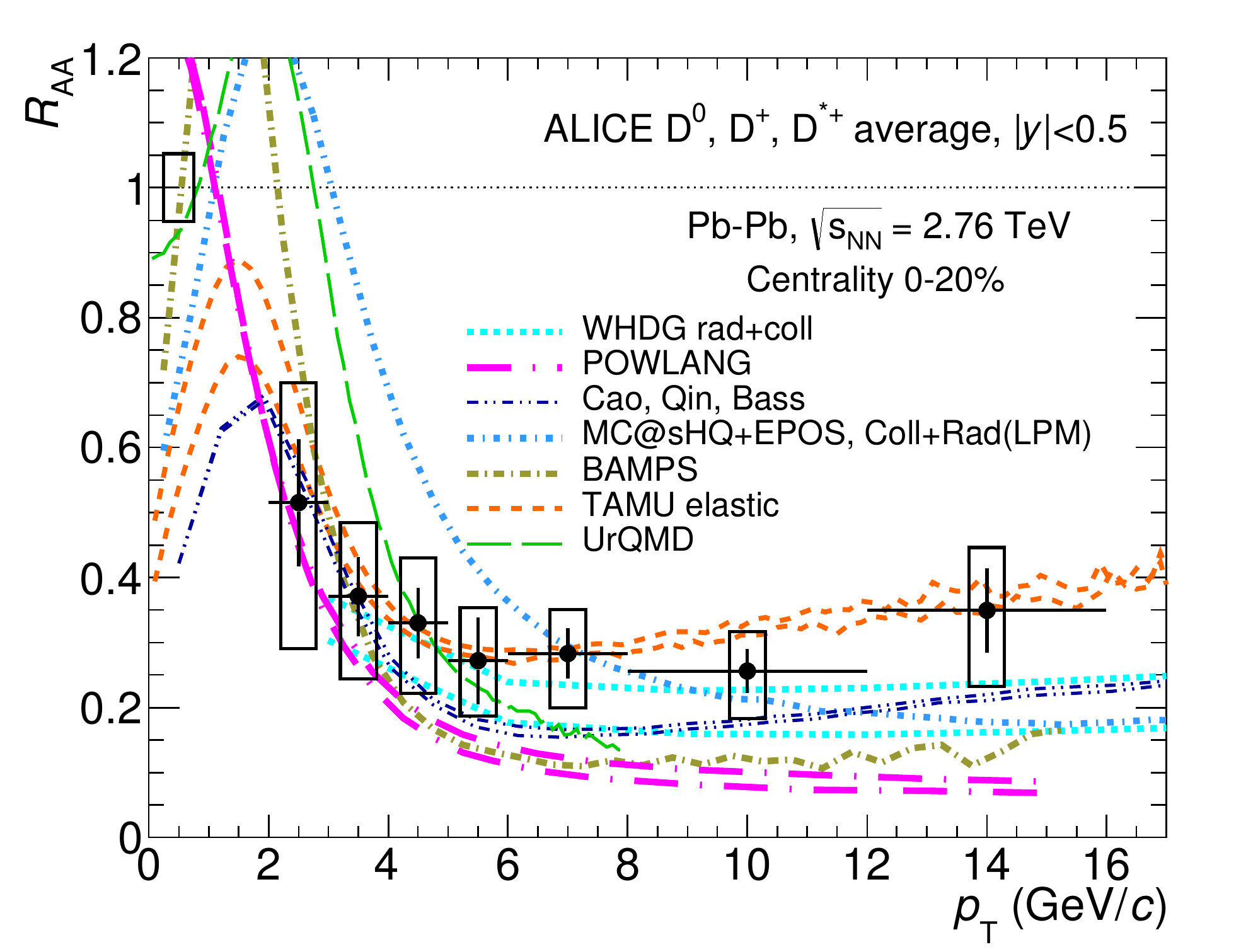}}
\caption{Left: the average $D$-meson $v_2$ in the 30-50\% centrality class, measured by ALICE~\cite{Abelev:2014ipa}. Right: the average $D$-meson $R_{\rm AA}$ on the 0-20\% centrality class, measured by ALICE~\cite{Abelev:2014ipa}.
 Both results refer to \mbox{PbPb} collisions at $\sqrt{_{\rm NN}}$ = 2.76 TeV, and are compared to theoretical models.} 
\label{fig:DRAAandv2}
\end{figure}

Further insight into the interaction processes of heavy quarks in the medium can be obtained by studying the $p_{\rm T}$-dependence of $R_{\rm AA}$ and $v_2$, as the two quantities are sensitive to the interplay of the various energy loss mechanisms. In Fig.~\ref{fig:DRAAandv2} we show (left panel) ALICE results on the average $D$-meson $v_2$ measured in PbPb collisions in the 30-50\% centrality class, and (right panel) the $R_{\rm AA}$ for the 0-20\% centrality class~\cite{Abelev:2014ipa}. The results are compared with various model calculations~\cite{Horowitz:2011gd,Nahrgang:2013xaa,He:2014cla,Alberico:2011zy,Uphoff:2012gb,Lang:2012cx,Cao:2013ita}. A significant ($>5\sigma$, for $p_{\rm T}$ integrated results) non-zero  $v_2$ is observed, an effect that can be related, at large $p_{\rm T}$, to the difference in the in-medium path length between $D$-mesons emitted in-plane or out-of-plane. At low $p_{\rm T}$ in-medium interactions of the produced heavy-quarks can also lead to a momentum anisotropy  and consequently to a non-zero $v_2$. Concerning $R_{\rm AA}$, a strong suppression, reaching a factor 3-4, is seen for $p_{\rm T}>5$ GeV/$c$, with a tendency towards a smaller suppression for decreasing $p_{\rm T}$. The comparison with the models shows that it is still challenging to describe quantitatively both observables over the full $p_{\rm T}$ range. Finally, it has to be noted that cold nuclear matter effects, and in particular parton shadowing in the nucleus~\cite{Eskola:2009uj,deFlorian:2011fp}, cannot account for the observed $D$-meson suppression, since the corresponding measurements for \mbox{pPb} collisions give nuclear modification factor values compatible with 1~\cite{Abelev:2014hha} in the whole $p_{\rm T}$-range explored ($p_{\rm T}> 1$ GeV/$c$).

\begin{figure}[htbp]
\centering
\resizebox{0.44\textwidth}{!}
{\includegraphics*{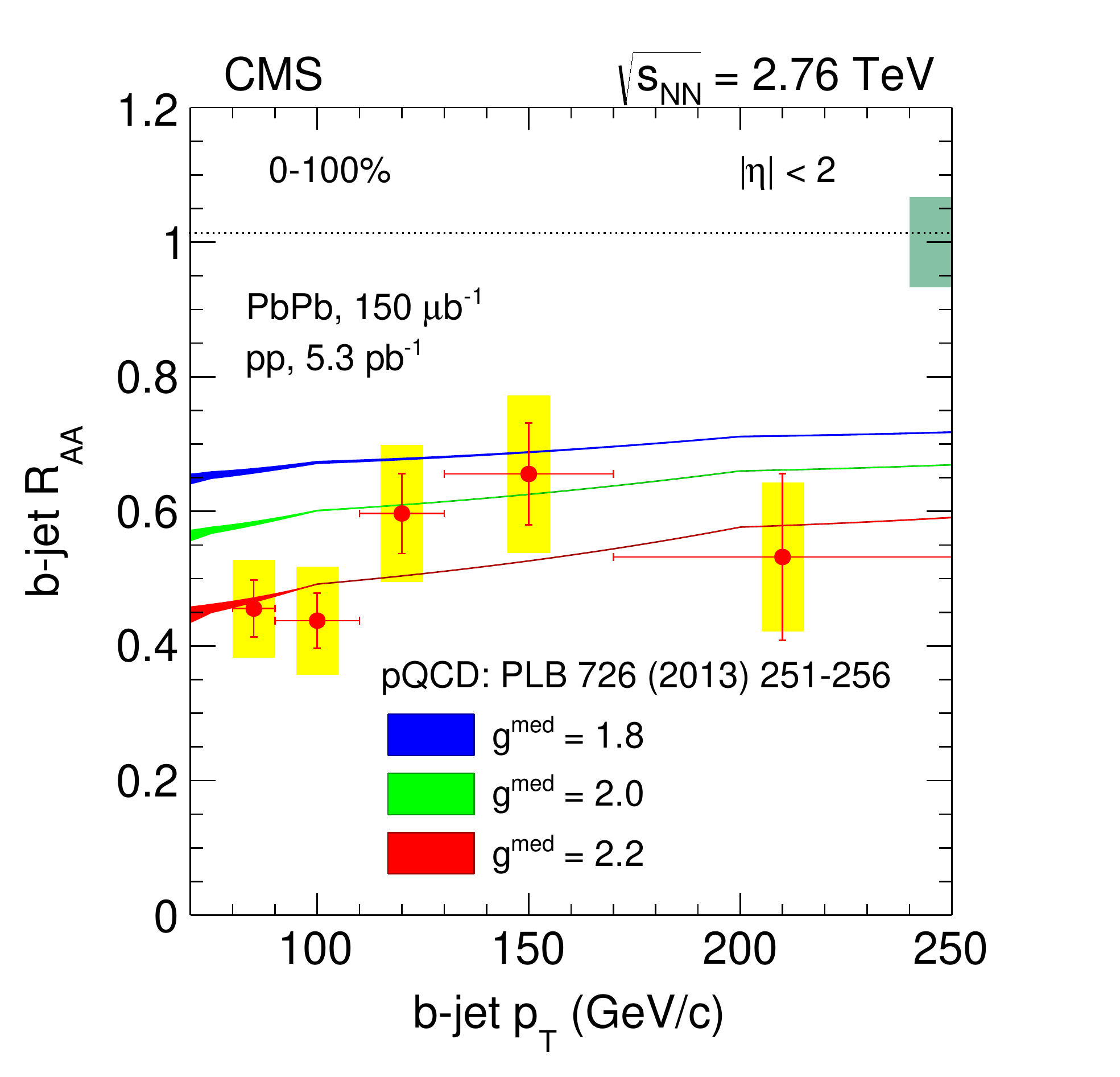}}
\resizebox{0.43\textwidth}{!}
{\includegraphics*{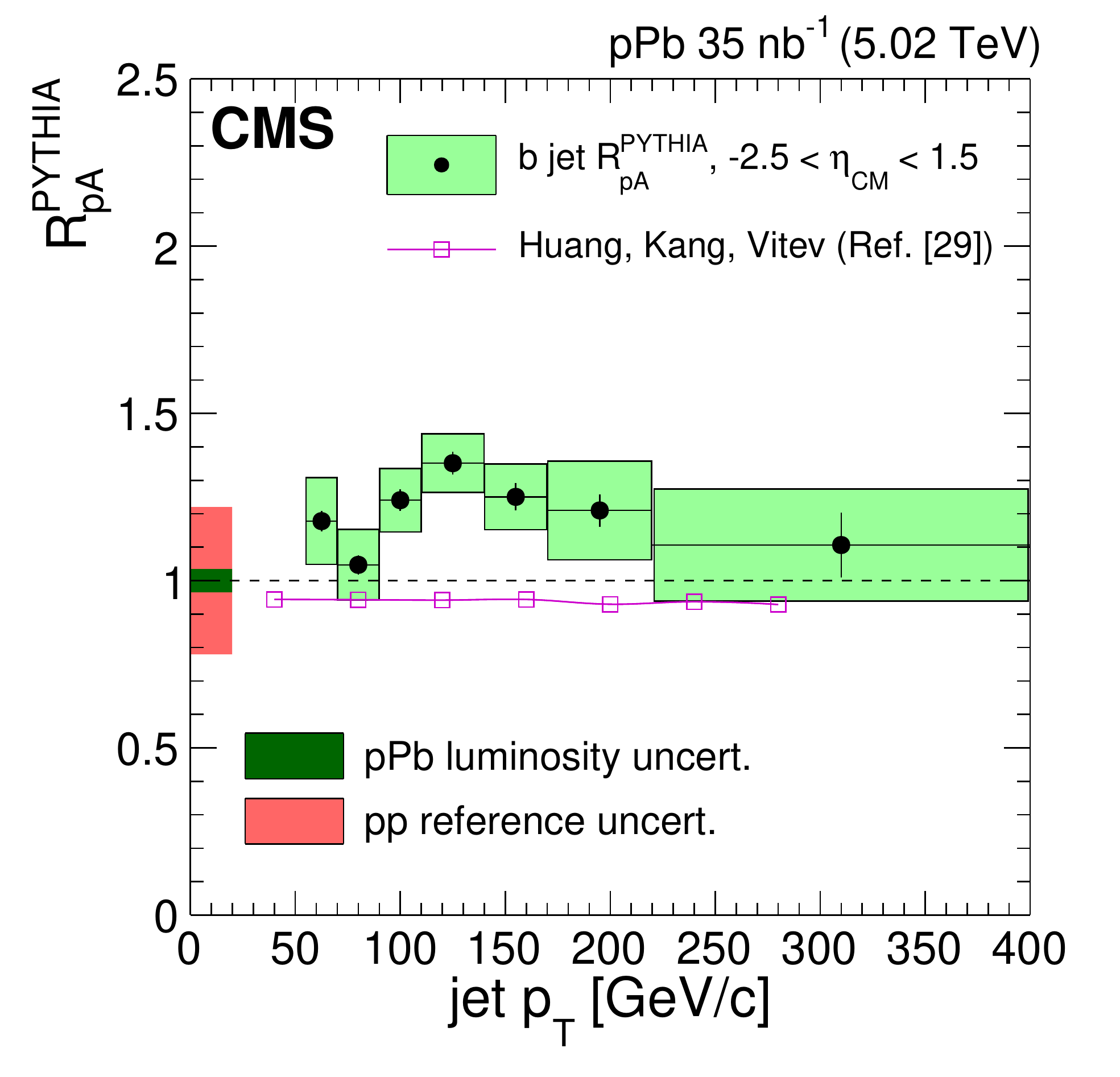}}
\caption{Left:the centrality integrated $b$-jet $R_{\rm AA}$ 
as a function of $p_{\rm T}$, from CMS~\cite{Chatrchyan:2013exa}. The data are compared to pQCD-based calculations from~\cite{Huang:2013vaa}. Right: the b-jet $R_{\rm pA}$ as a function of jet $p_{\rm T}$~\cite{Khachatryan:2015sva}. Filled boxes represent the  systematic uncertainties. The pp reference and integrated luminosity uncertainties are shown as red and green bands around unity, respectively. A pQCD prediction from Huang et al.~\cite{Huang:2013vaa} is also shown.} 
\label{fig:CMSbjet}
\end{figure}

The study of the behaviour of heavy quarks in the medium has been extended to very large $p_{\rm T}$ by studying $b$-jet production. Figure~\ref{fig:CMSbjet}(left) 
shows the centrality-integrated $R_{\rm AA}$ measured by CMS~\cite{Chatrchyan:2013exa}. A significant suppression is observed, qualitatively consistent with that of inclusive jets~\cite{Aad:2012vca}, showing that at high $p_{\rm T}$ a large mass/flavor dependence for parton energy loss is unlikely, as expected by theoretical models for large parton energies. Recent CMS results on b-jet production in pPb~\cite{Khachatryan:2015sva} (see Fig.~\ref{fig:CMSbjet}(right)) show a nuclear modification factor compatible with unity, indicating that the suppression observed in PbPb is not due to cold nuclear matter effects.

\subsection{Quarkonium}
\label{QQbar}

The study of bound states of heavy quarks, charmonia ($c\overline c$) and bottomonia ($b\overline b$), plays an important role in the understanding of the properties of the medium created in nuclear collisions. A suppression of quarkonium production, related to the screening of the quark-antiquark binding in a color-deconfined medium, was predicted long ago as a signature of the formation of a QGP~\cite{Matsui:1986dk}. In addition, states with increasing binding energies were predicted to melt for increasing temperatures of the medium~\cite{Digal:2001ue}. The suppression of the charmonium states J/$\psi$ and $\psi(2S)$ was indeed observed at SPS~\cite{Alessandro:2004ap,Arnaldi:2007zz,Scomparin:2009tg} and RHIC~\cite{Adare:2011yf,Abelev:2009qaa} energies, although the interpretation of the results proved to be not straightforward, due to competing mechanisms which also reduce the charmonium yield but are not related to deconfinement. These include the break-up of the charmonia by interaction with cold nuclear matter and with the hot hadronic gas produced when the medium cools down below the deconfinement temperature~\cite{Vogt:2001ky,Kopeliovich:1991pu,McGlinchey:2012bp}. Also initial state effects as nuclear shadowing~\cite{Eskola:2009uj,deFlorian:2011fp} have been proven to play a sizeable role on the quarkonium production rates. 
At the LHC, new features come into play. On the charmonium side, as a consequence of the large multiplicity of $c{\overline c}$ pairs ($>100$ for central \mbox{PbPb} collisions), recombination mechanisms in the deconfined phase and/or at hadronization were predicted to counterbalance the suppression effects~\cite{BraunMunzinger:2000px,Thews:2000rj}. In addition, bottomonium states are more copiously produced at the LHC with respect to lower energy machines, and an accurate study of the strongly bound $\Upsilon(1S)$ state, along with the corresponding 2S and 3S resonances, becomes possible.

Heavy quarkonium states in \mbox{pPb} and \mbox{PbPb} collisions were mainly  studied by ALICE~\cite{Abelev:2012rv,ALICE:2013xna,Abelev:2013ila,Abelev:2013yxa,Abelev:2014zpa,Adam:2015iga,Adam:2015rba,Adam:2015isa,Adam:2015jsa,Adam:2015gba,Abelev:2014nua,Abelev:2014oea} and CMS~\cite{Chatrchyan:2011pe,Chatrchyan:2012np,Chatrchyan:2012lxa,Khachatryan:2014bva,Chatrchyan:2013nza}, with contributions on selected topics also from ATLAS~\cite{Aad:2010aa,Aad:2015ddl} and LHCb~\cite{Aaij:2013zxa,Aaij:2014mza}. The experiments measure quarkonia in the dilepton decay mode (mainly $\mu^+\mu^-$, but also e$^+$e$^-$), both at central rapidity (CMS, with maximum coverage corresponding to $|y|<2.4$) and forward rapidity (ALICE, $2.5<y<4$). ALICE covers charmonium production down to zero $p_{\rm T}$ while CMS acceptance  starts at $p_{\rm T}\sim 6$ GeV/$c$ (down to a minimum of 3 GeV/$c$ for selected rapidity intervals). For bottomonium, all the experiments have coverage down to $p_{\rm T}=0$.

\begin{figure}[htbp]
\centering
\resizebox{0.45\textwidth}{!}
{\includegraphics*{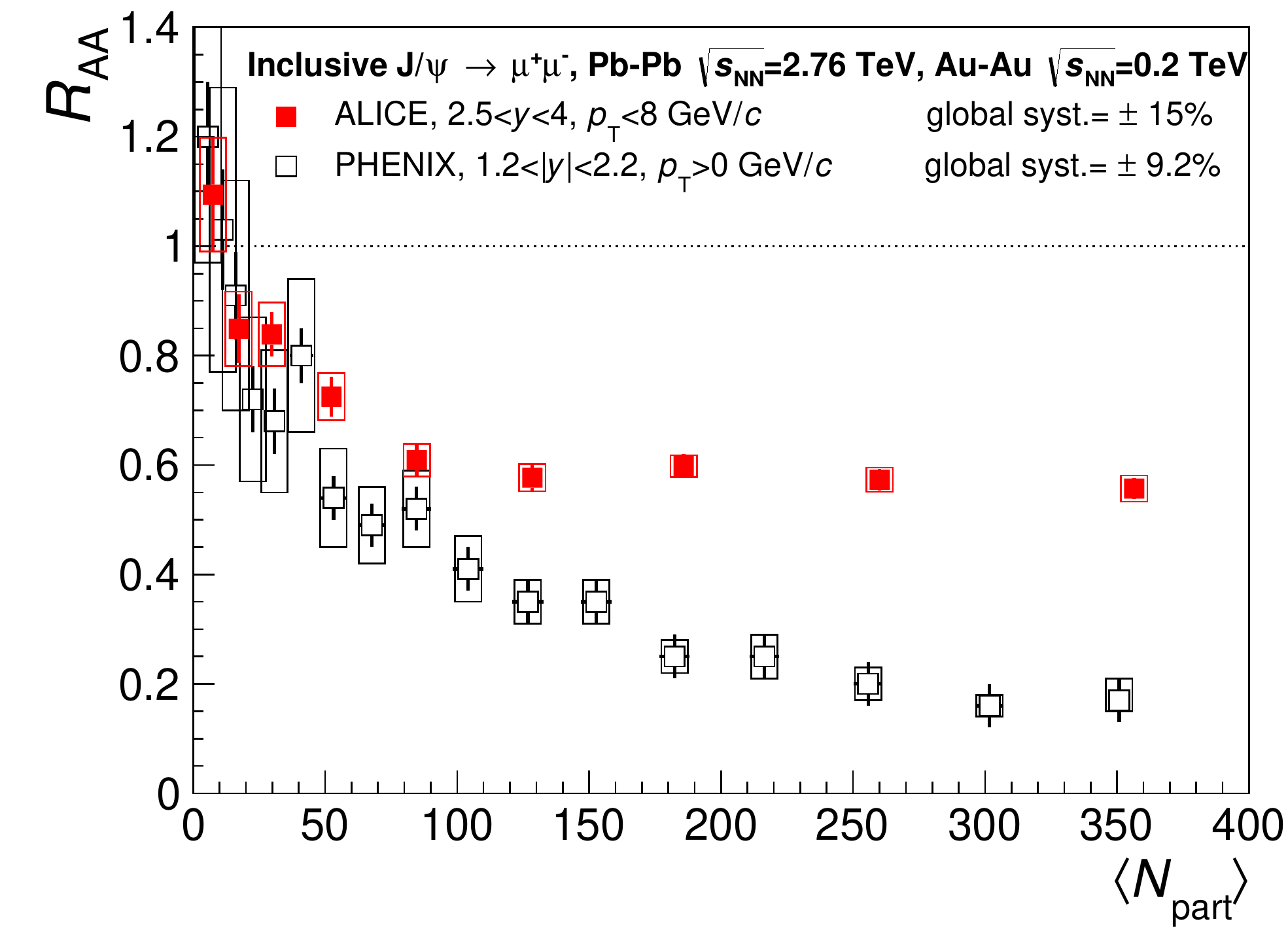}}
\resizebox{0.45\textwidth}{!}
{\includegraphics*{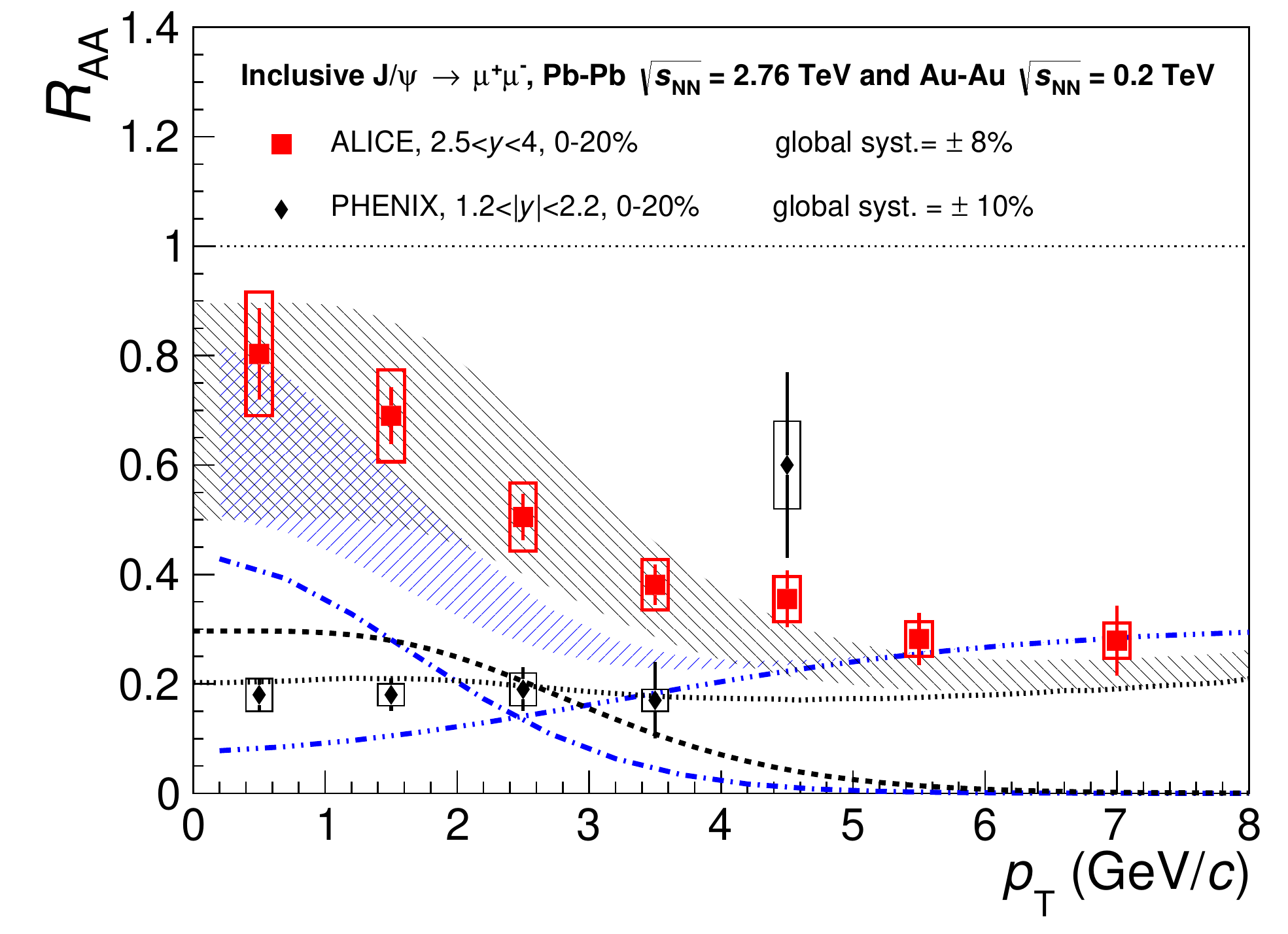}}
\caption{Left: the inclusive J/$\psi$ $R_{\rm AA}$ as a 
function of $\langle N_{\rm part}\rangle$, measured in PbPb collisions at forward-$y$ by ALICE~\cite{Adam:2015isa} and compared to PHENIX Au-Au results. Right: the $p_{\rm T}$-dependence of the inclusive J/$\psi$ $R_{\rm AA}$~\cite{Adam:2015isa}, compared to PHENIX results and to theoretical models.} 
\label{fig:JpsiRAAcentALICE}
\end{figure}

In Fig.~\ref{fig:JpsiRAAcentALICE} (left) the dependence of the inclusive J/$\psi$ $R_{\rm AA}$ on $\langle N_{\rm part}\rangle$, measured by ALICE at forward rapidity~\cite{Adam:2015isa}, is compared to the corresponding measurement by PHENIX (\mbox{Au-Au}, $1.2<|y|<2.2$)~\cite{Adare:2011yf}. For  $\langle N_{\rm part}\rangle > 100$ the suppression observed in ALICE data is clearly smaller than at RHIC energy. Such a behaviour, also seen at midrapidity~\cite{Abelev:2013ila}, is compatible with the presence of a significant regeneration of J/$\psi$. This mechanism is expected to enhance charmonium production mainly at low transverse momentum. When comparing the $p_{\rm T}$-dependence of $R_{\rm AA}$ for central collisions between PHENIX and ALICE, as in Fig.~\ref{fig:JpsiRAAcentALICE} (right)~\cite{Adam:2015isa}, one can clearly see that the difference is concentrated at low $p_{\rm T}$. Theoretical calculations (including transport and statistical models) qualitatively reproduce the ALICE data~\cite{Zhao:2011cv,Zhou:2014kka,Andronic:2011yq,Ferreiro:2012rq}. For transport models, which implement charmonium dissociation and regeneration in a thermally expanding fireball, more than 50\% of the measured J/$\psi$ is produced via regeneration. As for many observables, it is expected that cold nuclear matter effects may influence the measured $R_{\rm AA}$. Estimates based on the results from \mbox{pPb} collisions at $\sqrt{s_{\rm NN}} = 5.02$ TeV show that cold nuclear matter effects are negligible at large $p_{\rm T}$, while at low $p_{\rm T}$ their size becomes of the same order of magnitude as the observed suppression. Once such effects are corrected for, the low-$p_{\rm T}$ enhancement of the J/$\psi$ production becomes even stronger~\cite{Adam:2015iga}.
  
\begin{figure}[htbp]
\centering
\resizebox{0.3\textwidth}{!}
{\includegraphics*{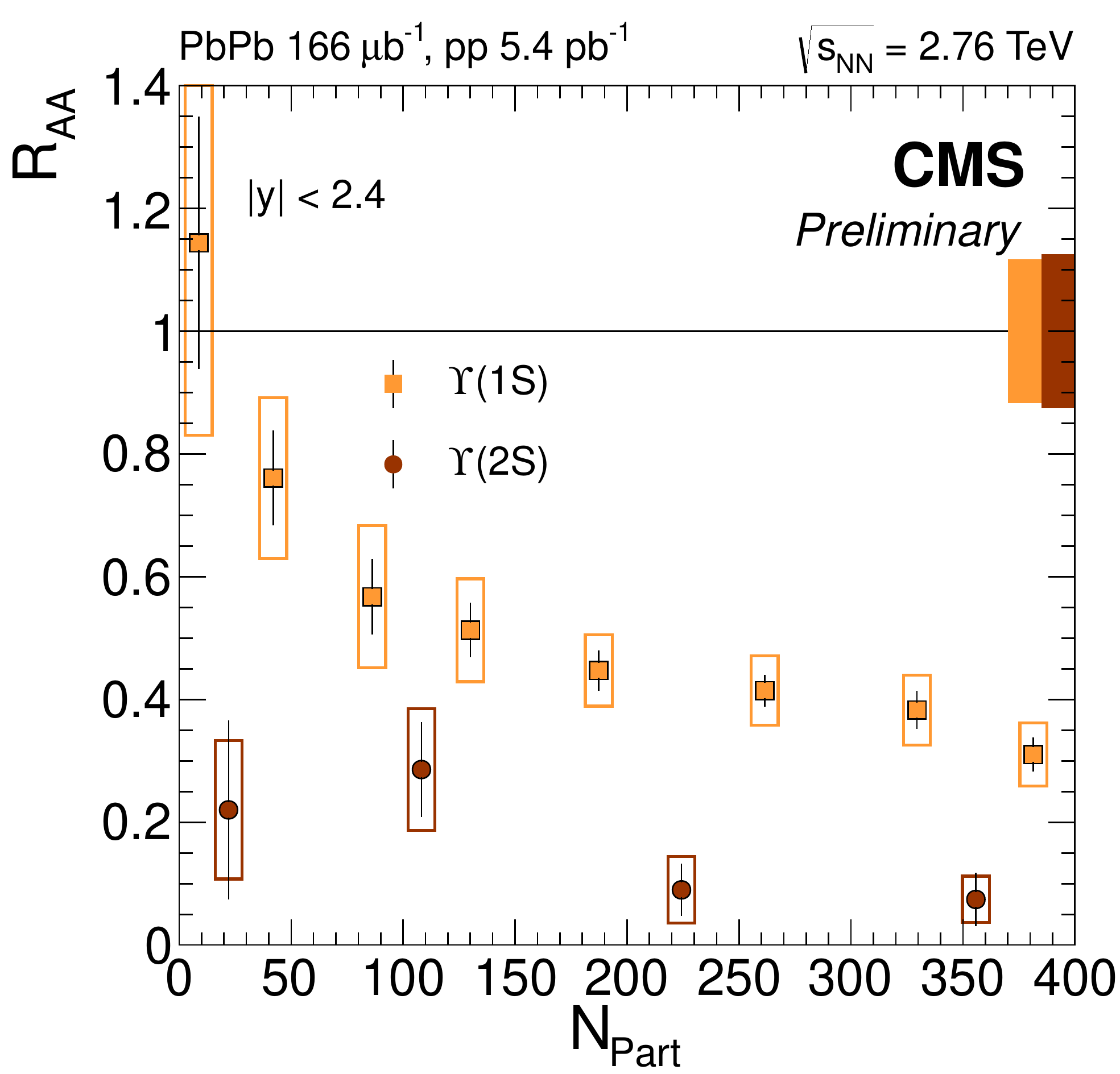}}
\resizebox{0.3\textwidth}{!}
{\includegraphics*{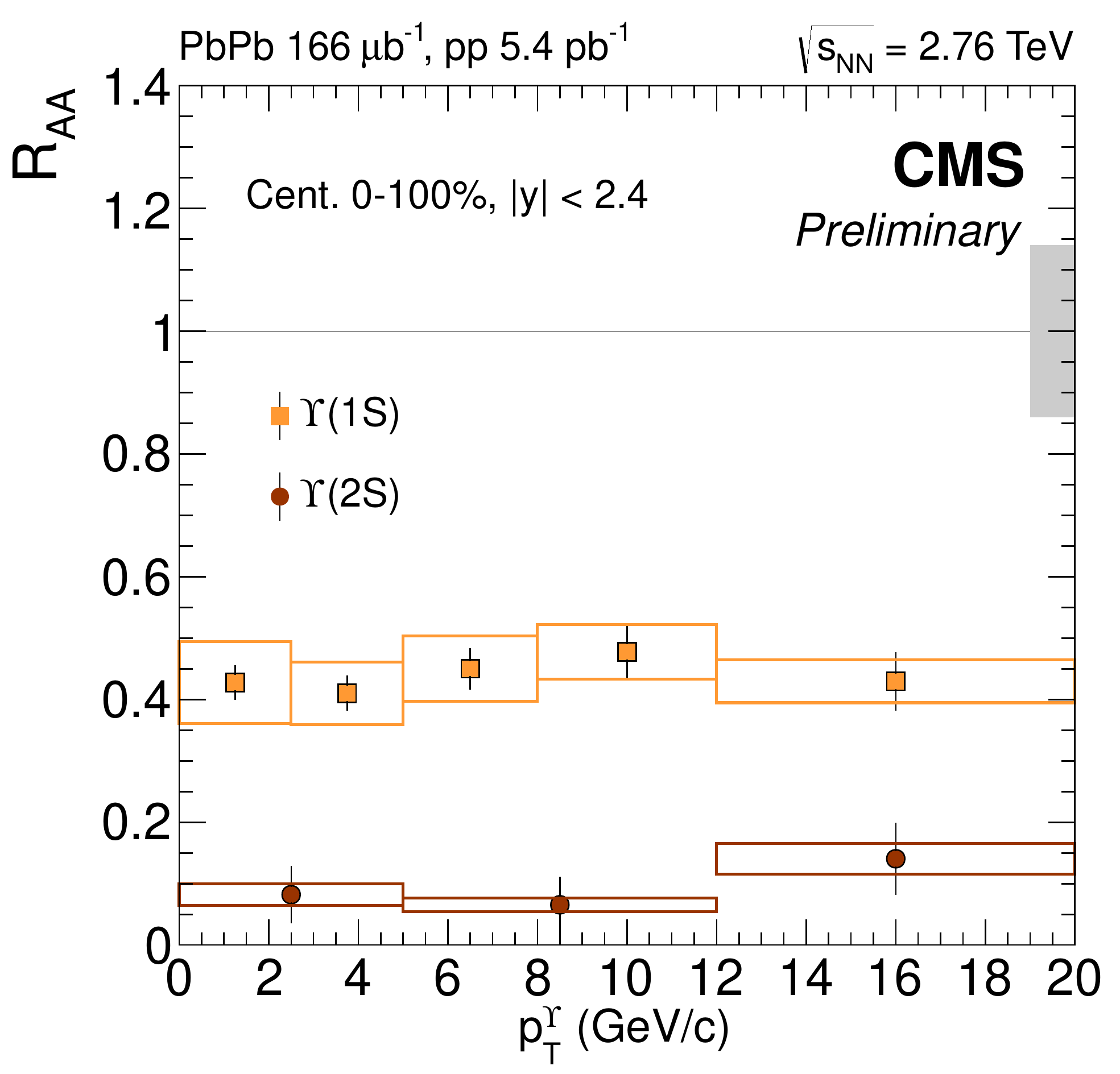}}
\resizebox{0.3\textwidth}{!}
{\includegraphics*{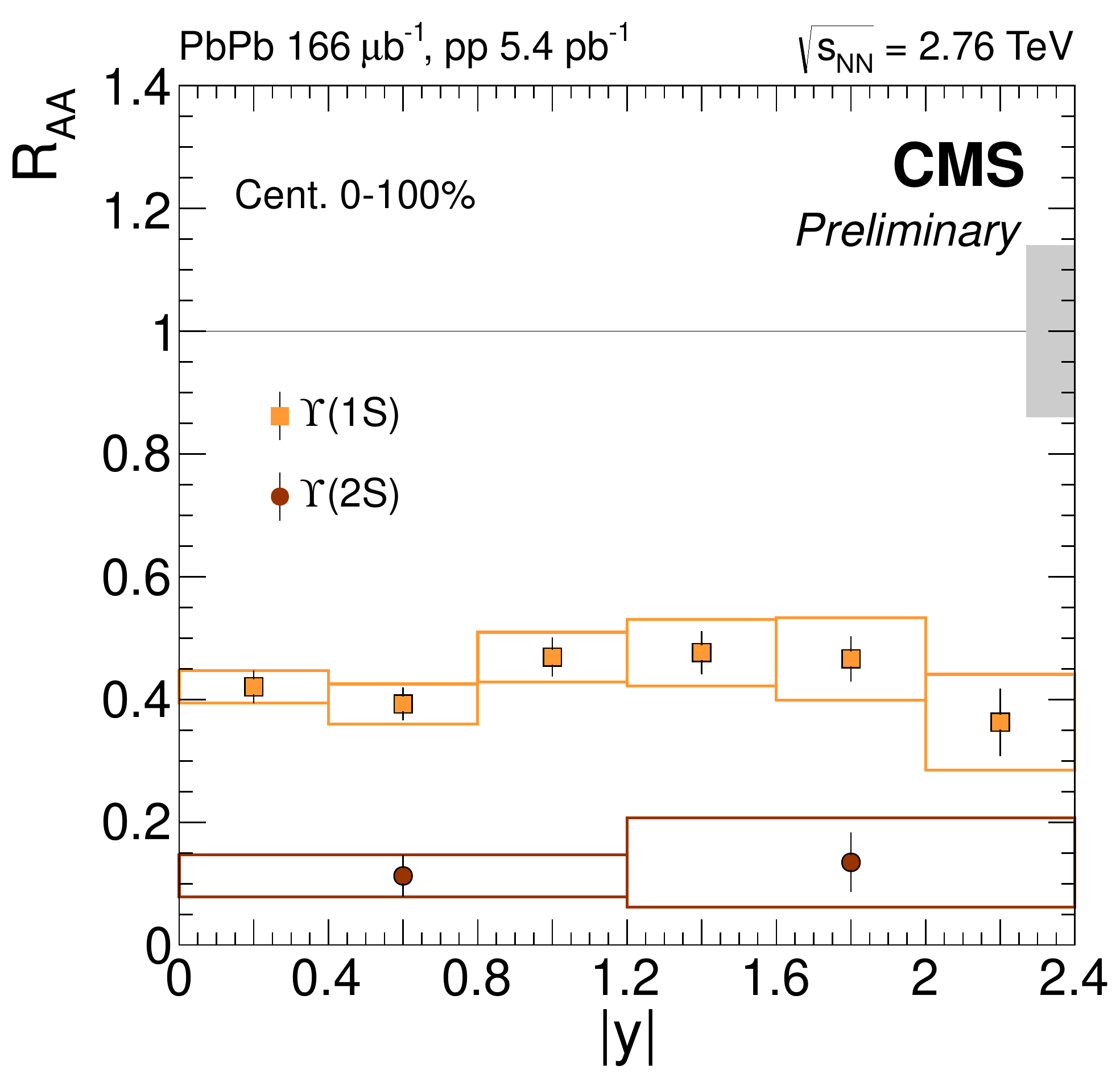}}
\caption{The $\Upsilon(1S)$ and $\Upsilon(2S)$ $R_{\rm AA}$ as a function of $N_{\rm part}$ (left), $y$ (center), $p_{\rm T}$ (right), measured by CMS~\cite{CMS:PASHIN15001}.} 
\label{fig:UpsilonCMS}
\end{figure}

In Fig.~\ref{fig:UpsilonCMS} (left) preliminary results on the nuclear modification factor for $\Upsilon(1S)$ and $\Upsilon(2S)$, measured by CMS in $|y|<2.4$, are shown as a function of $N_{\rm part}$~\cite{CMS:PASHIN15001}. A strong suppression, increasing with centrality, is seen, in particular for the relatively less bound 2S state. Such a behaviour is consistent with the observation of the {\it sequential} suppression of the quarkonium states according to their binding energy~\cite{Digal:2001ue}. One should note that a significant fraction (originally estimated to be $\sim 50$\% by CDF~\cite{Affolder:1999wm}, but likely  to be $\sim$30\% following more recent LHCb results~\cite{Aaij:2014caa}) of measured $\Upsilon(1S)$ states comes from the decay of higher-mass bottomonium resonances ($\Upsilon(2S)$, $\Upsilon(3S)$, $\chi_{\rm b}$) so that a large fraction of the observed 1S suppression is connected to such feed-down effects. The contribution of regeneration effects is expected to be much smaller in the bottomonium sector, due to the lower $b{\overline b}$ yields compared to charm. In Fig.~\ref{fig:UpsilonCMS} (center, right) the $p_{\rm T}$- and $y$-dependence of $R_{\rm AA}$ are shown~\cite{CMS:PASHIN15001}. No significant dependence on the two variables can be seen, within uncertainties. It is remarkable that also ALICE results on $\Upsilon(1S)$, extending the $y$-coverage up to $y=4$, exhibit the same level of suppression~\cite{Abelev:2014nua}. Theoretical models do not still reproduce quantitatively the size of the suppression and/or its rapidity dependence~\cite{Grandchamp:2005yw,Emerick:2011xu,Strickland:2012cq}.
   
\subsection{Electroweak bosons and large transverse momentum photons}
\label{EW}

High transverse momentum isolated photons have been measured at the LHC by CMS \cite{Chatrchyan:2012vq} and ATLAS \cite{Aad:2015lcb}. Electroweak bosons have been measured in PbPb by ATLAS \cite{Aad:2010aa,Aad:2012ew,Aad:2014bha} and CMS \cite{Chatrchyan:2011ua,Chatrchyan:2012nt,Chatrchyan:2014csa} and in pPb by LHCb \cite{Aaij:2014pvu}, CMS \cite{Khachatryan:2015hha} and ATLAS \cite{Aad:2015gta}.

\begin{figure}[htbp]
\begin{center}
\includegraphics[width=0.58\textwidth]{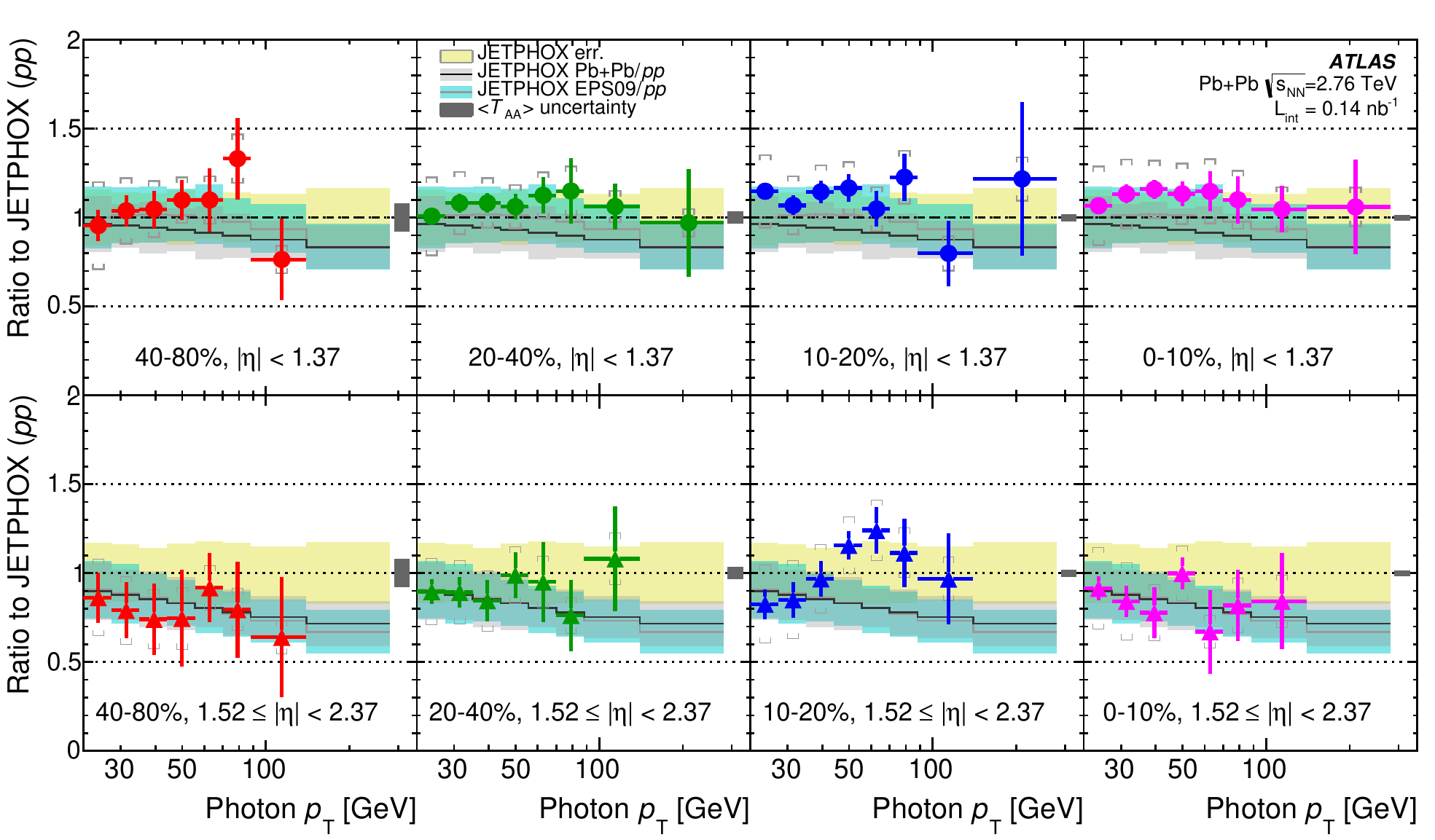}\hskip 0.3cm\includegraphics[width=0.38\textwidth]{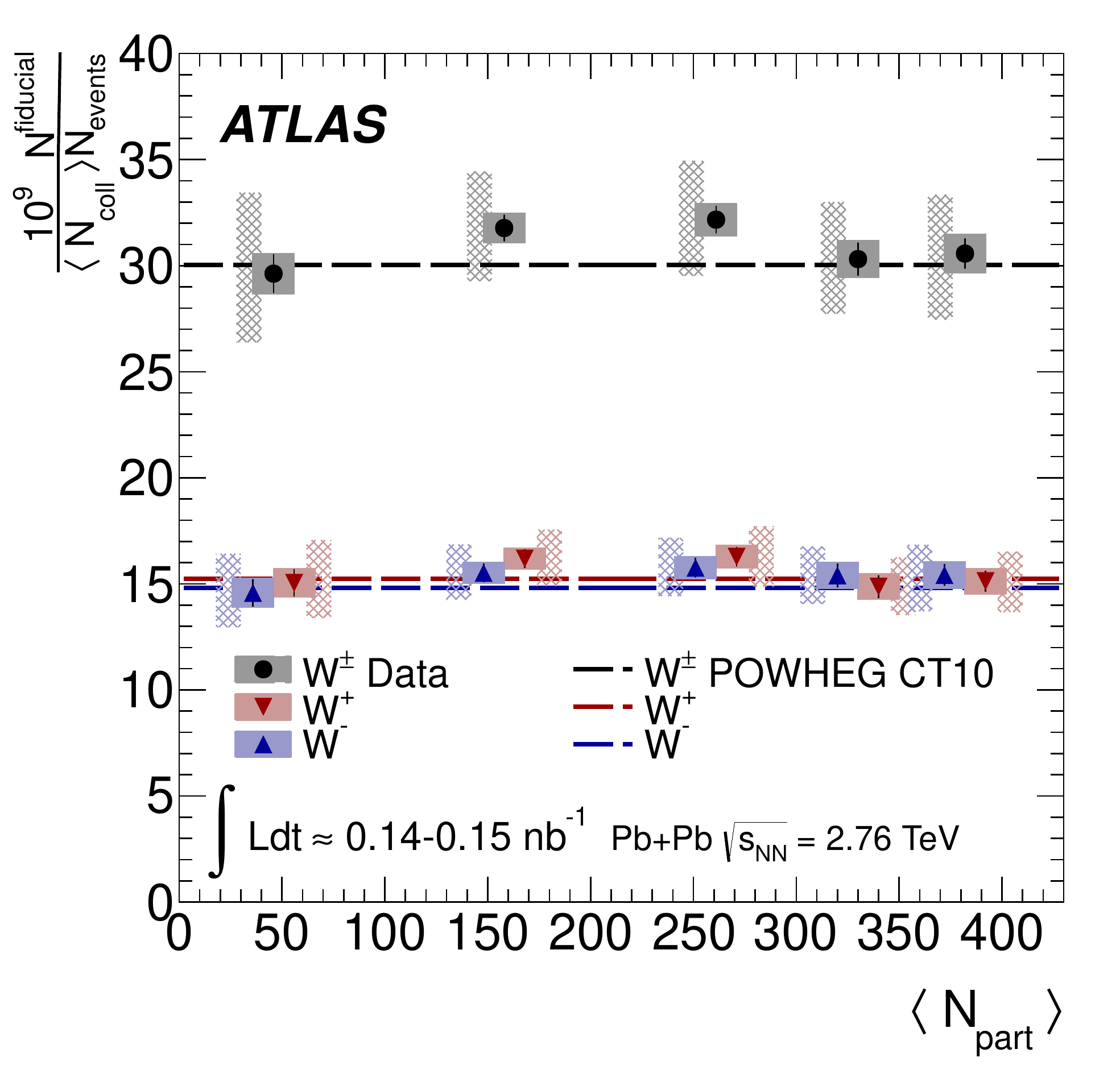}
\end{center}
\caption{Left: fully corrected normalized yields of prompt photons by ATLAS as a
  function of $p_{\rm T}$ in $|\eta|<1.37$ and $1.52\leq |\eta|<2.37$ using
  tight photon selection, isolation cone size $R_{iso} = 0.3$ and
  isolation transverse energy of less than 6 GeV, divided by
JETPHOX predictions for pp collisions, which implement the same
  isolation selection.  The combined scale and PDF uncertainty on the
  JETPHOX calculation is shown by the yellow area.  In addition two
  other JETPHOX calculations are shown, also divided by the
  pp results: PbPb collisions with no nuclear modification (solid
  black line), and PbPb collisions with EPS09 nuclear modifications
  (blue area).  Statistical uncertainties are shown by the bars.
  Systematic uncertainties on the photon yields are combined and shown
  by the upper and lower braces, and the uncertainty of the nuclear overlap in grey. Right: W boson production yield per binary collision as a function of
the mean number of participants $N_{part}$ for W$^+$, W$^-$
and W$^\pm$ bosons
for combined muon and electron channels, in PbPb collisions at 2.76 TeV/nucleon from ATLAS. Taken from \cite{Aad:2015lcb} and \cite{Aad:2014bha}.} 
\label{fig:ew1}
\end{figure}

In PbPb collisions, the main use of such measurements is as a control probe of our understanding of the calibration of the initial
flux of hard probes that goes through the medium. Such a flux is calculated in perturbative QCD without modification due to the 
presence of such dense partonic medium. Note that in perturbative QCD we assume that collinear factorisation works, with the scaling between pp and AA collisions given by the number of binary collisions (or the average nuclear thickness) computed in the Glauber model. This is illustrated in Fig. \ref{fig:ew1} where it is shown how the nuclear modification factor of high $p_{\rm T}$ photons is compatible with unity and with perturbative QCD calculations, and how $W$ production at different centralities scales very well with the number of collisions.
%
%\begin{figure}[htbp]
%\begin{center}
%\includegraphics[width=0.48\textwidth]{Figures/EWbosons/RAACentral}\hskip 0.3cm\includegraphics[width=0.48\textwidth]{Figures/EWbosons/fig_com_08_paper}
%\end{center}
%\caption{Left: nuclear modification factor for isolated photons versus photon transverse momentum, in central PbPb collisions at 2.76 TeV/nucleon from CMS, compared to theoretical calculations. Right: W boson production yield per binary collision as a function of
%the mean number of participants $N_{part}$ for W$^+$, W$^-$
%and W$^\pm$ bosons
%for combined muon and electron channels, in PbPb collisions at 2.76 TeV/nucleon from ATLAS. Taken from \cite{Chatrchyan:2012vq} and \cite{Aad:2014bha}.} 
%\label{fig:ew1}
%\end{figure}

\begin{figure}[htbp]
\begin{center}
\includegraphics[width=0.46\textwidth]{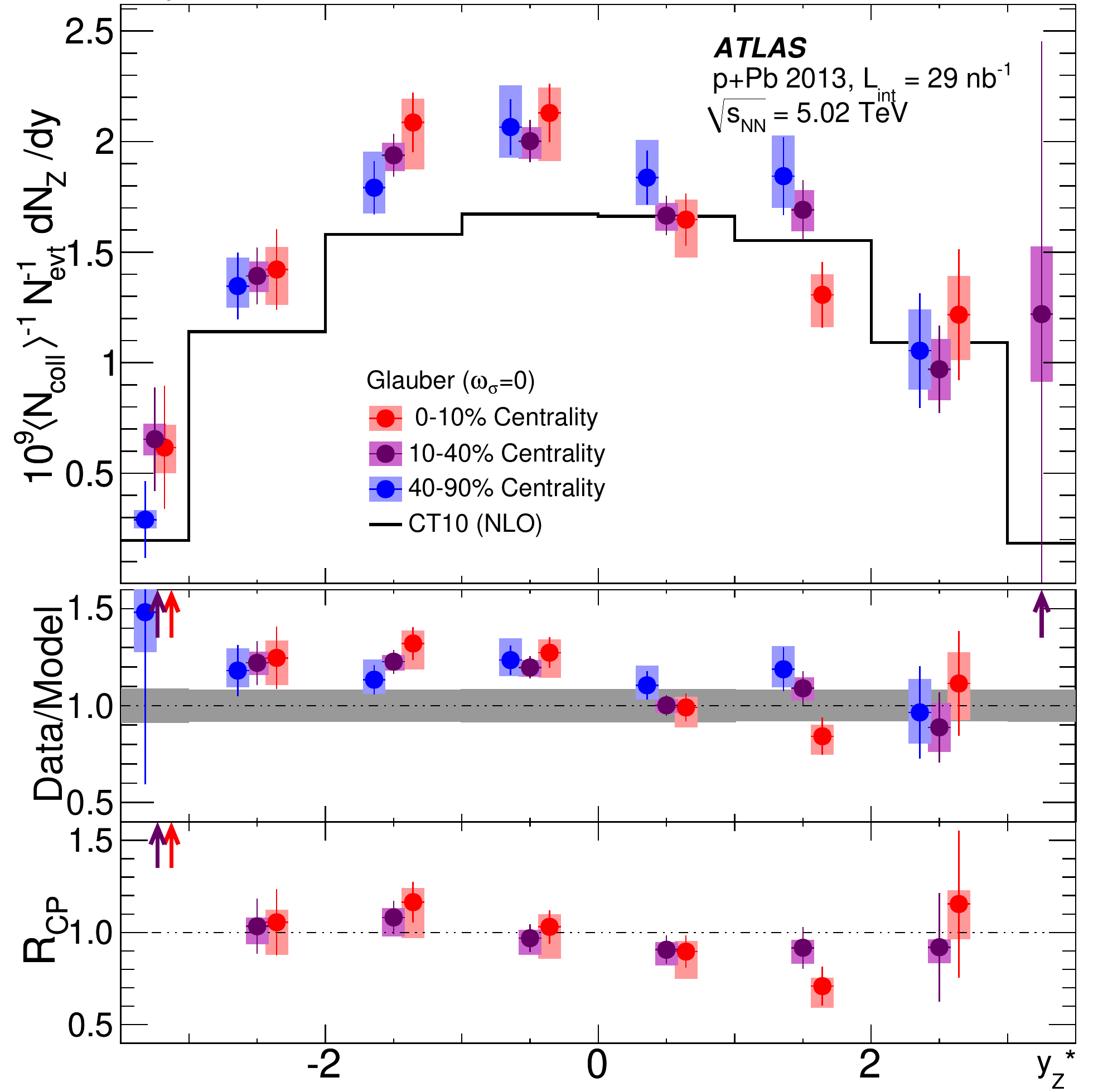}\hskip 0.3cm\includegraphics[width=0.50\textwidth]{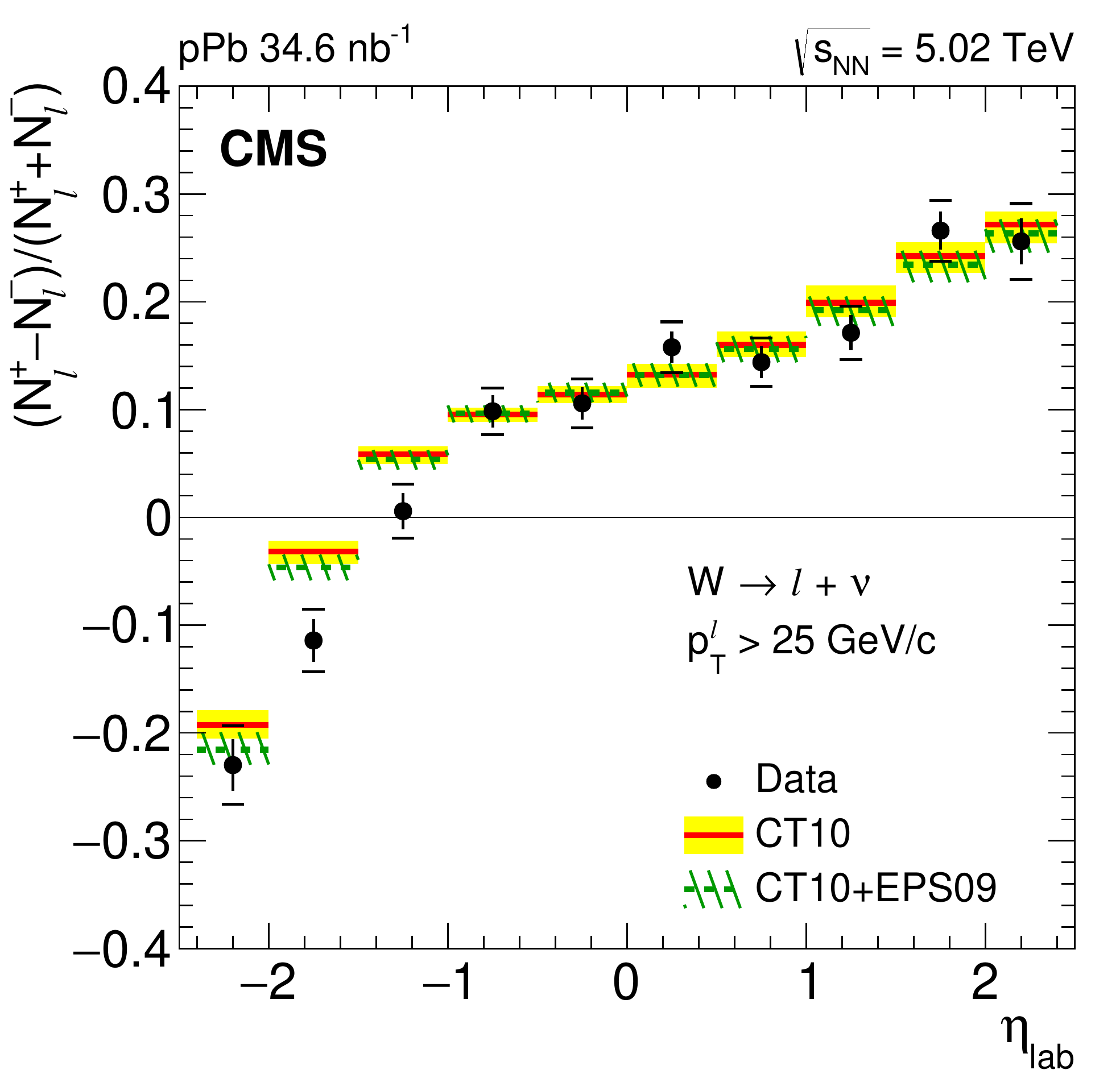}
\end{center}
\caption{Left: Top panel: The rapidity differential $Z$ boson yields by ATLAS, scaled by $\langle N_{coll}\rangle$, for three centrality ranges compared with the CT10 model calculation.  The bars indicate statistical uncertainty and the shaded boxes systematic uncertainty.   The scale uncertainty stemming from the centrality calculation for each bin is included in the systematic uncertainty.  The uncertainty associated with the model is not plotted.  Middle panel: The ratios of the data to the model.  The uncertainty of the model added in quadrature with the scale uncertainty due to uncertainty in the inclusive $NN$ cross section is shown as a band around unity.  Bottom panel: $R_{CP}$. The 0--10\% and 40--90\% centrality points are offset for visual clarity.  The arrows in the lower panels indicate values outside the plotted axes. Right: lepton charge asymmetry $(N_{l^+}-N_{l^-})/(N_{l^+}+N_{l^-})$ for $W\to l$, versus lepton pseudorapidity (with the proton going towards positive pseudorapidities), from CMS, compared to theoretical calculations. Taken from \cite{Aad:2015gta} and \cite{Khachatryan:2015hha}.} 
\label{fig:ew2}
\end{figure}

In pPb collisions, the main use of these observables is for constraining the nuclear modification of parton densities  \cite{Paukkunen:2010qg,Arleo:2011gc}. In Fig. \ref{fig:ew2} we show some LHC results compared to predictions of perturbative QCD containing the state-of-the-art nuclear modification of parton densities \cite{Eskola:2009uj}. In the backward (Pb-going) region, some discrepancies between theory and data can be observed, pointing to deficiencies in the current nuclear parton densities, most probably related with the assumption of isospin independence for their nuclear modifications. These data, together with those from jets, will be used for improving the nPDFs.
%
%\begin{figure}[htbp]
%\begin{center}
%\includegraphics[width=0.54\textwidth]{Figures/EWbosons/Summary}\hskip 0.3cm\includegraphics[width=0.42\textwidth]{Figures/EWbosons/gch}
%\end{center}
%\caption{Left: cross section for $Z$ production decaying into muons, in the backward (Pb-going) and forward (p-going) regions, in pPb collisions at 5.02 TeV/nucleon, from LHCb, compared to theoretical calculations. Right: lepton charge asymmetry $(N_{l^+}-N_{l^-})/(N_{l^+}+N_{l^-})$ for $W\to l$, versus lepton pseudorapidity (with the proton going towards positive pseudorapidities), from CMS, compared to theoretical calculations. Taken from \cite{Aaij:2014pvu} and \cite{Khachatryan:2015hha}.} 
%\label{fig:ew2}
%\end{figure}

\subsection{Low transverse momentum photons and dileptons}
\label{lowptphotons}

The measurement of the electromagnetic probes (real and virtual photons) plays an important role since these particles do not interact strongly with the medium 
and carry undistorted information on the various stages of the collision history including the early ones. In particular, low $p_{\rm T}$ thermal photons (or dileptons) 
from the equilibrated QGP phase can give information on the early temperature of the system. Dielectrons/dimuons, thanks to
their coupling with low-mass resonances and in particular with the $\rho$-meson, can also be sensitive to the partial restoration of chiral symmetry which may induce 
variations in the spectral functions of the $\rho$ and its chiral partner $a_1$~\cite{Koch:1997ei}. At SPS energy, the most accurate results come from the NA60 experiment which observed
a strong broadening of the $\rho$ peak~\cite{Arnaldi:2006jq}, with no mass shift and an enhancement, in the mass region between the $\phi$ and the J/$\psi$, which could be ascribed to thermal 
dimuon production corresponding to a temperature $T\sim 190$ MeV, above the critical temperature for deconfinement~\cite{Arnaldi:2007ru}. At RHIC energies, 
by comparing the measured 
invariant mass dilepton spectrum with the expected yield from hadronic decays (commonly referred to as the ``hadronic cocktail''), a clear excess has been singled out in the low-mass 
region below the $\rho$ and $\omega$. Results from STAR~\cite{Adamczyk:2013caa} are well reproduced by models~\cite{Rapp:2000pe,Linnyk:2011vx} which were able to describe the NA60 data. Very recent results from PHENIX~\cite{Adare:2015ila} are consistent with STAR measurements. By treating the PHENIX 
low-mass excess below $m_{\rm e^+ e^-}=0.03$ GeV/c$^2$ as photon internal conversions~\cite{Adare:2008ab}, the slope
of their transverse momentum distribution is compatible with $T\sim 220$ MeV and can be reproduced by hydrodynamical models~\cite{Turbide:2003si} with initial temperatures in the 
range 300-600 MeV.

\begin{figure}[htbp]
\centering
\resizebox{0.45\textwidth}{!}
{\includegraphics*{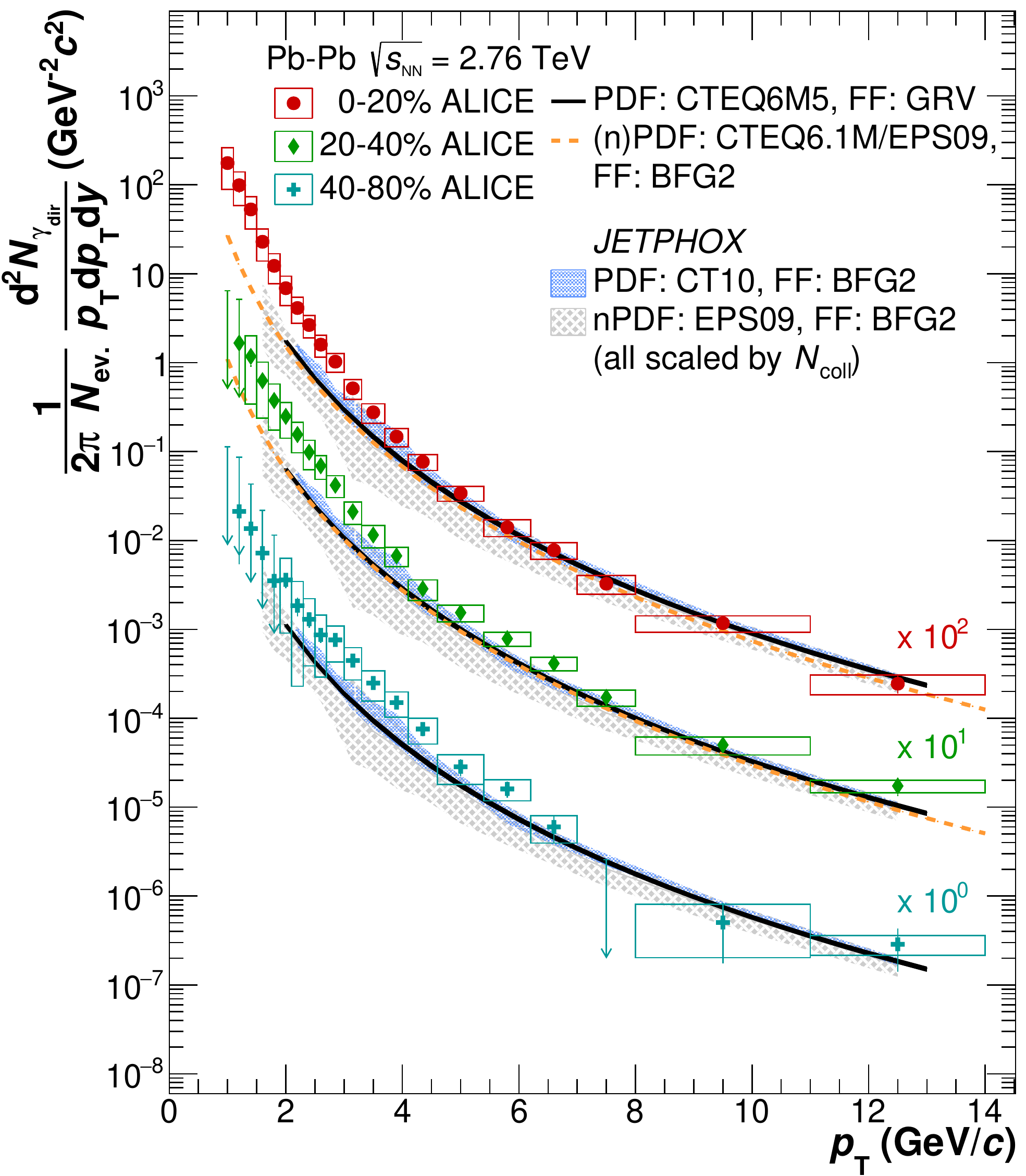}}
%\resizebox{0.43\textwidth}{!}
%{\includegraphics*{Figures/photons/DirectPhotonv2_prelim_withTheory}}
\caption{Direct photon spectrum measured by ALICE for PbPb collisions at $\sqrt{s_{\rm NN}}=2.76$ TeV~\cite{Adam:2015lda}. Comparisons with NLO pQCD calculations are shown.}
%Right: direct photon $v_2$ from ALICE for 0-40\% PbPb %collisions~\cite{Morreale:2014spa}.} 
\label{fig:photons}
\end{figure}

At the LHC, such measurements are made much more difficult by the large background levels, mainly connected with the large increase in the combinatorial decays of 
charged pions/kaons and heavy flavour for dilepton measurements, and from $\pi^0$ decays for real photons. For the moment, analyses of the dilepton invariant mass 
spectra have been carried out by ALICE in pp and \mbox{pPb} collisions, and the results have been compared with hadronic cocktail calculations~\cite{Koehler:2014dba}. 
Within uncertainties, an agreement is observed. In pp collisions, an extraction of the direct photon yield starting from the dilepton spectrum in the low-mass region, gives results
consistent with NLO pQCD calculations~\cite{Koehler:2014dba}. In PbPb collisions, no results have been provided up to now, and a significant improvement of the 
experimental conditions is expected for Run 3, due to the foreseen upgrades of the ALICE Inner Tracking System and Time Projection Chamber~\cite{Reichelt:2014uca}. 
Direct photon studies have been carried out by ALICE by using photon conversions to $e^+e^-$ or a calorimetric measurement~\cite{Adam:2015lda}. 
The direct photon yield $\gamma_{\rm direct}$ was extracted subtracting the decay contribution 
$\gamma_{\rm decay}$ from the inclusive spectrum $\gamma_{\rm incl}$ as $\gamma_{\rm direct} = \gamma_{\rm incl}\times (1-\gamma_{\rm decay}/\gamma_{\rm incl})$. NLO pQCD calculations~\cite{Klasen:2013mga,Gordon:1993qc,Vogelsang:1997cq,Paquet:2015lta} well describe the spectrum above $p_{\rm T}\sim 5$ GeV/$c$, while in the region $0.9<p_{\rm T}< 2.1$ GeV/$c$ a 2.6$\sigma$ excess is found for the 0-20\% centrality class (see Fig.~\ref{fig:photons}). It is compatible with a thermal slope with $T=297 \pm 12({\rm stat})\pm 41({\rm syst})$ MeV. Preliminary results~\cite{Morreale:2014spa} show a non-zero $v_2$ for direct photons, possibly reflecting the development of collective expansion at early stages. 

\subsection{Ultraperipheral collisions}
\label{UPC}

Ultraperipheral collisions \cite{Baltz:2007kq}, usually analysed for exclusive final states, offer unique possibilities for understanding the production mechanisms and for determining parton densities in a hadron collider that otherwise could only be studied in lepton-hadron/nucleus machines. In nucleus-nucleus collisions, one or both nuclei act as quasireal photon sources that collide with the other nucleus or photon. Therefore, $\gamma$A collisions can be studied, in which the nucleus remains intact (coherent production) or gets excited and dissolves but without filling the rapidity gap between the nucleus and the produced particle (incoherent production).

\begin{figure}[htbp]
\begin{center}
\includegraphics[width=0.48\textwidth]{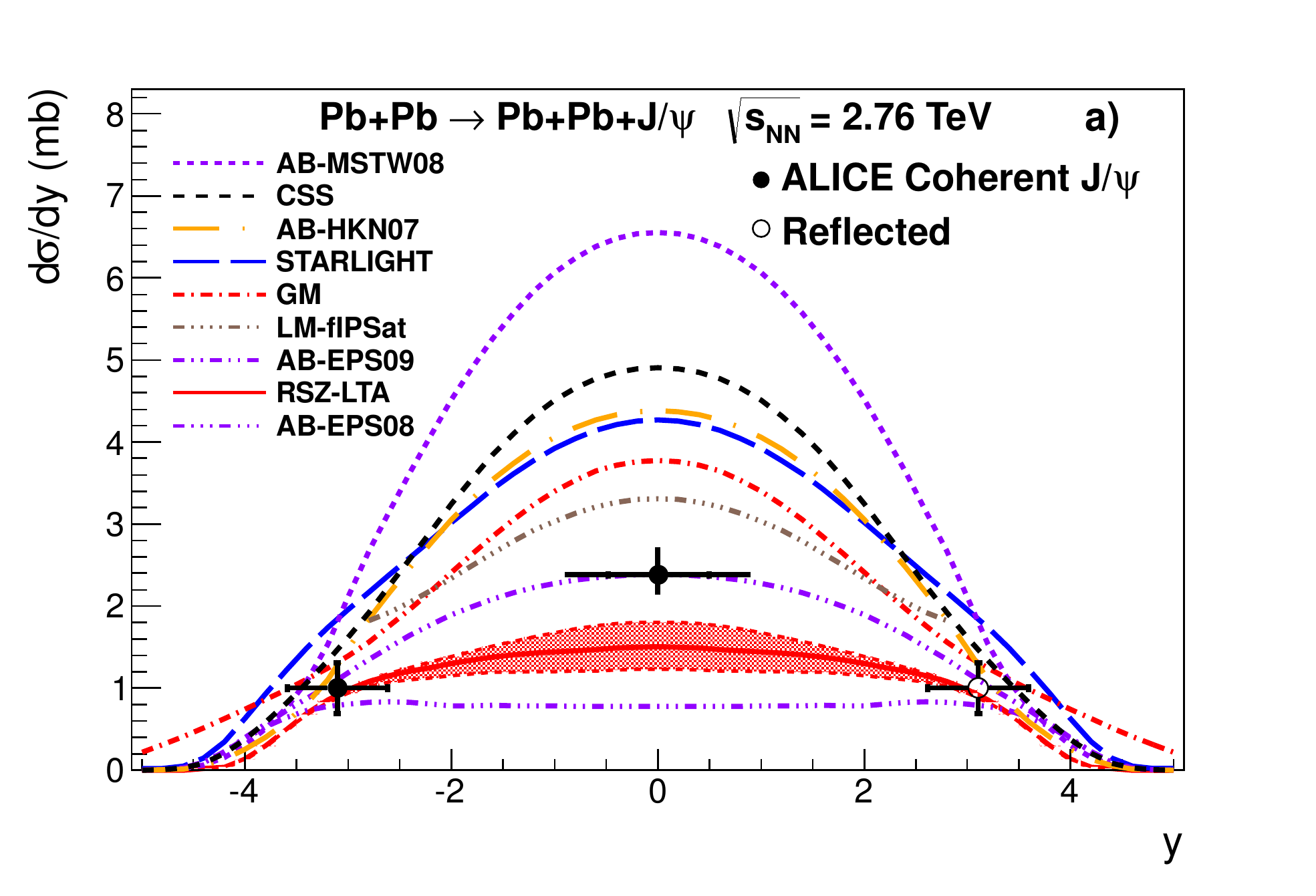}\hskip 0.3cm\includegraphics[width=0.48\textwidth]{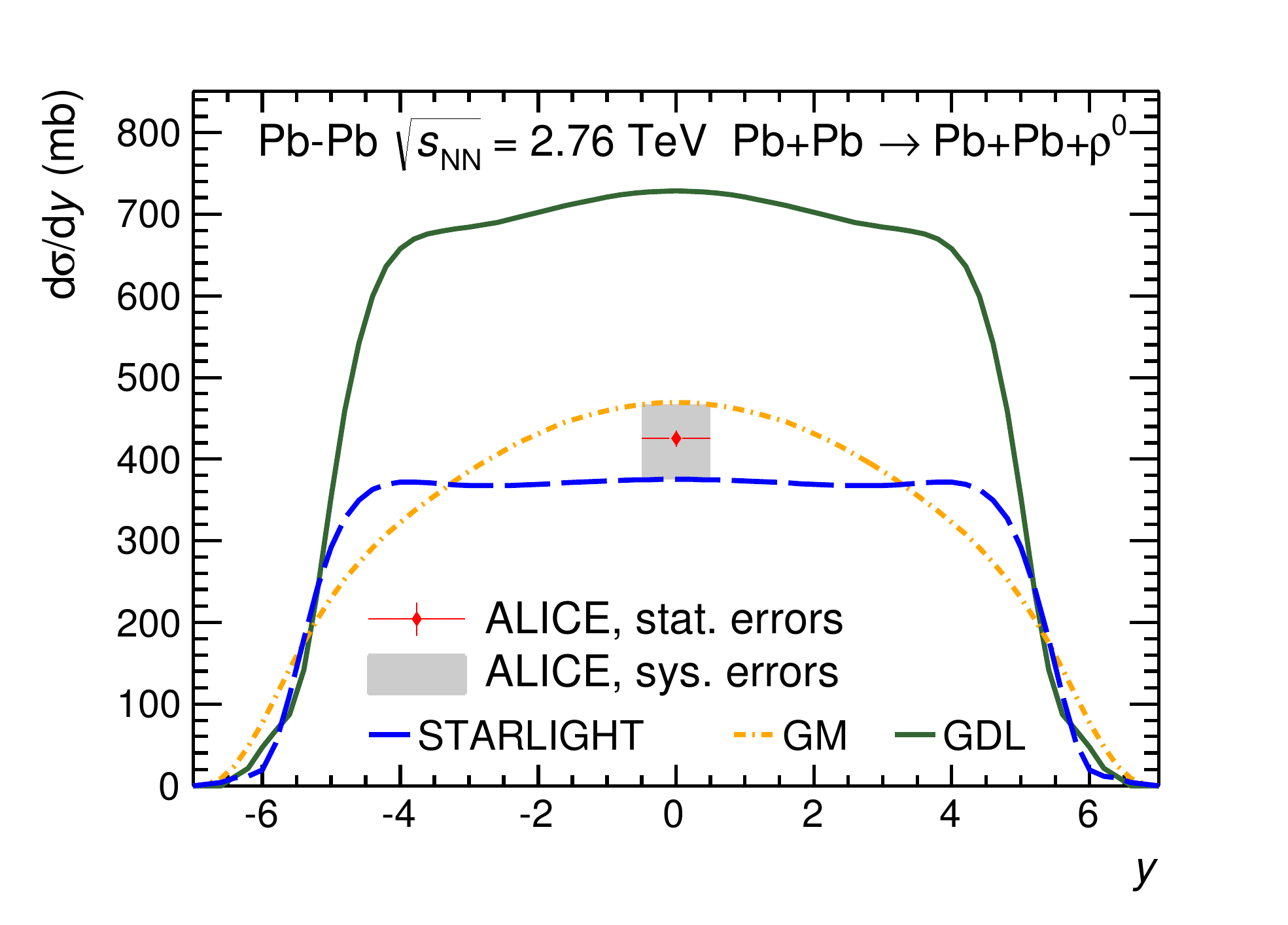}
\end{center}
\caption{Left: cross section for coherent exclusive $J/\psi$ production versus rapidity in ultraperipheral PbPb collisions at 2.76 TeV/nucleon from ALICE, compared to theoretical calculations. Right: cross section for coherent exclusive $\rho^0$ production versus rapidity in ultraperipheral PbPb collisions at 2.76 TeV/nucleon from ALICE, compared to theoretical calculations. Taken from \cite{Abbas:2013oua} and \cite{Adam:2015gsa}.} 
\label{fig:upc1}
\end{figure}

ALICE has measured exclusive coherent and incoherent charmonium production \cite{Abelev:2012ba,Abbas:2013oua,Adam:2015sia} and exclusive coherent $\rho^0$  production \cite{Adam:2015gsa}, see Fig. \ref{fig:upc1}. Data show a large discriminatory power on models\footnote{Due to lack of space we cannot provide a description of the different models. Let us simply note that STARLIGHT, a simulator widely used by the experimental collaborations, is a model that contains no shadowing of nPDFs.}  and, in the case of charmonium, have been taken as a first direct evidence of the existence of gluon shadowing in nuclei \cite{Guzey:2013xba}. They have also been used to test dipole models that describe data in lepton-hadron collisions \cite{Ducati:2015jfa,Lappi:2013am}.

 Finally, the photon flux from Pb has been used in pPb  \cite{TheALICE:2014dwa} to study the energy dependence of $J/\psi$ production in $\gamma$p collisions with results competitive with those obtained through pp \cite{Aaij:2013jxj,Aaij:2014iea}.

\section{Summary}
\label{summary}

In this manuscript we have overviewed the main results obtained in pPb and PbPb collisions at the Large Hadron Collider during Run 1. We have first discussed the results for soft probes of the medium created in the collisions and then turned to hard and electromagnetic probes.
We have focused on a selection of experimental results and, due to space limitations, the main current theoretical explanations have just been briefly presented.

Overall, the physical picture of the medium created in PbPb collisions that arises from the results in Run 1 is qualitatively similar to that obtained at RHIC: a system with a large energy density and strong collectivity, behaving very early close to a nearly ideal liquid and very opaque to energetic coloured particles that traverse it. But the larger collision energy provided by the LHC has led to a longer-lived medium with a larger energy density and to a more abundant rate of hard probes. In this situation, and thanks to the sophisticated detector systems and new experimental techniques, many of the observations at RHIC have been made again at the higher energy, with substantially increased precision. This is particularly true for azimuthal asymmetries where the higher energy density leads to higher multiplicities which, in association with the much larger detector acceptances, provides dramatically improved measurements of anisotropies. In particular, spectacular progress has been carried out in determining them on an event-by-event basis. Also for hard probes, where particles with higher masses and transverse energies can be precisely measured, new density effects appear that affect the production and in-medium dynamics of quarkonia, and new observables can be analysed due to the excellent performance of the detectors.

This new era of precision is triggering numerous theoretical efforts. First, data on soft and electromagnetic probes are stimulating new developments on: (i) the description of the initial state \cite{Albacete:2014fwa}, (ii) the understanding of the weak or strong dynamics leading to a stage that is isotropised and thermalised enough for relativistic hydrodynamics to be applicable  \cite{Chesler:2015lsa,Gelis:2015gza}, and (iii) the relation between transport and hydrodynamics and the extraction of the transport coefficients characterising the medium \cite{Jeon:2015dfa}. Second, measurements on hard probes are leading to new developments on our understanding of: (i) energy loss processes for both light and heavy partons in order to get a unified description of particle and jet inclusive and differential observables \cite{Ghiglieri:2015zma,Blaizot:2015lma,jqqgp5}, and (ii) the formation of quarkonium bound states and the different physical mechanisms through which the presence of a coloured medium may affect them \cite{Ding:2015ona,Andronic:2015wma}. All these elements are required for a quantitative characterisation of the new phase of matter produced in the collisions with controlled uncertainties. They go in parallel with new developments in lattice and perturbative QCD at finite temperature, and with our understanding of the benchmark for soft and hard probes.

Besides, the pPb runs have brought the bonus of the apparent collective behaviour of such collisions, that may also be at play in  high-multiplicity pp ones. These findings raise new questions about the nature, thermal or not, and onset of such collectivity, together with new opportunities for benchmarking for hard processes like the determination of nPDFs, and for understanding of QCD at large partonic densities. The latter point naturally links the studies at the LHC with the proposals of future electron-ion colliders \cite{Accardi:2012qut,AbelleiraFernandez:2012cc}.

With the increasing statistics and larger energy that will be achieved in Run 2 and, further in the future, with the detector upgrades~\cite{Abelevetal:2014cna,atlasupgrade,Hoepfner:2015jba,LHCb:2011dta} that will become operative in Run 3 and afterwards, heavy-ion physics will continue to play a very important role in the LHC programme, leading to a complete and quantitative comprehension of the high-temperature QGP phase of QCD matter.

\section*{Acknowledgements}

 We thank John Jowett for useful discussions and for providing Fig. \ref{fig:introduction1}, and Anton Andronic, Elena Ferreiro, Matthew Luzum,  Andreas Morsch, Matthew Nguyen and Peter Steinberg for a critical reading of the manuscript and most useful suggestions. The work of NA was supported by the People Programme (Marie Curie
Actions) of the European Union Seventh Framework Programme FP7/2007-2013/ under
REA grant agreement \#318921, the European Research Council
grant HotLHC ERC-2011-StG-279579, Ministerio de Ciencia e Innovaci\'on of Spain under
project FPA2014-58293-C2-1-P, Xunta de Galicia (Conseller\'{\i}a de Educaci\'on and Conseller\'{\i}a de Innovaci\'on
e Industria - Programa Incite), the Spanish Consolider-Ingenio 2010 Programme
CPAN and FEDER.

%
% For  figures use
%\begin{figure*}
%\vspace*{5cm}       % Give the correct figure height in cm
%\includegraphics{leer.eps}
%\end{figure*}
% or  this
%\begin{figure}
%\centering
%\resizebox{0.75\textwidth}{!}{%
%  \includegraphics{leer.eps}
%\vspace{5cm}       % Give the correct figure height in cm
%\end{figure}
%\begin{table}
%\centering
%\begin{tabular}{lll}
%\hline\noalign{\smallskip}
%\end{tabular}
% Or use
%\vspace*{5cm}  % with the correct table height
%\end{table}

%
% BibTeX users please use
% \bibliographystyle{}
% \bibliography{}

\end{document}